\documentclass[11pt,a4paper]{article}

\usepackage[T1]{fontenc} 
\usepackage{jheppub}

\usepackage[utf8]{inputenc}
\usepackage{graphicx,dcolumn,subcaption,mathtools}
\usepackage{xcolor}
\usepackage{braket}
\usepackage{siunitx}
\usepackage{appendix}
\usepackage[mathscr]{eucal}
\usepackage{bm}
\usepackage{bbm}

\usepackage{enumitem}

\usepackage{cancel}

\usepackage{footnotebackref}

\usepackage{rotating}

\usepackage[]{float}
\usepackage[font=small, labelfont=bf]{caption}

\usepackage{slantsc}
\usepackage{xspace}

\usepackage{setspace}
\usepackage{etoolbox}

\usepackage{marginnote}
\reversemarginpar  

\allowdisplaybreaks

\makeatletter

\newcounter{uncheckedcount}
\providecommand{\uncheckedtotal}{0} 

\AtEndDocument{%
  \immediate\write\@auxout{%
    \string\gdef\string\uncheckedtotal{\arabic{uncheckedcount}}%
  }%
}
\newcounter{checkedcount}
\providecommand{\checkedtotal}{0} 

\AtEndDocument{%
  \immediate\write\@auxout{%
    \string\gdef\string\checkedtotal{\arabic{checkedcount}}%
  }%
}
\makeatother

\allowdisplaybreaks

\definecolor{ReadableGreen}{RGB}{34,139,34} 
\definecolor{ReadableBlue}{RGB}{65,105,225} 
\definecolor{FixPink}{RGB}{236, 0, 140} 
\newcommand{\bluemath}[1]{\textcolor{ReadableBlue}{#1}}
\newcommand{\greenmath}[1]{\textcolor{ReadableGreen}{#1}}

\renewcommand{\thefigure}{\arabic{figure}} 
\newcommand{\rd}{\mathrm{d}}
\newcommand{\secref}[1]{Section~\ref{#1}}

\newcommand{\fix}[1]{
    \ifmmode
        \mathcolor{FixPink}{#1}
    \else
    \begingroup
        \color{FixPink}#1
    \endgroup
    \fi}

\newcommand{\mytext}[1]{\text{#1}}
\newcommand{\sumprime}{\mathop{\sum\raisebox{1.5ex}{\scriptsize\(\prime\)}}}

\DeclareMathOperator{\Tr}{Tr}

\newcommand\dki{[\rd k_i]}
\newcommand\dkj{[\rd k_j]}

\newcommand\dktj{[\rd \tilde{k}_j]}

\newcommand\krec{k_{\mathrm{rec}}}
\newcommand\brec{\beta_{\mathrm{rec}}}
\newcommand\nrec{n_{\mathrm{rec}}}
\newcommand\nbrec{\bar{n}_{\mathrm{rec}}}
\newcommand{\bbbone}{\boldsymbol{\mathbbm{1}}}
\newcommand{\deltabar}{\ensuremath{\,\delta\kern-0.52em\raisebox{0.45ex}{--}}}

\newcommand{\ktness}{{k_t^{\text{ness}}}}

\newcommand{\DeltaET}{\Delta E_t}

\newcommand{\qf}{\mathfrak{q}}
\newcommand{\qft}{\tilde{\mathfrak{q}}}
\newcommand{\qfcut}{\mathfrak{q}_{\mathrm{cut}}}

\newcommand{\as}{\alpha_S}
 
\newcommand\asbare{\alpha_{S}^u}

\newcommand\bF{\mathbf{F}} 

\newcommand{\BDelta}{\mathbf{\Delta}}

\newcommand\bTi{\mathbf{T}_i} 
\newcommand\bTj{\mathbf{T}_j}
\newcommand\bTk{\mathbf{T}_k}

\newcommand\amp{\mathcal{M}}

\newcommand{\Amunu}{ \A^{\mu\nu}}

\newcommand{\bJ}{\mathbf{J}}
\newcommand{\bJJ}{\mathbf{J\!\!J}}
\newcommand{\bcalJJ}{\bm{\mathcal{J}\!\!\!\!\mathcal{J}}}
\newcommand{\bcalPhat}{\bm{\hat{\mathcal{P}} }}
\newcommand{\bb}{\mathbf{b}}

\newcommand{\bT}{\boldsymbol{T}}

\def\bb{\mathbf{b}}

\newcommand{\A}{\mathcal{A}}

\newcommand{\F}{\mathcal{F}}

\newcommand{\J}{\mathcal{J}}

\newcommand{\calPhat}{\hat{\mathcal{P}} }

\newcommand{\BB}{\mathcal{B}\!\!\mathcal{B}}
\newcommand{\BS}{%
  \mathrel{\raisebox{0.0ex}{\scalebox{1.0}{$\mathcal{B}$}}%
  \mkern-3mu \mathcal{S}}}
\newcommand{\JS}{%
  \mathrel{\raisebox{0.21ex}{\scalebox{0.92}{$\mathcal{J}$}}%
  \mkern-6mu \mathcal{S}}}

\newcommand{\Af}{\mathfrak{A}}

\newcommand{\pf}{\mathfrak{p}}

\newcommand{\Hb}{\boldsymbol{\mathcal{H}}}
\newcommand{\Bb}{\boldsymbol{\mathcal{B}}}
\newcommand{\Sb}{\boldsymbol{\mathcal{S}}}

\newcommand{\MSbar}{\overline{\mathrm{MS}}}
\newcommand{\msbar}{%
    \ifmmode
        \overline{\mathrm{MS}}
    \else
        $\overline{\mathrm{MS}}$
    \fi
}
\newcommand{\SCETI}{%
    \ifmmode
        \mathrm{SCET}_{\mathrm{I}}
    \else
        $\mathrm{SCET}_{\mathrm{I}}$
    \fi
}
\newcommand{\SCETII}{%
    \ifmmode
        \mathrm{SCET}_{\mathrm{II}}
    \else
        $\mathrm{SCET}_{\mathrm{II}}$
    \fi
}
\usepackage{comment}

\allowdisplaybreaks

\renewcommand{\thefigure}{\arabic{figure}}

\title{
Dissecting Exclusive Multijet Cross Sections}

\author{J\"urg Haag}

\affiliation{Albert Einstein Center for Fundamental Physics, Institut für Theoretische Physik,
Universität Bern,
Sidlerstrasse 5, CH-3012 Bern, Switzerland}

\emailAdd{juerg.haag@unibe.ch}

\date{Received: date / Accepted: \today}

\abstract{
This paper studies multijet cross sections in the kinematic limit where one or more of the jets are unresolved. We study the asymptotic expansion of the cross section by using the method of regions. We explain in large generality how to identify the relevant phase space regions and derive the leading-power approximation of the phase space. The leading-power phase space factorizes into a hard phase space and a radiation phase space for each collinear sector and the soft sector. Using the infrared factorization properties of squared matrix elements at leading power, we derive a factorization formula for generic resolution variables describing the exclusive n-jet limit in terms of fully differential beam, jet, and soft functions. We show how the factorization formula can be simplified for specific resolution variables. We regularize rapidity divergences in the beam and jet functions using a time-like reference vector, a method we call the ``$z_N$-prescription'', and demonstrate how the associated zero-bin contributions combine with the soft function to define a soft subtracted function free of rapidity divergences and suitable for numerical evaluation. The $z_N$-prescription can be used both for $\SCETII$- and $\SCETI$-type resolution variables. As an application of our framework, we derive factorization formulas for several transverse-momentum-like variables, and we present a variable that factorizes into simple cumulant functions to all orders in perturbation theory.
}


\begin{document}


\maketitle
\flushbottom


\section{Introduction}
Jet production is central to testing Quantum Chromodynamics (QCD) at high-energy colliders. At the LHC, jet observables are essential for constraining parton distribution functions (PDFs) and determining the strong coupling constant $\as$. These measurements will become even more important in the High-Luminosity phase of the LHC and at future $e^+e^-$ colliders, where clean environments enable high-precision QCD studies.

The simplest jet processes at hadron colliders arise from $2 \rightarrow 2$ parton scattering. Next-to-next-to-leading-order (NNLO) QCD corrections have recently been computed for inclusive jet and dijet observables~\cite{Currie:2016bfm,Currie:2017eqf,Gehrmann-DeRidder:2019ibf,Czakon:2019tmo,Chen:2022tpk}. Among the most complex calculations is the NNLO prediction for three-jet production in $pp$ collisions~\cite{Czakon:2021mjy}. Extending NNLO precision to higher jet multiplicities is challenging. For $e^+e^-$ collisions, N$^3$LO results exist only for dijet production~\cite{Chen:2025kez}, while NNLO predictions are available for three-jet final states~\cite{Gehrmann-DeRidder:2007vsv,Weinzierl:2008iv,DelDuca:2016ily}. Achieving NNLO accuracy for higher jet multiplicities requires further methodological advances.

A useful class of observables for jet processes are \emph{$n$-jet resolution variables}, which quantify how well an event is described by exactly $n$ jets. These variables vanish on the Born phase space and are positive in the presence of additional hard radiation not collinear to a jet or beam. They are sensitive to QCD radiation patterns and provide input for precision QCD studies such as $\as$ determinations. Examples at lepton colliders include $y_{n(n+1)}$ in the Durham algorithm~\cite{Catani:1991hj}, thrust ($1-T$)~\cite{Farhi:1977sg}, and total broadening ($B_T$)~\cite{Catani:1992jc}. At hadron colliders, general $n$-jet resolution variables include $n$-jettiness ($\tau_n$)~\cite{Stewart:2010tn}, $k_T$-ness~\cite{Buonocore:2022mle}, and $q_T$-imbalance using the winner-take-all (WTA) scheme~\cite{Fu:2024fgj}. For 0-jet production, the transverse momentum of the colorless system, $q_T$, and $p_T$-veto are well studied examples of resolution variables.

When a dimensionful resolution variable $\qf$ is much smaller than the hard scale $Q$, the perturbative expansion in $\as$ contains large logarithms $\log(\qf/Q)$. These logarithms must be resummed to all orders to maintain predictive power. Resummation, or the calculation of large logarithms at fixed order, relies on the factorization properties of QCD matrix elements in soft and collinear limits. This leads to factorization formulas in terms of perturbative beam, jet, and soft functions. Leading-power factorization formulas exist for many $n$-jet resolution variables, including $n$-jettiness~\cite{Stewart:2010tn}, $q_T$~\cite{Collins:1984kg,Becher:2010tm}, and $p_T$-veto~\cite{Becher:2012qa}. These formulas enabled the resummation of large logarithms at, and beyond, next-to-next-to-leading-logarithmic${}^\prime$ (NNLL$^\prime$) accuracy~\cite{Bozzi:2005wk,Catani:2000vq,Bozzi:2010xn,deFlorian:2012mx,Stewart:2013faa,Jouttenus:2013hs,Chen:2018pzu,Becher:2012qa,Bizon:2017rah,Bizon:2018foh,Bizon:2019zgf,Alioli:2020fzf,Becher:2020ugp,Camarda:2022qdg,Billis:2021ecs,Ju:2021lah,Re:2021con,Chen:2022cgv,Neumann:2022lft,Campbell:2023cha,Camarda:2023dqn,Moos:2023yfa,Alioli:2023rxx,Piloneta:2024aac,Billis:2024dqq}.

The resolution variables resummed to NNLL$^\prime$ accuracy all have the property that they simplify drastically in the soft and collinear limits. Not all variables share this property. Even commonly used observables, such as the jet-resolution parameter $y_{23}$, retain a large level of complexity in the soft and collinear limits, making high-accuracy resummation challenging. General-purpose parton showers and numerical resummation tools allow resummation to next-to-leading-logarithmic (NLL) accuracy, but no general formalism exists for resummation beyond NLL for arbitrary resolution variables. The framework introduced in this paper provides a starting point for such a formalism.

Even when all-order resummation is not possible, the fixed-order expansion of the leading-power cumulant cross section in a small resolution variable $\qf$ remains useful. The fixed-order expansion can be used to construct slicing schemes for NNLO calculations of arbitrary $n$-jet processes and observables. The $q_T$ slicing method~\cite{Catani:2007vq} has been used to compute NNLO corrections for many color singlet processes~\cite{Grazzini:2008tf, Catani:2009sm, Ferrera:2011bk, Ferrera:2014lca, Ferrera:2017zex, deFlorian:2016uhr, Catani:2011qz, Grazzini:2013bna, Grazzini:2015nwa, Cascioli:2014yka, Grazzini:2015hta, Gehrmann:2014fva, Grazzini:2016ctr, Grazzini:2016swo, Grazzini:2017ckn, Campbell:2016yrh, Heinrich:2017bvg, Campbell:2017aul, Catani:2018krb} and for final states with heavy quarks~\cite{Catani:2019iny, Catani:2020kkl, Catani:2019hip, Catani:2022mfv, Buonocore:2022pqq, Buonocore:2023ljm}. For the simplest processes, even N$^3$LO results have been obtained using $q_T$ slicing~\cite{Cieri:2018oms, Camarda:2021ict, Billis:2021ecs, Chen:2022cgv, Neumann:2022lft, Chen:2022lwc}. Jettiness has been used for NNLO predictions of several benchmark hadron collider processes with colorless final states~\cite{ Gaunt:2015pea, Boughezal:2016wmq, Campbell:2016yrh, Heinrich:2017bvg, Campbell:2017aul}, and for processes with one jet~\cite{Boughezal:2015dva, Boughezal:2015aha, Boughezal:2015ded, Boughezal:2016dtm}. The fixed-order expansion also provides ingredients for NNLO matching to parton showers.

This paper presents a systematic method for studying cumulant multijet cross sections, where an arbitrary dimensionful resolution variable $\qf$ is constrained to be less than a cut $\qfcut$ with $Q\gg \qfcut \gg \Lambda_{\mathrm{QCD}}$. The analysis starts from the all-order perturbative cross section
\begin{equation}
  \label{eq_HadronicCrossSectionFactorization}
  \sigma_{\qfcut>\qf}=\sum_{b_1, b_2}\int \rd x_{1 }\rd x_2 f_{b_1/h_1}(x_1) f_{b_2/h_2}(x_2)\hat{\sigma}_{b_1b_2}(x_1, x_2) + \mathcal{O}\left[ \left( \frac{\Lambda_{\mathrm{QCD}}}{\qfcut} \right)^{p_\Lambda} \right]\, ,
\end{equation}
where we assume PDF factorization holds, i.e., $p_\Lambda>0$. We use the method of regions~\cite{Beneke:1997zp,Smirnov:2002pj} to identify the relevant phase space regions contributing for small $\qfcut$. The method of regions allows a systematic power expansion of the phase space and squared matrix elements, leading to a factorization formula involving fully differential beam, jet, and soft functions. While we do not use Soft-Collinear Effective Theory (SCET)~\cite{Bauer:2000yr,Bauer:2001yt} to derive our results, our beam, jet, and soft functions have close analogues in SCET. Some key technical challenges addressed in this paper, such as the general treatment of recoil in the phase space expansion, are also relevant for SCET factorization formulas.

This paper is structured as follows. Section~\ref{Gen_sec_MethodOfRegionsResolution} details how we use the method of regions to expand the cross section and specifies the conditions a resolution variable $\qf$ must satisfy for our framework. Section~\ref{sec_SingularLimits} reviews the singular limits of QCD amplitudes and introduces an all-order ansatz for the factorization of squared matrix elements. Our ansatz does not fully account for Glauber effects. Section~\ref{Gen_sec_FactorizationForResolutionVariables} derives a general leading-power phase space factorization and combines it with the factorized squared matrix elements to obtain a factorization formula in terms of differential beam, soft, and jet functions. Section~\ref{Gen_sec_RapidityDivergences} analyzes rapidity divergences that appear for certain resolution variables, using a time-like reference vector to regularize divergences in beam and jet functions—a method called the $z_N$-prescription. We show how zero-bin contributions combine with the soft function to define a rapidity-finite, soft subtracted function suitable for numerical evaluation. The $z_N$-prescription is applicable to both $\SCETII$- and $\SCETI$-type variables. Section~\ref{chap_FactorizationFormulaForKtNess} applies the framework to $n$-$\ktness$~\cite{Buonocore:2022mle} and presents factorization formulas for several definitions, including one that factorizes into simple cumulant functions to all orders in perturbation theory.

\section{The Method of Regions for Resolution Variables}
\label{Gen_sec_MethodOfRegionsResolution}
In this paper, we study $n_J$-jet processes, $h_1(P_1)+h_2(P_2)\to j(P_3)+\dots + j(P_{n_J+2})+F(p_F)+X$, where we require a minimum of $n_J$ energetic and well-separated jets in the final state. We allow an arbitrary color-singlet system $F$ with total momentum $p_F$. The partonic channels with LO kinematics for this process are of the form $a_1(p_1)+a_2(p_2)\to a_3(p_3)+\dots + a_{n_J+2}(p_{n_J+2})+F(p_F)$, i.e., they live on a phase space with $n_J$ massless QCD partons. 

This section explains how to use the method of regions~\cite{Beneke:1997zp, Smirnov:2002pj} to approximate the cumulant partonic cross section in a small resolution variable. The (bare) partonic cross section in \eqref{eq_HadronicCrossSectionFactorization} is defined as
\begin{equation}
  \label{Gen_eq_PartonicCrossSection}
  \rd x_1 \rd x_2\hat{\sigma}_{b_1b_2}(\qfcut)=\rd x_1 \rd x_2\sum_{n}\sum_{\Af_n}\int \rd\Pi_n  \frac{1}{S_{\Af_n}}\frac{\lvert \amp_{b_1b_2\Af_n} \rvert^2}{2\hat{s}}\F \theta\!\left(\qfcut -\qf\right)\, ,
\end{equation}
where $(b_1,x_1)$ and $(b_2,x_2)$ are the flavors and momentum fractions entering the bare parton distribution functions, $\Af_n$ is a multi-index containing the possible flavors of the final-state QCD partons, $S_{\Af_n}$ is the corresponding symmetry factor, $\rd\Pi_n$ is the $n-$particle phase space, $\F$ is an arbitrary infrared safe measurement function that does not introduce any other small scale, and $\hat{s}=2x_1 x_2 P_1\cdot P_2$ is the squared partonic center-of-mass energy. We have grouped the measure $\rd x_1 \rd x_2$ together with the partonic cross section for future convenience, and we suppressed the momentum dependence in the squared matrix element, the measurement function, and the resolution variable $\qf$. To set up an appropriate power counting, we assume that $\qfcut$ is much smaller than any other scale appearing in \eqref{Gen_eq_PartonicCrossSection}.  For example, we can define an infrared-safe (IR-safe) function of the momenta, Q, that reduces to the center-of-mass energy of the final state at the Born level. We can then define the power-counting parameter
\begin{equation}
  \lambda = \frac{\qfcut}{Q}\ll 1\, .
\end{equation}
We will always assume that resolution variables have mass dimension one. This comes with no loss of generality because we can always multiply $\qf$ by a suitable power of the hard scale $Q$.
Importantly, the resolution variable $\qf$ depends only on external momenta, not on loop momenta. Thus, we only need to assign power counting to the external momenta, which simplifies the method of regions. External momenta are always on-shell. Therefore, we only need to consider power countings that respect the on-shell condition, which, for example, forbids Glauber modes.

To select the relevant modes for the real emissions, we remember that $\qf>0$ almost everywhere except on phase space configurations that are degenerate with the Born phase space. As \mbox{$\qfcut \to 0$}, the available phase space becomes increasingly constrained until we are left with only collinear or soft emissions\footnote{Phase space configurations where multiple emissions conspire to produce a small $\qf$, like the emission of hard back-to-back partons in $q_T$-resummation, are phase space suppressed and need not be considered, at least at leading power.}. However, we need to be more precise. We must know exactly how soft and collinear particles scale with $\lambda$. Let us therefore assume that we are in a kinematic configuration in which the set of final-state colored parton momenta $\{k_i\}$ splits into two sets of momenta $I_1$ and $I_2$ collinear to the incoming hadrons with momenta $P_1$ and $P_2$ respectively; a set of soft momenta $S$; and, for each jet, a set $F_j$ of partons collinear to the final-state jet with momentum $P_j$, i.e.,
\begin{equation}
  \label{eq_decomposition_into_sectors}
  \{k_i\}={I_1}\cup {I_2}\cup {S} \bigcup_{j=3}^{n_J+2}{F_j}\, .
\end{equation}
We refer to the individual sets $I_1$, $I_2$, $\dots$ as \emph{sectors}.
For concreteness, let us denote the momenta in each sector as follows:
\begin{equation}
\begin{aligned}
    I_1&=\{k_{1_1},k_{1_2},\dots\},\quad I_2=\{k_{2_1},k_{2_2},\dots\},\quad  F_3=\{k_{3_1},k_{3_2},\dots\},\quad\dots,\\
    F_{n_J+2}&=\{k_{(n_J+2)_1},k_{(n_J+2)_2},\dots\},\quad S=\{k_{S_1},k_{S_2},\dots\}
\end{aligned}
\end{equation} 
Section~\ref{Gen_sec_PhaseSpaceFactorization} shows in detail how, given a composition into the sectors $I_1$, $I_2$, $S$ and $F_i$, we can factorize the phase space into a product of a Born phase space with hard momenta $p_1,p_2,\dots,p_{n_J+2},\tilde{p}_F$ and a radiation phase for each sector. To be precise, we will find that at leading power we can approximate each of the collinear momenta as
  \begin{equation}
    \label{Momentum_With_Homogeneous_Scaling}
    \begin{aligned}
      k_{1_j}^\mu&=\frac{\greenmath{z_{1_j}}}{z_1}p_1^\mu+\greenmath{k_{t,1_j}^\mu}+\frac{z_1\lvert \greenmath{k_{t,1_j}}\rvert^2}{\greenmath{z_{1_j}}Q^2}p_2^\mu\\
      k_{2_j}^\mu&=\frac{\greenmath{z_{2_j}}}{z_2}p_2^\mu+\greenmath{k_{t,2_j}^\mu}+\frac{z_2\lvert \greenmath{k_{t,2_j}}\rvert^2}{\greenmath{z_{2_j}}Q^2}p_1^\mu\\
      k_{i_j}^\mu&=\greenmath{z_{i_j}} p_i^\mu+k_{\perp,i_j}^\mu+\frac{Q^2\lvert k_{\perp,i_j}\rvert^2}{4\greenmath{z_{i_j}}\left( p_i\cdot q \right)^2}\bar{p}_i^\mu, \quad \text{for }i>2\\
      k_{\perp,i_j}&=\greenmath{\tilde{k}_{\perp,i_j}}-\greenmath{z_{i_j}} \frac{p_i\cdot q}{Q^2}k_\mathrm{rec,\perp}^\mu =\greenmath{\tilde{k}_{\perp,i_j}}-\greenmath{z_{i_j}} \frac{p_i\cdot q}{Q^2}\left( k_\mathrm{rec}^\mu-\frac{\bar{p}_i\cdot k_\mathrm{rec}}{p_i\cdot \bar{p}_i}p_i-\frac{p_i\cdot k_\mathrm{rec}}{p_i\cdot \bar{p}_i}\bar{p}_i \right)\\
      k_\mathrm{rec}&=\greenmath{k_{t,I_1}}+\greenmath{k_{t,I_1}}+\greenmath{k_S}=\sum_{j}\greenmath{k_{t,1_j}}+\sum_{j}\greenmath{k_{t,2_j}}+\sum_{j}\greenmath{k_{S_j}}\\
      \,
    \end{aligned}
  \end{equation}
  where $q=p_1+p_2$, we defined $\bar{p}_i=E_i \bar{n}_i=2\frac{p_i\cdot q}{Q}\frac{q}{Q}-p_i$, and we highlighted radiation phase space variables in {\color{ReadableGreen}{green}}. Note that $p_2=\bar{p}_1$ and $\bar{p}_2=p_1$. The transverse components are defined such that 
  \begin{equation}
    k_{t,1_j}\cdot p_1=k_{t,1_j}\cdot p_2=0,\quad k_{t,2_j}\cdot p_1=k_{t,2_j}\cdot p_2=0,\quad k_{\perp,i_j}\cdot p_i= k_{\perp,i_j}\cdot \bar{p}_i=0\, ,
  \end{equation} 
  i.e., they are transverse to the respective jet or beam in the center of mass frame where $q=(Q,0,0,0)$. The collinear momentum fractions $z_{i_j}$ in a given collinear sector add up to one and the transverse momenta ${\tilde{k}_{\perp,i_j}}$ in a given final-state collinear sector add up to zero, i.e.,
  \begin{equation}
    \sum_j z_{i_j}=1,\quad \sum_j \tilde{k}_{\perp,i_j}=0\, .
  \end{equation}
  All other radiation phase space variables are unconstrained.
  We point out that the momenta $k_{S_j}$ are directly used as radiation variables while we express the transverse momenta of the final state collinear momenta $k_{\perp,i_j}$ through the ``recoiled'' variables $\tilde{k}_{\perp,i_j}$. This will be explained in detail in Section~\ref{Gen_sec_PhaseSpaceFactorization}. We can now express a resolution variable $\qf$ in terms of the Born and radiation variables as
  \begin{equation}
    \qf=\qf\!\left(\!\{p_i\},\{z_{i_j}\},\{\Omega_{j}\},\{\phi_{i_j}\},\{k^0_{S_j}\},\{\lvert k_{t,i_j}\rvert\},\{\lvert \tilde{k}_{\perp,i_j}\rvert\} \!\right) ,
  \end{equation}
  where $\Omega_j$ specifies the direction of the soft momenta $k_{S_j}$, and $\phi_{i_j}$ specifies the direction of the transverse momentum $k_{t,i_j}$ (or $\tilde{k}_{\perp,i_j}$).
  We now assign positive scaling dimensions to each of the dimensionful radiation variables, i.e., 
  \begin{equation}
    \label{eq_most_general_scaling}
    \begin{aligned}
      \lvert k_{t,i_j}\rvert&\sim \lambda^{a_{i_j}} Q,\quad  \lvert\tilde{k}_{\perp,i_j}\rvert \sim \lambda^{b_{i_j}} Q,\quad k^0_{S_j}\sim \lambda^{c_j} Q\, .
    \end{aligned}
  \end{equation} 
Throughout this paper, a decomposition \eqref{eq_decomposition_into_sectors} into sectors, together with an assignment of scaling dimensions, will be referred to as a \emph{region}.
The choice of scaling dimensions will imply a scaling of the resolution variable $\qf\sim Q \lambda^{d}$ for some $d$ in the limit $\lambda\to 0$. Because $\qfcut \sim Q \lambda$, the scaling dimensions have to be chosen such that $d=1$.

If we instead choose the powers too small, such that $d<1$, we could make the approximation
\begin{equation}
  \theta(\qfcut-\qf)\approx \theta(-\qf)=0\, ,
\end{equation}
which would lead to a vanishing contribution. On the other hand, choosing the powers too large, we could make the approximation
\begin{equation}
  \theta(\qfcut-\qf)\approx \theta(\qfcut)=1\, ,
\end{equation}
which at first glance looks like a contribution we have to include. However, it turns out that such a choice would lead to integrals where the phase space for the collinear and soft particles is unconstrained. This leads to scaleless integrals, which are zero in dimensional regularization. 

It turns out that there are infinitely many ways to choose the scaling dimensions while still satisfying the condition $d=1$. However, most of these choices still lead to a scaleless integral. For instance, for 
\begin{equation}
    \qf=\lvert q_T \rvert =\lvert{k_{t,I_1}}+{k_{t,I_1}}+{k_{t,S}}\rvert=\lvert\sum_{j}{k_{t,1_j}}+\sum_{j}{k_{t,2_j}}+\sum_{j}{k_{t,S_j}}\rvert
\end{equation}
in color singlet production, we have $d=1$ as long as $\min\{a_{1_i},b_{1_j},c_{1_k}\}=1$. However, as soon as one of our dimensional variables vanishes in the approximation for $\qf$ as $\lambda\to 0$ this leads to a scaleless integral in the cross section. The only choice that yields non-scaleless integrals (at leading power) is $a_{1_i}=a_{2_j}=c_k=1$ for all $i,j,k$. In general, all power counting parameters should be chosen as small as possible while still satisfying the condition $d=1$. To be more precise, for a given choice of the hard configuration $\{p_i\}$ and dimensionless radiation variables $\{z_{i_j}\},\{\Omega_{j}\},\{\phi_{i_j}\}$ the set 
\begin{equation}
  R:=\{(a_{1_1},a_{1_2},\dots,a_{2_1},\dots,b_{3_1},\dots,b_{(n_J+2)_1},\dots,c_1,\dots)\vert d\geq 1 \}
\end{equation}
is a polytope in $\mathbb{R}^n$ (with some corners at infinity). We conjecture that the finite corners, i.e., the corners of $R$ where all $a_{i_j},b_{i_j},c_{j}$ are finite, correspond to all relevant regions contributing to the cross section at leading power. We also conjecture that the IR safety of $\qf$ implies that the scaling dimensions are the same for all particles belonging to the same sector, i.e., \footnote{A region where this does not hold would yield an integral where the matrix element can be approximated by a strongly ordered limit. The poles of this region would mismatch the ones in the virtual hard amplitudes. Thus, if there are regions where the scaling dimensions in a given sector differ, then the spurious poles must cancel when summing over all these regions.}
\begin{equation}
  a_{i_j}=a_i, \quad b_{i_j}=b_i, \quad c_j=c\, .
\end{equation}
To simplify the notation, we will also only consider variables $\qf$ that treat the incoming hadrons equally, so $a_1=a_2=a$ and all hard jets equally, i.e., $b_i=b$. Thus, for the IR-safe resolution variables considered in this paper, the relevant polygon for a fixed hard configuration and dimensionless radiation variables is
\begin{equation}
  R:=\{(a,b,c)\vert d\geq 1 \}\, .
\end{equation}

For completely generic resolution variables it can be quite complicated to find all relevant regions that contribute to the cross section. In this paper, we want to restrict the possible choices to a handleable, but still very general class. In particular, we want to avoid the following two scenarios. First, we discard resolution variables where the power-counting parameters, i.e., $R$, vary as functions of the dimensionless variables $z_{i_j},\phi_{i_j},\Omega_{j}$ and $\phi_{i_j}$\footnote{$R$ is allowed to change for special configurations of the dimensionless variables, as long as these special configurations are of zero measure.}. We conjecture that for IR-safe variables this is equivalent to requiring \emph{continuous globalness}~\cite{Dasgupta:2002bw, Dasgupta_2002}. Second, we require that there is only one contributing region, i.e., only one finite corner of $R$, for fixed sectors $I_1,I_2,\dots$, and that they are independent of the number of particles in the sectors $I_1,I_2,\dots$. We conjecture that this is equivalent to requiring \emph{recursive IR safety}~\cite{Banfi:2004yd}. For these types of variables one can determine $a,b$ and $c$ for all phase space multiplicities by looking at the possible configurations with only one extra emission. E.g., $c$ can be determined by extracting the behavior of $\qf$ in the presence of a single soft emission. We also assume that the power counting parameters do not depend on the hard configuration $p_1,p_2,\dots$. The last restriction is not necessary for our framework, but it makes the notation easier to handle. In the remainder of this paper we will use the term ``resolution variable'' for variables that fulfill all these constraints unless stated otherwise.

Before presenting several examples, we comment on how our exponents relate to similar power counting parameters defined, for example, in~\cite{Banfi:2004yd}. In~\cite{Banfi:2004yd}, the scaling of a general observable in the limit where an emitted parton becomes soft and collinear to leg $\ell$ is parametrized as 
\begin{equation}
  \label{Gen_eq_CeasarPowerCounting}
  \qf\sim d_{\ell}\left(\frac{k_t^{(\ell)}}{Q}\right)^{\tilde{a}_{\ell}} e^{-\tilde{b}_{\ell} \eta^{(\ell)}} g_{\ell}\left(\phi^{(\ell)}\right)\, ,
\end{equation}
where $k_t^{(\ell)}$, $\eta^{(\ell)}$ and $\phi^{(\ell)}$ are the transverse momentum, rapidity, and azimuthal angle of the emitted parton with respect to leg $\ell$. The constants $d_{\ell}$ and functions $g_{\ell}$ depend on the observable details and do not influence the power counting.
If we restrict ourselves to observables that are continuously global, then all $\tilde{a}_\ell$ are the same, and we have the relation
\begin{equation}
  \label{Gen_eq_CeasarPowerCountingRelation}
  \begin{aligned}
    \frac{1}{\tilde{a}_\ell}&=c \\
    \frac{1}{ \tilde{a}_\ell+\tilde{b}_\ell} &=\begin{cases}
      a & \mytext{if }\ell \mytext{ is an initial-state leg} \\
      b & \mytext{if }\ell \mytext{ is a final-state leg.} \\
     \end{cases}
  \end{aligned}
\end{equation}
IR safety requires $\tilde{a}_\ell+\tilde{b}_\ell > 0$ and $\tilde{a}_\ell>0$. It follows that IR safety requires $a>0$, $b>0$ and $c>0$, as we have assumed above. 

Finally, we point out that the power-counting considerations made in this section were only based on the kinematic dependence of the resolution variable. Not all regions we identified contribute to leading power. For example, regions where the soft sector $S$ contains net quark flavor (e.g., a single quark, or a quark and a gluon) do not contribute at leading power.

\subsection{Examples}
\label{Gen_sec_PCExamples}
\begin{itemize}
  \item ${\DeltaET}$: A possible resolution variable for color-singlet production accompanied by a resolved jet  is the so-called $\DeltaET$. It is defined as
  \begin{equation}
    \DeltaET=\sum_{i=1}^{n}\lvert k_{i,t}\rvert-\lvert p_{F,t} \rvert=\sum_{i=1}^{n}\lvert k_{i,t}\rvert-\lvert \sum_{i=1}^{n}k_{i,t} \rvert\, ,
  \end{equation}
  where the sum goes over the QCD partons in the final state, and $p_F$ is the momentum of the color singlet. The paper~\cite{Buonocore:2023rdw} studied the leading power expansion of the cumulant cross section for small $\DeltaET$ at NLO. Expanding on the analysis in~\cite{Buonocore:2023rdw}, we find that for a given decomposition of the phase space,
  \begin{equation}
    \{k\}={I_1}\cup {I_2}\cup {S} \cup F\, ,
  \end{equation}
  where $F$ is the set of momenta collinear to the jet, there is only one contributing region. All momenta in each sector have the same scaling dimension, independent of particle multiplicity, hard configuration and dimensionless radiation variables, as we require. They are given by $a=1,b=1/2,c=1$ and the leading power approximation of $\DeltaET$ in this region reads
  \begin{equation}
    \label{eq_generalET}
    \DeltaET \sim \sum_{k_{j}\in I_1 \cup I_2 \cup  S }\lvert k_{t,j}\rvert (1+\cos\phi_{j})+\frac{1}{2\lvert p_{J,t}\rvert }\sum_{k_j\in F}\frac{\lvert \tilde{k}_{\perp,j}\rvert^2 }{z_{j}}\sin^2\varphi_{j}\, , 
  \end{equation}
  where $p_{J,t}=-p_{F,t}$ is the transverse momentum (WRT to the beam) of the jet, $\phi_j$ is the azimuthal angle between the transverse momentum $k_{t,j}$ and the transverse momentum of the color singlet $p_{F,t}$ and $\varphi_j$ is the azimuthal angle between the transverse momentum (WRT the jet) $k_{\perp,j}$ and the transverse momentum of the incoming hadron $p_{1,\perp}$.
  It is straightforward to see that with the scaling dimensions $a=1,b=1/2,c=1$ the right-hand side of \eqref{eq_generalET} reproduces the required scaling $d=1$.
  \item ${\tau}$: A well-known observable for dijet production at lepton colliders is the thrust~\cite{Farhi:1977sg} 
  \begin{equation}
    \label{eq_ThrustDef}
    T=\frac{1}{Q}\max_{\vec{n}}\sum_{i=1}^{n}\lvert\vec{k}_i\cdot \vec{n} \rvert\, ,
  \end{equation}
  where $\vec{n}$ is a unit vector and the sum runs over all final-state momenta. Using $T$, we can define a resolution variable with mass dimension one as
  \begin{equation}
    \tau = (1-T)Q\,. 
  \end{equation}
  For a given phase space decomposition $\{k\}=F_3\cup F_4\cup S$, there is only one region that contributes at leading power. The power counting parameters are $b=\frac{1}{2}$ and $c=1$ independent of multiplicity and dimensionless radiation variables, as we require. The leading power approximation of $\tau$ in this region reads
  \begin{equation}
    \label{eq_tauexpansion}
    \tau \sim \! \sum_{\tilde{k}_j\in F_3}\frac{\lvert \tilde{k}_{\perp,j}\rvert^2 }{z_{j}2Q}+\sum_{k_j\in F_4}\frac{\lvert \tilde{k}_{\perp,j}\rvert^2  }{z_{j}2 Q}+\sum_{k_j\in S}\frac{\lvert (p_3-p_4)\cdot k_j \rvert}{Q}\sim\frac{s_3}{2Q}+\frac{s_4}{2Q}+\sum_{k_j\in S}\frac{\lvert (p_3-p_4)\cdot k_j \rvert}{Q}\, ,
  \end{equation}
  where $s_i$ is the total invariant mass of the momenta in sector $F_i$.
  \item ${B_T}$: Another common resolution variable defined for dijet production at lepton colliders is the total broadening~\cite{Catani:1992jc} 
  \begin{equation}
    B_T=\sum_i\left|\vec{k}_i \times \vec{n}_T\right|\, ,
  \end{equation}
  where $\vec{n}_T$ is the thrust axis, i.e., the unit vector that maximizes the thrust in \eqref{eq_ThrustDef}. Note that we do not include a normalization with two times the total energy in our definition of $B_T$. We also define the purely spatial four-vector thrust axis $n_T$ that is defined such that $n_T^2=-1,n_T\cdot q=0$ and $\vec{k}_i\cdot \vec{n}_T=-k_i\cdot n_T$. We again fix a phase space decomposition $\{k\}=F_3\cup F_4\cup S$. It is useful to split the set of soft momenta into a left-moving and a right-moving set, $S_L$ and $S_R$. $S_L$ contains the soft momenta $k$ with positive rapidities,
  \begin{equation}
    \eta =\frac{1}{2}\log\left(\frac{p_4\cdot k}{p_3 \cdot k}\right)\, ,
  \end{equation}
  while $S_R$ contains the soft momenta with negative rapidities. To leading power, the thrust axis can now be expressed as
\begin{equation}
  n_T=\frac{k_{\perp,L}-k_{\perp,R}}{Q}+\frac{p_3-p_4}{Q}\sqrt{1-\frac{\lvert k_{\perp,L}-k_{\perp,R}\rvert^2}{Q^2}}\, ,
\end{equation}
where
\begin{equation}
  k_{\perp,L}=\sum_{k_j\in S_L}k_{\perp,j}\, , \quad \text{and} \quad k_{\perp,R}=\sum_{k_j\in S_R}k_{\perp,j}\, ,
\end{equation}
are the total transverse momenta (WRT the hard configuration $p_3$ and $p_4$) of the left and right moving soft particles respectively. Using this, we find that there is only one contributing region at leading power, and again all momenta in the same sector have the same power counting independent of multiplicity and dimensionless radiation variables, as we require. We find $b=c=1$ and $B_T$ approximates as
\begin{equation}
  \label{eq_BTexpansion}
  B_T\sim \sum_{k_i\in F_3}  \lvert \tilde{k}_{\perp,i}-z_i k_{\perp,L} \rvert+\sum_{k_i\in F_4}  \lvert \tilde{k}_{\perp,i}-z_i k_{\perp,R} \rvert+\sum_{k_i\in S}\lvert k_{\perp,i}\rvert\, .
\end{equation}

\item $y_{n(n+1)}$:
For $n-$jet production at lepton colliders, we can define the well-known resolution variable $y_{n(n+1)}$.  To define $y_{n(n+1)}$, we use the distance
\begin{equation}
  \label{eq_dijynnp1}
  d^2_{ij}=\frac{E_i^2E_j^2}{(E_i+E_j)^2}2(1-\cos\theta_{ij})
\end{equation}
which is defined both for single partons and pseudojet momenta $k_i$ and $k_j$, which can be sums of parton momenta. For lepton colliders there are no initial-state collinear modes, and, because \eqref{eq_dijynnp1} is a $k_t$-like distance, one can show that the relevant scaling of the final-state collinear and soft momenta is given by $b=1$ and $c=1$ respectively, where $\lambda=\frac{y_{n(n+1)}}{Q}$ is the small parameter. If both pseudojets only contain partons from the soft sector $S$, $d_{ij}$ is already homogeneous in the power counting and cannot be approximated further. If, on the other hand, $k_i$ contains only partons from $S$ while $k_j$ contains partons from the collinear sector $F_k$ and $S$, then $d_{ij}$ can be approximated as
\begin{equation}
  d_{ij}^2\sim E_i^22\left(1-\cos\theta_{i, \mathrm{jet}}\right)+\mathcal{O}(\lambda^3)=2\frac{\left( k_i\cdot q \right)^2}{Q^2}\left(\! 1-\frac{1}{\sqrt{1-\frac{k_i^2Q^2}{\left( k_i\cdot q \right)^2}}} \left( 1-\frac{k_i\cdot p_k Q^2}{p_k\cdot q k_i\cdot q} \right)\!\!\right)\!\, ,
\end{equation}
where $\theta_{i, \mathrm{jet}}$ is the angle between the parton $i$ and the hard momentum $p_k$, i.e., the soft partons only resolve the direction of the jets and not the individual collinear partons. Sometimes, one must determine which exact collinear parton the soft parton $i$ is clustered with. In such cases, one must expand the distance to higher orders in $\lambda$. If both $k_i$ and $k_j$ contain partons from the same collinear sector (as well as potentially soft partons), then $d_{ij}$ approximates to
\begin{equation}
  \label{Gen_eq_dij}
  d_{ij}^2\sim \frac{(k_i^+k_j^+)^2}{(k_i^++k_j^+)^2}\biggl\lvert \frac{k_{i, \perp}}{k_{i}^+}-\frac{k_{j, \perp}}{k_{j}^+}\biggr\rvert^2\sim\frac{(\tilde{k}_i^+\tilde{k}_j^+)^2}{(\tilde{k}_i^++\tilde{k}_j^+)^2}\biggl\lvert \frac{\tilde{k}_{i, \perp}}{\tilde{k}_{i}^+}-\frac{\tilde{k}_{j, \perp}}{\tilde{k}_{j}^+}\biggr\rvert^2\, ,
\end{equation}
where the soft momenta can be dropped in the $+$-components, but not in the transverse components of the momenta.
 To be specific, let us write down the partonic momenta for the case where $k_i$ and $k_j$ are in $F_3$, and we assume that a soft momentum $k_k\in S$ has clustered with $k_i$ at some previous state. Then the resulting distance can be approximated as 
\begin{equation} 
  \label{eq_dijwithkrec}
  d_{(i+k)j}^2\sim \frac{(k_i^+k_j^+)^2}{(k_i^++k_j^+)^2}\biggl\lvert \frac{k_{i, \perp}+k_{k, \perp}}{k_{i}^+}-\frac{k_{j, \perp}}{k_{j}^+}\biggr\rvert^2\sim  \frac{(z_iz_j)^2}{(z_i+z_j)^2}\biggl\lvert \frac{\tilde{k}_{i, \perp}+k_{k, \perp}}{z_i}-\frac{\tilde{k}_{j, \perp}}{z_j}\biggr\rvert^2\, ,
\end{equation}
where we dropped the $+$-component of the soft momentum $k_k$. The fact that the soft momentum $k_k$ cannot be dropped in \eqref{eq_dijwithkrec} massively complicates factorization formulas for the cross section at small $y_{n(n+1)}$. We will analyze this issue in detail, when we discuss the factorization properties of $n$-$\ktness$, a generalization of $y_{n(n+1)}$, in Section~\ref{chap_FactorizationFormulaForKtNess}.

\item $\sqrt{B_T\tau}$: It is well known that the product of thrust and total broadening is an example of an IR-safe observable that is not recursively IR-safe~\cite{Banfi:2004yd}. It is thus an example that we cannot treat within the framework presented in this paper. Let us see what goes wrong when performing a region analysis for the observable. We will first perform the analysis for all situations relevant for the one-emission phase space and then look at the case of two emissions.

If there were only one collinear emission along $p_3$, i.e., $F_3=\{k_1,k_2\}$, $F_4=\{k_3\}$ and $S=\emptyset$, we would find 
\begin{equation}
  \sqrt{B_T\tau} \sim \sqrt{\frac{2 \lvert \tilde{k}_{\perp,1}\rvert^3}{Qz_1(1-z_1)}}\, ,
\end{equation}
and we would deduce that the relevant scaling dimension for the collinear modes is $b=\frac{2}{3}$. 

Similarly, if we look at only a single soft emission, i.e.,  $F_3=\{k_1\}$, $F_4=\{k_2\}$ and $S=\{k_3\}$, we would find
\begin{equation}
  \sqrt{B_T\tau} \sim \lvert k_{\perp,3}\rvert \sqrt{2 \lvert \sinh\eta_3\rvert } \, ,
\end{equation}
and we would extrapolate the relevant scaling dimensions for the soft modes to be $c=1$.

Now let us consider the two-emission phase space. To be specific, we consider the case where we have one collinear emission along $p_3$ and a soft emission, i.e., $F_3=\{k_1,k_2\}$, $F_4=\{k_3\}$ and $S=\{k_4\}$. Using approximations \eqref{eq_tauexpansion} and \eqref{eq_BTexpansion}, we would find
\begin{equation}
  \label{eq_BTtaunonhomog}
\begin{aligned}
    B_T\tau &\sim \Biggl(  \lvert \tilde{k}_{\perp,1}-z_1 \Theta(\eta_4)k_{\perp,4} \rvert+\lvert \tilde{k}_{\perp,1}+(1-z_1) \Theta(\eta_4)k_{\perp,4} \rvert \\
   &+\Theta(-\eta_4)\lvert k_{\perp,4}\rvert+\lvert k_{\perp,4}\rvert \Biggr)\left(\frac{\lvert \tilde{k}_{\perp,1}\rvert^2}{2z_1(1-z_1)Q} +\lvert k_{\perp,4}\sinh \eta_4\rvert  \right)\, ,
\end{aligned}
\end{equation}
where the $\Theta$-function distinguishes between the cases where the soft emission is left or right moving. Equation \eqref{eq_BTtaunonhomog} does not have homogeneous scaling whatever power counting we assign to the dimensionful variables. Worse, if we insert the scaling dimensions $b=\frac{2}{3}$ and $c=1$ we deduced from the NLO example, we find 
\begin{equation}
  \label{eq_BTscaleless}
\begin{aligned}
    B_T\tau &\sim 2 \lvert \tilde{k}_{\perp,1}\rvert\lvert k_{\perp,4}\rvert\sinh\eta_4 \, ,
\end{aligned}
\end{equation}
which leads to the wrong scaling dimension $d=\frac{5}{6}$.

Performing a more thorough analysis\footnote{ A similar analysis was performed in \cite{Makris:2020ltr}.}, we show the polygon $R$ of scalings $(b,c)$ that lead to $d\geq 1$ in Figure~\ref{fig_scalings}. We see that there are two corners of $R$ which give us two regions to consider, namely $(b,c)=(1,1)$ leading to 
\begin{equation}
  \begin{aligned}
    B_T\tau &\sim \Biggl(  \lvert \tilde{k}_{\perp,1}-z_1 \Theta(\eta_4)k_{\perp,4} \rvert+\lvert \tilde{k}_{\perp,1}+(1-z_1) \Theta(\eta_4)k_{\perp,4} \rvert \\
   &+\Theta(-\eta_4)\lvert k_{\perp,4}\rvert+\lvert k_{\perp,4}\rvert \Biggr)\lvert k_{\perp,4}\sinh \eta_4\rvert  \, ,
\end{aligned}
\end{equation}
and $(b,c)=(2/3,4/3)$ leading to 
\begin{equation}
  \begin{aligned}
    B_T\tau &\sim 2 \lvert \tilde{k}_{\perp,1}\rvert\left(\frac{\lvert \tilde{k}_{\perp,1}\rvert^2}{2z_1(1-z_1)Q} +\lvert k_{\perp,4}\sinh \eta_4\rvert  \right)\, .
\end{aligned}
\end{equation} 
In our framework we cannot account for multiple regions because we want to factorize cumulant cross sections into hard, beam, jet and soft functions. If we had multiple regions, we would need multiple different beam, jet and soft functions and perform sums over them. Furthermore, it turns out that the presence of multiple regions for a given phase space decomposition $I_1,I_2,F_3,\dots S$ can lead to complications because the expansions in the individual regions might no longer be regularized by dimensional regularization, even after rapidity regularization.

\begin{figure}[ht]
  \centering
  \includegraphics[width=0.8\textwidth]{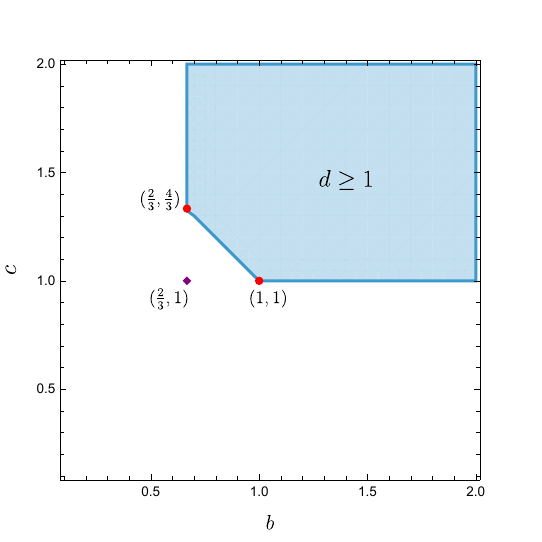}
  \caption{The figure shows the polytope $R$ of scalings $(b,c)$ that lead to $d\geq 1$ for the case where we have one collinear emission along $p_3$ and a soft emission. The red points show the scaling dimensions that need to be included in the region expansion for the cross section. The purple diamond shows the scaling dimensions that one would have deduced from studying the one-emission phase space.}
  \label{fig_scalings}
\end{figure}
\end{itemize}

\section{Singular Limits of Squared QCD Amplitudes}
\label{sec_SingularLimits}
Having identified the regions relevant for the asymptotic expansion of the cross section in the previous section is helpful because the QCD matrix elements have well-known factorization properties in these regions. This section summarizes the factorization properties of QCD matrix elements in the singular limits. Readers familiar with these factorization properties can directly skip to Section~\ref{sec_MixedIRLimits}, where we summarize the final results in our notation.
\subsection{Collinear Limits of QCD Amplitudes}\label{sec_CollinearLimits}
This section analyzes the behavior of squared QCD amplitudes in the limit where two or more partons become collinear. We will, in particular, focus on the limits relevant to perturbative QCD (pQCD) calculations up to NNLO. At NLO, one only needs to consider limits where two partons become collinear in tree-level matrix elements. At NNLO, one also needs to consider the so-called \emph{triple-collinear} limit where three partons become collinear in tree-level matrix elements, and the limit where two partons become collinear in one-loop matrix elements. Additionally, we also need to consider the so-called \emph{double-collinear} limits where two separate pairs of partons become collinear to each other in tree-level matrix elements. For the following discussion, we will closely follow the notation of~\cite{Catani:1999ss}.

To define the collinear limit of $n$ final-state partons with momenta $k_1, \dots, k_n$, we parametrize them as 
\begin{equation}
  \label{eq_SudakovParametrization}
  k_i^\mu=z_i p^\mu+k_{\perp i}^\mu-\frac{k_{\perp i}^2}{z_i} \frac{\bar{n}^\mu}{2 p \cdot \bar{n}}\, ,
\end{equation} 
where the light-like vector $p^\mu$ denotes the collinear direction and the auxiliary light-like vector $\bar{n}^\mu$ specifies how the collinear direction is approached $\left(k_{\perp i} \cdot p=k_{\perp i} \cdot \bar{n}=0\right)$. Note that no other constraint (in particular $\sum_i z_i \neq 1$ and $\sum_i k_{\perp i} \neq 0$) is imposed on the longitudinal and transverse variables $z_i$ and $k_{\perp i}$. Thus, we can easily consider any (asymmetric) collinear limit simultaneously. The collinear limit is defined by taking the limit $k_{\perp i}\to \lambda k_{\perp i}$, $\lambda\to 0$ at fixed $z_i$ for all $i$ simultaneously. The matrix element in the collinear limit only depends on the boost invariant variables
\begin{equation}
  \label{eq_BoostInvariantVariables}
  \begin{aligned}
    \tilde{z}_i & =\frac{z_i}{\sum_{j=1}^n z_j}, \\
    \widetilde{k}_{\perp i}^\mu & =k_{\perp i}^\mu-\frac{z_i}{\sum_{k=1}^n z_k} \sum_{j=1}^n k_{\perp j}^\mu\, ,
    \end{aligned}
\end{equation}
which satisfy $\sum_i \tilde{z}_i=1$ and $\sum_i \widetilde{k}_{\perp i}=0$. 

The all-order squared matrix element in the collinear limit of the final-state partons $k_1, \dots, k_n$ factorizes as
\begin{equation}
  \label{eq_GeneralCollinearLimit}
  \left|\amp_{a_1 \ldots a_n \ldots}\left(k_1, \ldots, k_n, \ldots\right)\right|^2 \simeq\left(\frac{8\pi\asbare}{s_{1 \ldots n}}\right)^{n-1} \mathcal{T}_{a \ldots}^{s s^{\prime}}(p, \ldots) \hat{P}_{a\to a_1 \ldots a_n}^{s s^{\prime}}\, ,
\end{equation}
where we drop subleading terms, i.e., terms less divergent than $\lambda^{-2(n-1)}$ on the right-hand side, and $s_{1\dots n}$ is the square of the total invariant mass of the collinear particles. The spin polarization tensor $\mathcal{T}_{a \ldots}^{s s^{\prime}}(p, \ldots)$ is obtained by replacing the partons $a_1, \dots a_n$ with a single parton with the unique flavor $a$ that can split into $a_1, \dots a_n$ through the QCD Feynman rules. In \eqref{eq_GeneralCollinearLimit}, $\asbare=\frac{g_S^2}{4\pi}$ is the (dimensionful) bare coupling. The splitting kernel $\hat{P}$ depends on the variables defined in \eqref{eq_BoostInvariantVariables}, and it mixes the spins of the parent parton $a$. In this work, we write \eqref{eq_GeneralCollinearLimit} using operators in color-spin space as
\begin{equation}
  \label{eq_GeneralCollinearLimitCS}
  \left|\amp_{a_1 \ldots a_n \ldots}\left(k_1, \ldots, k_n, \ldots\right)\right|^2 \simeq\left(\frac{8\pi\asbare}{s_{1 \ldots n}}\right)^{n-1}\Tr\left[ \ket{\amp_{a\dots}(p, \dots)}\bra{\amp_{a\dots}(p, \dots)} \hat{P}_{a\to a_1, \dots a_n}\right]\, ,
\end{equation}
where $\hat{P}$ is now an operator acting on the spin space of parton $a$ and $\Tr$ denotes a trace in the color-spin space of all external QCD partons. The \emph{color-spin density matrix} of the hard amplitude
\begin{equation}
  \ket{\amp_{a\dots}(p, \dots)}\bra{\amp_{a\dots}(p, \dots)}
\end{equation} 
contains a color and spin averaging factor for the initial state partons.
The splitting kernel has a loop expansion
\begin{equation}
  \hat{P}_{a\to a_1 \ldots a_m}^{s s^{\prime}}=\sum_{n=0}^{\infty}\left( \frac{\alpha_0}{\pi} \right)^n \left(\frac{\mu^2}{s_{1\dots m}}  \right)^{n \epsilon}\hat{P}_{a\to a_1 \ldots a_m}^{(n)s s^{\prime}}\, ,
\end{equation} 
where we defined $\alpha_0=\alpha_0(\mu)$ through \begin{equation}
  \asbare \frac{\left( 4\pi \right)^\epsilon}{e^{\epsilon\gamma_E}}=\alpha_0(\mu)\mu^{2\epsilon}\, .
\end{equation}
Collinear factorization is well understood for tree-level matrix elements~\cite{Berends:1987me, Mangano:1990by, Feige:2013zla}, and it has been studied to all orders in perturbation theory~\cite{Feige:2014wja}. It has been rigorously proven to leading color~\cite{Kosower:1999xi}, and it holds for all explicit cases worked out so far. The splitting kernels are known for the triple-collinear limit at tree level~\cite{Campbell:1997hg, Catani:1998nv, Catani:1999ss}, the two-particle collinear limit at one loop~\cite{Bern:1993qk, Bern:1994zx, Bern:1995ix, Bern:1998sc, Bern:1999ry, Kosower:1999rx} and several ingredients required at even higher orders are also known~\cite{DelDuca:1999iql, Catani:2003vu, Badger:2004uk, Bern:2004cz, Sborlini:2013jba, Sborlini:2014mpa, Badger:2015cxa, DelDuca:2019ggv, DelDuca:2020vst, Czakon:2022fqi, Guan:2024hlf}. We list the splitting kernels relevant to this paper in Appendix~\ref{sec_APkernels}. We point out that the collinear splitting kernels can also be written in terms of matrix elements of collinear Wilson lines---an introduction to this formalism can be found in~\cite{Agarwal:2021ais}.

So far, we have assumed that all collinear particles are in the final state---such limits are called \emph{time-like} collinear limits. However, scattering amplitudes are also singular if one (or more) of the collinear partons $k_1, \dots, k_n$, say $k_1$, is in the initial state---such a limit is called a \emph{space-like} collinear limit. We will, in particular, be interested in the physical process, where an incoming parton splits into $n$ collinear particles, of which one then enters a hard interaction cross section. The collinear direction $p$ will also have negative energy in this situation. We can still use the formalism presented in this section to describe space-like limits. However, there are some major differences. As a naive guess, one could assume that one can use crossing relations separately on the spin polarization tensor and the splitting kernel in \eqref{eq_GeneralCollinearLimit} to relate the space-like limit to the time-like limit. If this were true, then \eqref{eq_GeneralCollinearLimit} would essentially be unchanged in the space-like limit\footnote{Up to possible minus signs appearing when Fermions are crossed from the final to the initial state and a constant correction factor because the full matrix element is averaged over the spin and color of the particle with momentum $k_1$, while the spin-polarization tensor in \eqref{eq_GeneralCollinearLimit} is averaged over the spin and color of the (virtual) particle with momentum $p$.}. However, while this approach works for tree-level amplitudes, it breaks down at higher orders~\cite{Catani:2011st}. In fact, at the amplitude level there are already differences between the space-like and time-like collinear limits at one loop that cancel out at the cross section level. The origin of this difference is due to virtual soft gluons that contribute to the matrix elements. For the time-like limit, the soft gluon only resolves the total color charge of the collinear particles, and thus, the soft gluon loop factorizes. For the space-like limit, some collinear particles are in the initial state, which leads to a breakdown of color coherence, and thus, the soft gluon loop does not factorize. It turns out that color coherence can be restored in cases where there are no hard (non-collinear) colored particles in the final state, which is why $\mytext{N}^3\mytext{LO}$ calculations for color singlet processes do not suffer from the subtleties of space-like limits. In~\cite{Catani:2011st} it is proposed that the collinear factorization formula \eqref{eq_GeneralCollinearLimit} can be generalized to space-like limits by allowing for a more general splitting kernel that is also an operator in the color space of the non-collinear particles and distinguishes between initial- and final-state hard particles. The collinear approximation for an incoming particle with momentum $\tilde{p}$ and flavor $b$ that splits into $n-1$ collinear final-state particles with momenta $k_1, \ldots, k_{n-1}$ and flavors $a_1, \ldots, a_{n-1}$ can then be written as 
\begin{equation}
  \label{eq_SpaceLikeSplittingKernles}
 \begin{aligned}
   &\left|\amp_{b;a_1 \ldots a_{n-1} \ldots}\left(\tilde{p};k_1 \ldots k_{n-1}, \ldots\right)\right|^2 \simeq\frac{1}{z}\left(\frac{8\pi\asbare}{-(k_1+\dots+k_{n-1}-\tilde{p})^2}\right)^{n-1}\\
   &\hspace{4cm}\Tr\left[ \ket{\amp_{a;\dots}(p, \dots)}\bra{\amp_{a;\dots}(p, \dots)} \mathbf{\hat{P}}_{ab;a_1a_{n-1}}\right]\, ,
 \end{aligned}
\end{equation}
where $p$ and $a$ are the flavor and momentum of the hard parton that can be reached by the splitting $b\to a, a_1, \ldots, a_{n-1}$, and $\mathbf{\hat{P}}_{ab;a_1a_{n-1}}$ is an operator in the color space of all the hard partons and an operator in the spin space of the hard parton with momentum $p$. The factor $\frac{1}{z}=\frac{p\cdot \bar{n}}{\tilde{p}\cdot \bar{n}}$ is factored out, such that it cancels a factor $z$ that appears in the definition of the beam functions introduced in \eqref{Gen_eq_RadiativeFunctionsDifferential}, which originates from a difference in the flux factor between the hard and partonic cross-section. 

Finally, we introduce the notation 
\begin{equation}
  \label{eq_CollinearKernelsWithas}
  \begin{aligned}
      \calPhat_{a\to a_1, \dots a_n} &\equiv \left(\frac{8\pi\asbare}{s_{1 \ldots n}}\right)^{n-1}\hat{P}_{a\to a_1, \dots a_n}\\
      \bcalPhat_{ab;a_1a_{n-1}} &\equiv \frac{1}{z}\left(\frac{8\pi\asbare}{-(k_1+\dots+k_{n-1}-\tilde{p})^2}\right)^{n-1} \mathbf{\hat{P}}_{ab;a_1a_{n-1}}\, 
  \end{aligned} 
\end{equation}
as well as the trivial kernels for the case $n=1$:
\begin{equation}
  \begin{aligned}
      \calPhat_{a\to a_1} &\equiv \delta_{aa_1} \bbbone\\
      \bcalPhat_{ab;} &\equiv\delta_{ab}\delta(1-z) \bbbone\, ,
  \end{aligned} 
\end{equation}
where $\bbbone$ is the unit matrix in color-spin space. The kernels \eqref{eq_CollinearKernelsWithas} will be particularly useful when defining beam and jet functions. 

\subsection{Soft Limits of QCD Amplitudes}\label{sec_SoftLimits}
The next IR limit we will consider is the soft limit, where multiple QCD partons with momenta $k_1, \dots, k_m$ and flavors $b_1\cdots b_m$ become soft simultaneously. In contrast, the other external QCD partons $p_1, \dots, p_n$ with flavors $a_1, \dots, a_n$ remain hard and at wide angles. We also allow for the presence of additional colorless particles which do not affect the soft limit. The soft limit is defined by simultaneously rescaling the momenta $q_i\to \lambda q_i$ and taking the limit $\lambda\to 0$. In this paper, we assume that the all-order squared matrix element in the soft limit factorizes as 
  \begin{equation}
    \label{eq_GeneralSoftLimitSquared}
   \begin{aligned}
     &\left|\amp_{b_1 \ldots b_m a_1\ldots}\left(q_1 \ldots q_m, p_1 \ldots\right)\right|^2 \simeq\\
     &\hspace{3cm} (4\pi\asbare)^m\Tr\left[ \ket{\amp_{a_1, \dots}(p_1, \dots)}\!\bra{\amp_{a_1, \dots}(p_1, \dots)} \bJJ_{b_1 \ldots b_m}(q_1 \ldots q_m)\right]\, ,
   \end{aligned}
  \end{equation}
where $\bJJ_{b_1 \ldots b_m}(q_1 \ldots q_m)$ is a function of the soft momenta $q_1, \dots, q_m$ and the directions of the hard momenta $p_1, \dots, p_n$. $\bJJ_{b_1 \ldots b_m}(q_1 \ldots q_m)$ is a color operator in the color space of all hard partons and unit-diagonal in all spin spaces. The $\simeq$-sign in \ref{eq_GeneralSoftLimitSquared} indicates that we have dropped subleading terms in the soft limit, i.e., terms that are less singular than $\lambda^{-2m}$. At leading power, the soft limit is only non-vanishing if the total flavor of the partons $b_1, \dots, b_m$ vanishes, i.e., soft (anti-)quarks only contribute through $q\bar{q}$ pairs. The functions $\bJJ_{b_1 \ldots b_m}$ can be power expanded in $\as$, and we define the expansion coefficients as
\begin{equation}
  \bJJ_{b_1 \ldots b_m}(q_1 \ldots q_m)=\sum_{n=0}^{\infty}\left( \frac{\alpha_0}{\pi} \right)^n \bJJ_{b_1 \ldots b_m}^{(n)}(q_1 \ldots q_m)\, .
\end{equation}
The soft limit also factorizes at the amplitude level, and we summarize the corresponding formulas for some specific cases below. 

The validity of the factorization formula \ref{eq_GeneralSoftLimitSquared} has been studied at the tree level~\cite{Berends:1988zn, Feige:2013zla} and to all orders\cite{Feige:2014wja, Ma:2023hrt}. All explicitly worked out examples relevant for pQCD calculations at NLO and NNLO~\cite{Berends:1988zn, Bern:1993qk, Bern:1995ix, Bern:1998sc, Bern:1999ry, Catani:1999ss, Catani:2000pi} and higher order~\cite{Anastasiou:2013srw, Duhr:2013msa, Li:2013lsa, Catani:2019nqv, Zhu:2020ftr, Catani:2021kcy, Catani:2022hkb, Czakon:2022dwk, DelDuca:2022noh, Herzog:2023sgb, Chen:2023hmk, Chen:2024hvp} respect \eqref{eq_GeneralSoftLimitSquared}. We now discuss the cases relevant to pQCD calculations up to NNLO.

\paragraph{Single Soft Emission}
Tree-level matrix elements in the limit where a single gluon with momentum $q$, color $c$ and polarization index $\mu$ becomes soft satisfy the factorization formula
\begin{equation}
  \left\langle c ; \mu \mid \amp^{(0)}_{g, a_1, \ldots, a_n}\left(q, p_1, \ldots, p_n\right)\right\rangle \simeq g_S  \bJ^{(0), c ; \mu}(q)\ket{\amp^{(0)}_{a_1, \ldots, a_n}\left(p_1, \ldots, p_n\right)}\, ,
  \end{equation}
  where 
  \begin{equation}
    \bJ^{(0), c; \mu}(q)=\sum_{i=1}^n \boldsymbol{T}_i^c \frac{p_i^\mu}{p_i \cdot q}
  \end{equation} is the so-called \emph{eikonal current} (or \emph{soft current}), which is an operator on the color space of the amplitude $\ket{\amp^{(0)}_{a_1, \ldots, a_n}\left(p_1, \ldots, p_n\right)}$. The color operators $\boldsymbol{T}_i^c$ are the usual ones introduced in \cite{Catani:1996vz}.

Squaring the eikonal current and introducing the gluon polarization tensor \\\hbox{$d_{\mu \nu}(q)=(-g_{\mu \nu}+ \text{gauge terms} )$}, we find
\begin{equation}
\label{eq_bJJ1gTree}
\bJJ_g^{(0)}(q)\equiv\bJJ^{(0)}(q)\equiv \left[\boldsymbol{J}^{(0), \mu}(q)\right]^{\dagger} d_{\mu \nu}(q) \boldsymbol{J}^{(0),\mu}(q)=-2\sum_{i, j=1}^n \boldsymbol{T}_i \cdot \boldsymbol{T}_j
  \mathcal{S}_{ij}(q)
  \ldots\, ,
\end{equation}
where we defined 
\begin{equation}
\mathcal{S}_{ij}(q)\equiv \frac{p_i \cdot p_j}{2\left(p_i \cdot q\right)\left(p_j \cdot q\right)}\, ,
\end{equation} 
and we typically drop the flavor index of $\bJJ^{(0)}(q)$ because, at leading power, single soft emissions only contribute if the soft particle is a gluon. The notation $\bTi\cdot \bTj$ is defined as 
\begin{equation}
\boldsymbol{T}_i \cdot \boldsymbol{T}_j\equiv  \sum_a \bT_i^a \bT_j^a\, .
\end{equation} 
 The dots on the right-hand side of \ref{eq_bJJ1gTree} denote terms proportional to the total color charge $\sum_{i=1}^n \boldsymbol{T}_i$, which vanish in \eqref{eq_GeneralSoftLimitSquared} due to color conservation.

\paragraph{Gluon Emission at One Loop}
The one-loop matrix element in the limit where a single gluon becomes soft satisfies the factorization formula~\cite{Catani:2000pi}
\begin{equation}
  \label{eq:OneLoopSoftFac}
  \begin{aligned}
  \left\langle c;\mu \mid \amp^{(1)}\left(q, p_1, \ldots, p_m\right)\right\rangle \simeq g_S  \Biggl[ & \bJ^{(0),c;\mu}(q)\ket{\amp^{(1)}\left(p_1, \ldots, p_n\right)} \\
  & +\frac{\alpha_0}{\pi} \bJ^{(1),c;\mu}(q)\ket{\amp^{(0)}\left(p_1, \ldots, p_n\right)}\Biggr]\, ,
  \end{aligned}
  \end{equation}
  where the one-loop current is given by
\begin{equation}
  \label{eq_oneloopSoftCurrent}
  \begin{aligned}
 \bJ^{(1),a;\mu}(q)& =-\mu^{2\epsilon} \frac{c_S}{4\epsilon^2} \\
  & \times i f_{a b c} \sum_{i \neq j} \bT_i^b \bT_j^c\left(\frac{p_i^\mu}{p_i \cdot q}-\frac{p_j^\mu}{p_j \cdot q}\right)\left[\mathcal{S}_{ij}\frac{  e^{-i \sigma_{i j} \pi}}{e^{-i \sigma_{i q} \pi} e^{-i \sigma_{j q} \pi}}\right]^\epsilon .
  \end{aligned}
\end{equation}
In our notation, all the incoming and outgoing momenta are in the physical region (any $p_i$ has positive energy and $p_i \cdot p_j>0$). The complex factors $e^{-i \pi \sigma_{i j}}$ ($\sigma_{i j}=+1$ if $i$ and $j$ are both incoming or outgoing and $\sigma_{ij}=0$ otherwise) in \eqref{eq_oneloopSoftCurrent} are the unitarity phases related to the analytic continuation from unphysical to physical momenta. The constant $c_S$ is given by
\begin{equation}
  \label{eq_cs_definition}
  c_S=\frac{e^{\epsilon\gamma_E}\Gamma^3(1-\epsilon) \Gamma^2(1+\epsilon)}{\Gamma(1-2 \epsilon)}=1 +\frac{\pi ^2 }{12}\epsilon ^2-\frac{7 \zeta_3 }{3}\epsilon ^3-\frac{13 \pi ^4}{480} \epsilon ^4+ \mathcal{O}(\epsilon^5)\, .
\end{equation}
Using the eikonal current, we can write
  \begin{equation}
    \bJJ(q)= \left[\boldsymbol{J}^{\mu}(q)\right]^{\dagger} d_{\mu \nu}(q) \boldsymbol{J}^{\nu}(q)\, ,
  \end{equation}
  and in particular
  \begin{equation}
    \label{eq_bJJ1gOneLoop}
    \begin{aligned}
      \bJJ^{(1)}&=  \left[\boldsymbol{J}^{\mu(0)}(q)\right]^{\dagger} d_{\mu \nu}(q) \boldsymbol{J}^{\mu(1)}(q)+\left[\boldsymbol{J}^{\mu(1)}(q)\right]^{\dagger} d_{\mu \nu}(q) \boldsymbol{J}^{\mu(0)}(q)\\
       &=\mu^{2\epsilon}\frac{c_S}{\epsilon^2} \Biggl\{C_A \cos (\pi \epsilon) \sumprime_{i, j}\left[\mathcal{S}_{i j}(q)\right]^{1+\epsilon}\bTi\cdot \bTj \\
        &+2 \sin (\pi \epsilon) \sumprime_{i, j, k} \mathcal{S}_{k i}(q)\left[\mathcal{S}_{i j}(q)\right]^\epsilon\left(\sigma_{i j}-\sigma_{i q}-\sigma_{j q}\right)f^{abc}\bTk^a \bTi^b \bTj^c \Biggr\},
    \end{aligned}
  \end{equation}
  where the notation $\sumprime$ stands for the sum over the different values of the indices $(i \neq j, j \neq k, k \neq i)$.
  \paragraph{Double-Soft Emissions}
  For the double-soft limit, where two partons become soft at the same time, there are leading-power contributions both for the emission of a $q\bar{q}$ pair and for the emission of two gluons. The factorization formula for the squared matrix element for the emission of a $q\bar{q}$ pair reads~\cite{Catani:1999ss}
  \begin{equation}
    \begin{aligned}
    & \left|\amp^{(0)}_{q, \bar{q}, a_1, \ldots, a_n}\left(q_1, q_2, p_1, \ldots, p_n\right)\right|^2 \simeq\left(4 \pi \asbare\right)^2 \\
    & \quad \times\braket{\amp^{(0)}_{a_1, \ldots, a_n}\left(p_1, \ldots, p_n\right)|\bJJ^{(0)}_{q \bar{q}}\left(q_1, q_2\right)| \amp^{(0)}_{a_1, \ldots, a_n}\left(p_1, \ldots, p_n\right)}\, .
    \end{aligned}
  \end{equation}
  $\bJJ^{(0)}_{q \bar{q}}\left(q_1, q_2\right)$ can be written as
  \begin{equation}
    \begin{aligned}
    \bJJ^{(0)}_{q \bar{q}}\left(q_1, q_2\right) & =\left[\boldsymbol{J}^{(0)}_\mu\left(q_1+q_2\right)\right]^{\dagger} \Pi^{\mu \nu}\left(q_1, q_2\right) \boldsymbol{J}^{(0)}_\nu\left(q_1+q_2\right)+\ldots\, ,
    \end{aligned}
  \end{equation}
  where the dots denote terms that vanish due to color conservation, and we defined
  \begin{equation}
    \Pi^{\mu \nu}\left(q_1, q_2\right)=\frac{T_R}{\left(q_1 \cdot q_2\right)^2}\left\{-g^{\mu \nu} q_1 \cdot q_2+q_1^\mu q_2^\nu+q_2^\mu q_1^\nu\right\}\, .
    \end{equation}
    Performing the Lorentz algebra, we can also write

    \begin{equation}
      \label{eq_bJJqqbarTree}
       \bJJ^{(0)}_{q \bar{q}}\left(q_1, q_2\right)= -T_R\sum_{i, j}\bTi\cdot\bTj \frac{p_i\cdot p_j}{p_i\cdot \left( q_1+q_2 \right)p_j\cdot \left( q_1+q_2 \right) } \mathcal{S}_{q\bar{q}}(p_i, p_j;q_1, q_2)+\dots\, ,
    \end{equation}
    where we defined
    \begin{equation}
      \label{eq_Sqqbar}
      \mathcal{S}_{q\bar{q}}(p_i, p_j;q_1, q_2)=\frac{1-\frac{\left[\left( p_i\cdot q_2 \right) \left( p_j\cdot q_1 \right)-\left( p_i\cdot q_1 \right) \left( p_j\cdot q_2 \right)\right]^2}{\left( p_i\cdot p_j \right) \left( q_1\cdot q_2 \right)\, p_i\cdot (q_1+q_2)\, p_j\cdot (q_1+q_2)}}{q_1\cdot q_2}\, ,
    \end{equation}
    and we again dropped terms in \eqref{eq_bJJqqbarTree} that vanish due to color conservation. Using color conservation, $\mathcal{S}_{q\bar{q}}$ could be defined in many alternative ways. Equivalent expressions for \eqref{eq_Sqqbar} can be found in~\cite{Catani:1999ss}---here we chose this particular form because it makes it obvious that only pairs with $i\neq j$ contribute, and because, we will later see that the soft limit of the corresponding collinear splitting kernel looks very similar in this form.

    The tree-level matrix elements for the emission of two gluons in the soft limit factorize as~\cite{Catani:1999ss}
    \begin{equation}
      \label{eq_TwoGluonSoftLimitAmp}
     \begin{aligned}
       \left\langle a_1, a_2 ; \mu_1, \mu_2 \mid \amp_{g, g, a_1, \ldots, a_n}\left(q_1, q_2, p_1, \ldots, p_n\right)\right\rangle &\simeq \\
       &\hspace{-1.4cm}(4\pi\asbare)^2 \bJ_{a_1 a_2}^{\mu_1 \mu_2}\left(q_1, q_2\right)\ket{\amp_{a_1, \ldots, a_n}\left(p_1, \ldots, p_n\right)}\, ,
     \end{aligned}
    \end{equation}
    where the tree-level two-gluon soft current reads
    \begin{equation}
      \label{eq_twogluonCurrent}
      \begin{aligned}
      &  \bJ_{a_1 a_2}^{\mu_1 \mu_2}\left(q_1, q_2\right)=\frac{1}{2}\left\{\bJ_{a_1}^{\mu_1}\left(q_1\right), \bJ_{a_2}^{\mu_2}\left(q_2\right)\right\} \\
      & +i f_{a_1 a_2 a} \sum_{i=1}^n \bTi^a\left\{\frac{p_i^{\mu_1} q_1^{\mu_2}-p_i^{\mu_2} q_2^{\mu_1}}{\left(q_1 \cdot q_2\right)\left[p_i \cdot\left(q_1+q_2\right)\right]}-\frac{p_i \cdot\left(q_1-q_2\right)}{2\left[p_i \cdot\left(q_1+q_2\right)\right]}\left[\frac{p_i^{\mu_1} p_i^{\mu_2}}{\left(p_i \cdot q_1\right)\left(p_i \cdot q_2\right)}+\frac{g^{\mu_1 \mu_2}}{q_1 \cdot q_2}\right]\right\}\, .
      \end{aligned}
      \end{equation}
In \eqref{eq_TwoGluonSoftLimitAmp} and \eqref{eq_twogluonCurrent}, we put the color indices $a_1$ and $a_2$ as subscripts, and we left the loop counting, $(0)$, implicit everywhere to save space.
The reader is referred to~\cite{Catani:1999ss} for a list of the most important properties of $\bJ_{a_1 a_2}^{\mu_1 \mu_2}$. We define the leading order squared current as
\begin{equation}
    \bJJ^{(0)}_{gg}=
  {\left[\bJ_{a_1 a_2}^{\mu \rho}\left(q_1, q_2\right)\right]^{\dagger} d_{\mu \nu}\left(q_1\right) d_{\rho \sigma}\left(q_2\right) \bJ_{a_1 a_2}^{\nu \sigma}\left(q_1, q_2\right) } =\bJJ^{(0), \mathrm{ab}}_{gg}+\bJJ^{(0), \mathrm{nab}}_{gg} +\dots
  \end{equation}
  where 
  \begin{equation}
    \begin{aligned}
      \bJJ^{(0), \mathrm{ab}}_{gg}&=\frac{1}{2}\left\{\bJJ^{(0)}\left(q_1\right), \bJJ^{(0)}\left(q_2\right)\right\} \\
       \bJJ^{(0), \mathrm{nab}}_{gg}&=-C_A \sum_{i, j=1}^n \boldsymbol{T}_i \cdot \boldsymbol{T}_j \frac{p_i\cdot p_j}{p_i\cdot \left( q_1+q_2 \right)p_j\cdot \left( q_1+q_2 \right)}\mathcal{S}_{gg}\left(p_i, p_j;q_1, q_2\right)+\ldots\, ,
    \end{aligned}
  \end{equation}
  and the dots denote terms that vanish due to color conservation. We have split the squared soft current into an Abelian part $\bJJ^{(0), \mathrm{ab}}_{gg}$ and a non-Abelian part $\bJJ^{(0), \mathrm{nab}}_{gg}$. The function $\mathcal{S}_{gg}$ is again defined in a way that makes it clear that only two-particle terms contribute in the non-Abelian part, and we will see that $\mathcal{S}_{gg}$ also shows up in the double-soft limit of the triple-collinear limit. The explicit form of $\mathcal{S}_{gg}$ is
  \begin{equation}
    \label{eq_Sgg}
    \begin{aligned}
      \mathcal{S}_{gg}\left(p_i, p_j;q_1, q_2\right)&=\frac{(1-\epsilon )\left[  \left( p_i\cdot q_2 \right) \left( p_j\cdot q_1 \right)-\left( p_i\cdot q_1 \right)\left(  p_j\cdot q_2 \right) \right]^2}{\left( p_i\cdot p_j \right) (q_1\cdot q_2)^2 p_i\cdot (q_1+q_2) p_j\cdot (q_1+q_2)}\\
      &\hspace{-0.3cm}+\frac{\frac{p_i\cdot (q_1+q_2) p_j\cdot (q_1+q_2)}{\left( p_i\cdot q_2 \right) \left( p_j\cdot q_1 \right)}+\frac{p_i\cdot (q_1+q_2) p_j\cdot (q_1+q_2)}{\left( p_i\cdot q_1 \right) \left( p_j\cdot q_2 \right)}+\frac{p_i\cdot q_1}{p_i\cdot q_2}+\frac{p_i\cdot q_2}{p_i\cdot q_1}+\frac{p_j\cdot q_2}{p_j\cdot q_1}+\frac{p_j\cdot q_1}{p_j\cdot q_2}-4}{2 q_1\cdot q_2}\\
      &\hspace{-0.3cm}-\frac{\left( p_i\cdot p_j \right) (p_i\cdot (q_1+q_2) p_j\cdot (q_1+q_2)+\left( p_i\cdot q_1 \right) \left( p_j\cdot q_1 \right)+\left( p_i\cdot q_2 \right) \left( p_j\cdot q_2 \right))}{2 \left( p_i\cdot q_1 \right) \left( p_i\cdot q_2 \right) \left( p_j\cdot q_1 \right) \left( p_j\cdot q_2 \right)}\, .
    \end{aligned}
  \end{equation}

  As a final remark, we mention that all the soft currents defined in this paragraph could also be written in terms of matrix elements of soft Wilson lines of either QCD or SCET fields. An introduction to this formalism can be found, for example, in \cite{Agarwal:2021ais} for the case of QCD, or \cite{Becher:2014oda} for the case of SCET.

  Finally, we also define the shorthand notation
  \begin{equation}
    \bcalJJ_{b_1, \dots, b_m}\left(q_1, \dots, q_m\right)\equiv \left(4\pi \asbare  \right)^m\bJJ_{b_1, \dots, b_m}\left(q_1, \dots, q_m\right)
  \end{equation}
  as well as the trivial kernel for $m=0$,
  \begin{equation}
    \bcalJJ_{\emptyset}\equiv \bbbone\, ,
  \end{equation}
  which will be useful for the definition of soft functions later on.

  \subsection{Mixed Soft and Collinear Limits}\label{sec_MixedIRLimits}
  So far, we have only considered the singular limits where all unresolved partons were either soft or collinear along the same direction. However, in general, we can also have mixed soft and collinear limits, where some partons become soft, others become collinear along a direction $p$, and yet others go collinear along other directions. We must precisely define how these simultaneous limits are taken in these cases. For example, considering the amplitude $\amp_{g, a_1, \ldots, a_n}\left(q, p_1, \ldots, p_n\right)$ in the limit where the gluon becomes soft ($q \to 0$) and $p_1$ and $p_2$ become collinear along a direction $p$ ($k_\perp \to 0$), one has to decide how quickly $q$ and $k_\perp$ go to zero. For instance, one could keep the ratios $q/k_\perp$ or $q/k^2_\perp$ fixed. These mixed limits have not been studied as extensively as the standard soft or collinear limits at the level of the (squared) matrix element. The mixed double-unresolved limits were studied for color-ordered amplitudes in~\cite{Campbell:1997hg}, and an extensive analysis of mixed collinear and soft limits at tree level was performed in~\cite{Catani:1999ss}. An extensive analysis for an arbitrary number of timelike collinear emissions and soft particles was performed at tree level in~\cite{Feige:2013zla} and to all orders in~\cite{Feige:2014wja}. The study of the behavior of the squared matrix elements in these mixed limits is closely related to traditional studies of infrared factorization of hard scattering cross sections~\cite{Collins:1989gx,Bauer:2002nz}. 
  
This paper assumes a factorized ansatz for squared matrix elements with arbitrarily many collinear and soft emissions with arbitrary relative scalings. To be specific, for hadron colliders we consider the limit, where the set of all massless QCD partons in the partonic matrix element $\{k_i\}$ with flavors $\Af$ splits into sectors 
\begin{equation}
  \{k_i\}={I_1}\cup {I_2}\cup {S} \bigcup_{j=3}^{n_J+2}{F_j}\, ,
\end{equation} each carrying the flavors $\Af_1, \Af_2, \Af_S, \Af_3, \ldots $ respectively. Here, the partons in $I_1$ and $I_2$ are collinear to the two incoming partons, the partons in $F_i$ go collinear to some massless momentum $p_i$, and the partons in $S$ become soft. The sectors $F_i$ each contain at least one parton, while $I_1, I_2$, and $S$ can be empty. The simplest leading-power factorization formula for the all-order squared matrix element that would be consistent with what we have reviewed above reads
\begin{equation}
  \label{eq_GeneralSoftCollinearFactorization}
  \begin{aligned}
    &\left|\amp_{b_1, b_2, \Af}\left(\{k_i\}\right)\right|^2 \simeq \Tr\bigg[ \ket{ \amp_{\A}\left(p_1, \ldots, p_{2+n_J}\right)}\bra{  \amp_{\A}\left(p_1, \ldots, p_{2+n_J}\right)}\\
   & \times \bcalPhat_{a_1b_1;\Af_1}(\{k_i\}_{I_1}, z_1)\bcalPhat_{a_2b_2;\Af_2}(\{k_i\}_{I_2}, z_2)\bcalJJ_{\Af_S}(\{k_i\}_{S})\prod_{j=3}^{n_J+2}\calPhat_{a_j\to \Af_j}(\{k_i\}_{F_j}) \bigg]\, ,
  \end{aligned}
\end{equation}
where $\A=\{a_1, a_2, \dots, a_{n_J+2}\}$ is the set of hard parent parton flavors that can be determined by requiring the splittings $a_j\to \Af_j$ and $b_{1(2)}\to a_{1(2)}\cup \Af_{1(2)}$ to be flavor conserving. All $\mathcal{\hat{P}}$ are spin operators acting on the spin space of the respective parent partons, and all bold symbols represent operators on the color space of the hard amplitude $\ket{\amp_\A}$. In \eqref{eq_GeneralSoftCollinearFactorization}, we also account for the fact that naive collinear factorization is not valid at $\mytext{N}^3\mytext{LO}$, i.e., the collinear splitting kernels also depend on the color charges and directions of the non-collinear partons. However, we suppressed the momentum dependence to keep the formula short. The ansatz \eqref{eq_GeneralSoftCollinearFactorization} is broken at higher orders due to Glauber effects. There are preliminary works~\cite{Cieri:2024ytf} that attempt a generalization of \eqref{eq_GeneralSoftCollinearFactorization}. In the framework developed in Section~\ref{Gen_sec_FactorizationForResolutionVariables}, we will work under the assumption that \eqref{eq_GeneralSoftCollinearFactorization} holds. We note that standard versions of SCET without Glauber modes imply the ansatz in \eqref{eq_GeneralSoftCollinearFactorization}. Thus, the factorization formulas presented in this paper are in no way less rigorous or general than typical SCET-based factorization formulas like the one for $n-$jettiness derived in~\cite{Stewart:2010tn}. 

A systematic method to derive factorization formulas that go beyond the ansatz \eqref{eq_GeneralSoftCollinearFactorization} is SCET with Glauber modes~\cite{Rothstein:2016bsq}. A phenomenon that is closely related to the breakdown of \eqref{eq_GeneralSoftCollinearFactorization} due to Glauber modes is the appearance of super-leading logarithms (SLLs)~\cite{Forshaw:2006fk}. In the context of SLLs, the breakdown of \eqref{eq_GeneralSoftCollinearFactorization} in the case where one space-like collinear emission and a soft emission are present at one loop has been tracked down to a specific set of Feynman diagrams in~\cite{Becher:2024kmk}.

Presently, our framework is used only for NNLO calculations in pQCD, for which \eqref{eq_GeneralSoftCollinearFactorization} holds. Using our framework to higher orders in perturbation theory and/or performing high logarithmic accuracy resummation building on our framework, one should remember that the full result might be missing terms due to a breakdown of \eqref{eq_GeneralSoftCollinearFactorization}.

\section{Factorization for Resolution Variables}
\label{Gen_sec_FactorizationForResolutionVariables}

\subsection{Phase Space Factorization}
\label{Gen_sec_PhaseSpaceFactorization}
This section shows how the Lorentz invariant partonic phase space can be systematically expanded in the unresolved region. Together with the expansion of the squared matrix elements discussed in Section~\ref{sec_SingularLimits} this will allow use to factorize cross sections of the form \eqref{Gen_eq_PartonicCrossSection} into soft and collinear parts. 

If one is only interested in the leading-power behavior of the cross section, one could simplify and shorten the following analysis. However, we believe the results in this section will be useful for potential subleading power applications in the future. We also point out that already at leading power there are subtleties that could easily be missed if one is not precise enough with the phase space expansion. Such subtleties arise particularly for regions with collinear and soft sectors with comparable transverse momenta, i.e., if $a\geq c$ or $b\geq c$. In this case the recoil of the soft sector against the collinear sectors is not negligible, even at leading power, and has to be taken into account. We have seen an example of this in Section~\ref{Gen_sec_PCExamples} when considering the total broadening $B_T$ whose leading power expansion crucially depends on the fact that the soft momenta deflect the thrust axis away from the hard direction.

The partonic $n-$particle phase space relevant for hadroproduction of a colorless system accompanied by $n_J$ jets can be written as
\begin{equation}
  \frac{1}{2\hat{s}}\rd x_1\, \rd x_2 \, \rd\Pi_n=\frac{\rd x_1\, \rd x_2}{2\hat{s}}\prod_{i=1}^n \dki \, \rd p_F \rd \Pi_F \deltabar^{\, d}\biggl[ \sum_i k_i+p_F-x_1 P_1-x_2 P_2\biggr]\,  \F\, ,
\end{equation}
where 
\begin{equation}
  \dki=\frac{\rd^d k_i\delta_+(k_i^2)}{(2\pi)^{d-1}}\, , \quad \deltabar^{\, d}(k)=(2\pi)^d\delta^d(k)\, , \quad \frac{1}{2\hat{s}}=\frac{1}{4x_1x_2P_1\cdot P_2}\, .
\end{equation}
The $k_i$ are the momenta of the massless colored particles, $x_1P_1$ and $x_2P_2$ are the momenta of the initial-state partons extracted from the (bare) PDFs, $p_F$ is the total four-momentum of the colorless system and $\rd \Pi_F$ is the Lorentz invariant phase space for the decay of the colorless system. We also included an IR-safe measurement function $\F$ in the phase space. It ensures that we have at least $n_J$ jets, which are well separated and have a minimum energy or transverse momentum. We also included the flux factor in the phase space and the integration measure for the partonic momentum fractions $\rd x_1 \, \rd x_2$.

We now want to rewrite the phase space in a way that allows for a systematic expansion in the region where each of the $k_i$ is associated with one of the sectors $I_1, I_2, F_3, \dots, F_{n_J+2}$ or $S$. To proceed, we introduce the total momenta $k_{I_1}, k_{I_2}, K_3, \dots, K_{n_J+2}, k_S$ in the respective sectors. Here, we chose capital letters for the momenta we eventually want to approximate with a corresponding hard momentum in the Born phase space. We can now rewrite the phase space as
\begin{equation}
  \begin{aligned}
     \frac{1}{2\hat{s}}\rd x_1\, \rd x_2 \, \rd\Pi_n&=\frac{\rd x_1\, \rd x_2}{2\hat{s}} \rd^d k_{I_1}\rd \Pi_{I_1}\rd^d k_{I_2}\rd \Pi_{I_2}\rd^d k_{S}\rd \Pi_{S} \prod_{i=3}^{n_J+2}\rd^d\!K_{i}\rd \Pi_{i}\, \rd p_F \rd \Pi_F\\
    &\quad \times \deltabar^{\, d}\biggl[ \sum_{i=3}^{n_J+2} K_i+k_{I_1}+k_{I_2}+k_S+p_F-x_1 P_1-x_2 P_2\biggr]\,  \F\, ,
  \end{aligned}
\end{equation}
where 
\begin{equation}
  \rd \Pi_{i}=\prod_{j\in F_i}\dkj\delta^d(K_i-\sum_j k_j)
\end{equation}
and $\, \rd \Pi_{I_{1(2)}}$, and $\rd \Pi_S$ are defined analogously. We should somehow be able to approximate the momentum conservation $\delta$-function by only including the large collinear momentum fractions of the initial-state momenta $k_{I_{1(2)}}$. Additionally, there should be a way to ignore the soft momenta in the $\delta$ function. We will see in a moment that this intuition is almost correct. However, first, we need to understand the subtlety we have alluded to above. If one considers, for example, dijet production at hadron colliders with a resolution variable that requires $a=b=c$, then the soft momenta are not negligible in the direction that is both transverse to the beam and the dijet pair---the momenta in all sectors scale as $\lambda$ in this direction. Similarly, one cannot ignore the transverse components of the initial-state collinear momenta. Therefore, we should be more careful when decoupling the soft and initial-state collinear momenta.

To extract the collinear momentum fractions of the initial-state collinear emissions, we insert  a ``one'' in the form 
\begin{equation}
  1=\int_0^1 \rd \bar{z}_1\delta\left( \bar{z}_1-\frac{k_{I_1}\cdot P_2}{x_1P_1\cdot P_2} \right)\int_0^1 \rd \bar{z}_2\delta\left( \bar{z}_2-\frac{k_{I_2}\cdot P_1}{x_2P_1\cdot P_2} \right)\, ,
\end{equation}
and we split the total initial-state collinear momenta into a large and a small part as
\begin{equation}
  k_{I_i}\equiv\bar{z}_i x_i P_i+\tilde{k}_{I_i}\, .
\end{equation}
The phase space can now be recast into the form
\begin{equation}
  \begin{aligned}
     \frac{1}{2\hat{s}}\rd x_1\, \rd x_2 \, \rd\Pi_n&=\frac{\rd \tilde{x}_1\, \rd \tilde{x}_2}{2Q^2} \frac{\rd z_1\rd z_2}{z_1z_2}\left[ \rd^d k_{I_1}\rd \Pi_{I_1}z_1\delta\left( \bar{z}_1-\frac{k_{I_1}\cdot P_2}{x_1P_1\cdot P_2} \right) \times \left( 1\leftrightarrow 2 \right) \right] \\
    &\quad \times \prod_{i=3}^{n_J+2}\rd^d K_{i}\rd \Pi_{i}\rd^d k_{S}\rd \Pi_{S} \, \rd p_F\deltabar^{\, d}\biggl[ \sum_{i=3}^{n_J+2} K_i+p_F+\krec-q\biggr]\, \rd \Pi_F \F\, ,
  \end{aligned}
\end{equation}
where we defined 
\begin{equation}
  \krec\equiv \tilde{k}_{I_1}+\tilde{k}_{I_2}+k_S\, , \quad z_i\equiv 1-\bar{z}_i\, , \quad \tilde{x}_i=z_i x_i\, , \quad q=\tilde{x}_1 P_1+\tilde{x}_2 P_2 , \quad Q^2=q^2\, .
\end{equation}
Next, we will remove the recoil momentum $\krec$ from the momentum conservation $\delta$-function. To do this, we define a family of Lorentz boosts $\mathbb{B}[\krec]$ such that 
\begin{equation}
  \mathbb{B}[\krec]\left(q-\krec\right)=\sqrt{\frac{(q-\krec)^2}{Q^2}}q\, , \quad \mathbb{B}[0]=\bbbone\, .
\end{equation} 
One possible choice for $\mathbb{B}[\krec]$ can be derived by boosting along the direction of $q-\krec$ in the $q$ rest frame. In order to write an explicit formula for the action of $\mathbb{B}[\krec]$ on a four-vector $k$ we define 
\begin{equation}
  \begin{aligned}
    \brec&=\frac{\sqrt{\left(q\cdot k_{\mathrm{rec}}\right){}^2-Q^2 k_{\mathrm{rec}}^2}}{Q^2-q\cdot k_{\mathrm{rec}}}\quad
    y=\sqrt{\frac{Q^2-q\cdot\krec-\sqrt{(q\cdot\krec)^2-Q^2\krec^2}}{Q^2-q\cdot\krec+\sqrt{(q\cdot\krec)^2-Q^2\krec^2}}}\\
    \nrec&=\frac{Q \left(q-k_{\text{rec}}\right)}{\beta _{\text{rec}}\left(Q^2-q\cdot k_\mathrm{rec}\right) }+\left(1-\frac{1}{\beta _{\text{rec}}}\right)\frac{q}{Q} \quad
    \nbrec=-\frac{Q \left(q-k_{\text{rec}}\right)}{\beta _{\text{rec}}\left(Q^2-q\cdot k_\mathrm{rec}\right) }+\left(1+\frac{1}{\beta _{\text{rec}}}\right)\frac{q}{Q}\, .
  \end{aligned}
\end{equation}
An arbitrary four-vector $k$ then transforms as
\begin{equation}
  \label{Gen_eq_BoostedMomentum}
  \begin{aligned}
    \mathbb{B}[\krec]k&=k+\nrec\cdot k \left( \frac{1}{y}-1 \right)\frac{\nbrec}{2}+\nbrec\cdot k \left( y-1 \right)\frac{\nrec}{2}\, .
  \end{aligned}
\end{equation}
For any given momentum $k$, we then define the boosted and rescaled momentum $\tilde{k}$ as 
\begin{equation}
  \label{Gen_eq_BoostedRescaledMomentum}
  \tilde{k}\equiv \frac{Q}{\sqrt{(q-\krec)^2}}\mathbb{B}[\krec]k\, .
\end{equation}
We can now rewrite the phase space as
\begin{equation}
  \label{Gen_eq_PhaseSpacewithTildeMomenta}
  \begin{aligned}
     \frac{1}{2\hat{s}}\rd x_1\, \rd x_2 \, \rd\Pi_n&\sim\frac{\rd \tilde{x}_1\, \rd \tilde{x}_2}{2Q^2} \frac{\rd z_1\rd z_2}{z_1z_2}\left[ \rd^d k_{I_1}\rd \Pi_{I_1}z_1\delta\left( \bar{z}_1-\frac{k_{I_1}\cdot P_2}{x_1P_1\cdot P_2} \right)\times\left( 1\leftrightarrow 2 \right) \right]\\
    &\times \rd^d k_{S}\rd \Pi_{S}\prod_{i=3}^{n_J+2}\rd^d \tilde{K}_{i}\rd \tilde{\Pi}_{i}\, \rd \tilde{p}_F \deltabar^{\, d}\biggl[ \sum_{i=3}^{n_J+2} \tilde{K}_i+\tilde{p}_F-q\biggr]\, \rd \tilde{\Pi}_F \F\, ,
  \end{aligned}
\end{equation}
where we also defined 
\begin{equation}
  \rd \tilde{\Pi}_{i}=\prod_{j\in F_i}\dktj\delta^d(\tilde{K}_i-\sum_j \tilde{k}_j)
\end{equation}
as well as the analogous definition for $\rd \tilde{\Pi}_F$, and we dropped a subleading power correction in the Jacobian due to the rescaling in Eq.~\eqref{Gen_eq_BoostedRescaledMomentum}.

Now the second line in \eqref{Gen_eq_PhaseSpacewithTildeMomenta} looks like it contains the phase space for the Born process, except that the jet momenta $\tilde{K}_i$ are massive. We could introduce a map between the configuration with massive $\tilde{K}_i$ and a massless configuration to make the phase space look even more like the Born phase space. However, since we are currently only interested in the leading-power expansion, we will not pursue this further in this paper. 

The next step is to expand the phase space to leading power. We can make the following two approximations.
\begin{enumerate}
  \item The integration variables $k_{I_i}$ and $k_S$ in \eqref{Gen_eq_PhaseSpacewithTildeMomenta} are in principle constrained by the condition that the boost $B[\krec]$ exists. However, at leading power, this constraint is satisfied, and $k_{I_i}$ and $k_S$ can be any non-space-like four-vectors with positive energy. 
  \item One can replace the $\tilde{K}_i$ with massless vectors $p_i$ in the momentum conservation $\delta$-function.
\end{enumerate}
To make the second point more systematic, we now define the reference vectors
\begin{equation}
  \label{eq_nivecs}
  \begin{aligned}
    n_i&=\frac{Q}{\beta_i \tilde{K}_i\cdot q}\tilde{K}_i+\left( 1-\frac{1}{\beta_i} \right)\frac{q}{Q}\\
    \bar{n}_i&=-\frac{Q}{\beta_i \tilde{K}_i\cdot q}\tilde{K}_i+\left( 1+\frac{1}{\beta_i} \right)\frac{q}{Q}
  \end{aligned}\, ,
\end{equation}
where
\begin{equation}
  \label{eq_betaidef}
  \beta_i=\sqrt{1-\frac{\tilde{K}_i^2Q^2}{(\tilde{K}_i\cdot q)^2}}\, .
\end{equation}

The reference vectors satisfy $n^2_i=\bar{n}^2_i=0$, $n_i\cdot \bar{n}_i =2$ and $q=\frac{Q}{2}\left( n_i+\bar{n}_i\right)$. In any rest frame of $q$, $n_i$ and $\bar{n}_i$ are back-to-back unit vectors, i.e., they have unit spatial length. We now define massless four-vectors $p_i=p_i^+ \frac{n_i}{2}$ such that
\begin{equation}
  \label{eq_pidef}
  \tilde{K}_i=p_i+\frac{\tilde{K}_i^2}{4p_i\cdot q}Q\bar{n}_i\, .
\end{equation}
The phase space measure for $\tilde{K}_i$ can then be written as
\begin{equation}
  \label{eq_PhaseSpaceMeasureTildeKi}
  \rd^d\tilde{K}_i=\rd^dp_i\delta_+(p_i^2)\rd \tilde{K}_i^2\left( 1-\frac{\tilde{K}_i^2 Q^2}{4 \left( p_i\cdot q \right)^2} \right)^{d-2}\sim \rd^dp_i\delta_+(p_i^2)\rd \tilde{K}_i^2\, .
\end{equation}
It is now clear that $\tilde{K}_i$ can be approximated with $p_i$ in the momentum conserving $\delta$-function---the $\bar{n}_i$ component of $q$ is much larger than the $\bar{n}_i$ component of $\tilde{K}_i$ because $\tilde{K}_i^2\sim \lambda^{2b}$. Finally, we note that the $\delta-$function in $\tilde{\Pi}_i$ can be used to remove the $\tilde{K}_i^2$ integration. We find
\begin{equation}
  \begin{aligned}
    \rd \tilde{K}_i^2\rd \tilde{\Pi}_i&=\rd \tilde{K}_i^2\prod_{j\in F_i}\dktj\delta^d(p_i+\frac{\tilde{K}_i^2}{4p_i\cdot q}Q\bar{n}_i-\sum_j \tilde{k}_j)\\
    &=2\, \rd \tilde{K}_i^2\prod_{j\in F_i}\dktj\delta\left( p_i^+-\sum_j\tilde{k}_j^+\right)\delta \left( \frac{\tilde{K}_i^2}{2p_i\cdot q}Q-\sum_j\tilde{k}_j^- \right)\delta^{d-2}\left( \tilde{k}_{j, \perp} \right)\\
    &=2\prod_{j\in F_i}\dktj\delta(1-\sum_j\tilde{z}_j)\delta^{d-2}\left(\sum_j \tilde{k}_{j, \perp} \right)\, ,
  \end{aligned}
\end{equation}
where we defined the longitudinal momentum fractions in the sector $F_i$ as
\begin{equation}
  \tilde{z}_j=\frac{\tilde{k}_j\cdot \bar{n}_i}{p_i\cdot\bar{n}_i}\, ,
\end{equation}
and the light-cone coordinates are defined such that 
\begin{equation}
  \tilde{k}_j=\tilde{k}_j^+ \frac{n_i}{2}+\tilde{k}_j^- \frac{\bar{n}_i}{2}+\tilde{k}_{j, \perp}\, .
\end{equation}
We can conclude that the leading-power phase space reads
\begin{equation}
  \label{Gen_eq_PhaseSpaceLeadingPower}
  \begin{aligned}
     \frac{1}{2\hat{s}}\rd x_1\, \rd x_2 \, \rd\Pi_n&\sim\frac{\rd \tilde{x}_1\, \rd \tilde{x}_2}{2Q^2}\prod_{i=3}^{n_J+2} [\rd p_i]\, \rd \tilde{p}_F \deltabar^{\, d}\biggl[ \sum_{i=3}^{n_J+2} p_i+\tilde{p}_F-q\biggr]\, \rd \tilde{\Pi}_F \F(\{p_i\}, F)\\
    &\hspace{-2cm}\quad \times\frac{\rd z_1\rd z_2}{z_1z_2}\rd^d k_{I_1}\rd \Pi_{I_1}z_1\delta\left( \bar{z}_1-\frac{k_{I_1}\cdot P_2}{x_1P_1\cdot P_2} \right) \rd^d k_{I_2}\rd \Pi_{I_2}z_2\delta\left( \bar{z}_2-\frac{k_{I_2}\cdot P_1}{x_2P_1\cdot P_2} \right)\rd^d k_{S}\rd \Pi_{S} \\
    &\hspace{-2cm}\quad\times\prod_{i=3}^{n_J+2} 2(2\pi)^{d-1}\prod_{j\in F_i}\dktj\delta(1-\sum_j\tilde{z}_j)\delta^{d-2}\left( \sum_j \tilde{k}_{j, \perp} \right)\, ,
  \end{aligned}
\end{equation}
where the first line is the Born phase space together with the measurement function $\F$, which now only depends on the Born-level jet momenta $p_i$ and the colorless final-state $F$. The second line contains the phase space for initial-state collinear splittings and soft emissions, and the last line contains the phase space for emissions collinear to the jets. We remind the reader that $q=p_1+p_2$, where we defined the parton momenta entering the hard cross section as 
\begin{equation}
  p_{1(2)}\equiv \tilde{x}_{1(2)}P_{1(2)}\, .
\end{equation}

Now that we have completely factorized the Born phase space from the radiation phase space, let us summarize how one can write the various reference vectors we defined in terms of the Born momenta. We have the following relations
\begin{equation}
  \begin{aligned}
    n_i&=\frac{Q}{p_i\cdot q}p_i\\
    \bar{n}_i&=-\frac{Q}{p_i\cdot q}p_i+\frac{2}{Q}q=\frac{Q}{p_i\cdot q}\bar{p}_i\, \quad \, ,
  \end{aligned}
\end{equation}
where we also extend the definition of the reference vectors to the initial-state legs $i=1, 2$. 

Finally, we must stress that the matrix element and the resolution variable are functions of the original momenta $k_i$ rather than the boosted and rescaled momenta $\tilde{k}_i$. However, it turns out that the leading-power matrix elements are more easily expressed in terms of the $\tilde{k}_i$, as we will see shortly. To express the resolution variable in terms of the $\tilde{k}_i$ rather than the $k_i$, we must expand relation \eqref{Gen_eq_BoostedRescaledMomentum}. The leading-order relation for each light cone component of a collinear momentum $k_j\in F_i$ reads
\begin{equation}
  \label{Gen_eq_LP-relations-for-ktilde-LC}
  \begin{aligned}
    k_j^+&=\tilde{k}_j^+ + \dots \\
    k_j^-&=\tilde{k}_j^-+\frac{\tilde{k}_{j\perp}\cdot k_{\mathrm{rec}, \perp}}{Q}-\frac{\tilde{k}_j^+ k_{\mathrm{rec}, \perp}\cdot k_{\mathrm{rec}, \perp}}{4Q^2}+\dots\\
    k_{j, \perp}&=\tilde{k}_{j, \perp}-\frac{\tilde{k}_{j}^+}{2Q}{\krec}_{\perp}+\dots\, ,
  \end{aligned}
\end{equation}
where the light-cone coordinates are given with respect to $n_i$ and $\bar{n}_i$. The $\krec$ terms in \eqref{Gen_eq_LP-relations-for-ktilde-LC} are subleading if $a>b$ and $c>b$, as is typically the case for \SCETI type resolution variables like $n-$jettiness. Note in particular that the sum of all transverse momenta in the collinear sector is
\begin{equation}
  \sum_{j\in F_i} k_{j, \perp}=-\frac{p^+_i}{2Q}{\krec}_{\perp}+\dots=-\frac{p_i\cdot q}{Q^2}{\krec}_{\perp}+\dots
\end{equation}
and thus
\begin{equation}
  \label{eq_kjperptild_in_terms_of_kjperp_LP}
  \tilde{k}_{j, \perp}=k_{j, \perp}-z_j\sum_{j^\prime\in F_i} k_{j^\prime, \perp}+\dots\, ,
\end{equation}
where we also defined 
\begin{equation}
  \label{eq_zjtildsimzj}
  z_j\equiv\frac{k_j\cdot n_i}{p_i\cdot n_i}=\frac{k_j^+}{p_i^+}\sim \tilde{z}_j\, .
\end{equation}
\secref{sec_CollinearLimits} explained that the collinear matrix elements only depend on the boost invariant variables defined in Eq.~\eqref{eq_BoostInvariantVariables}. From \eqref{eq_kjperptild_in_terms_of_kjperp_LP} and \eqref{eq_zjtildsimzj}, we see that the light-cone components of the boosted and rescaled momenta $\tilde{k}_j$ directly gives us the boost invariant variables introduced in \eqref{eq_BoostInvariantVariables} (which is also the reason why we used the notation $\tilde{z}_i$ and $\tilde{k}_{\perp i}$ in  \eqref{eq_BoostInvariantVariables} in the first place). 

 In conclusion, this section has shown that, to leading power, the partonic momenta can indeed be written as anticipated in \eqref{Momentum_With_Homogeneous_Scaling}. We have found a way to express all momenta in terms of a Born configuration, the soft momenta, collinear momentum fractions, and transverse momenta with respect to the collinear direction. All these variables have a well-defined power counting, allowing us to systematically power expand kinematic functions as we showed for several examples in Section~\ref{Gen_sec_PCExamples}. The technical details on how such expansions can be performed in general are summarized in Appendix~\ref{sec_LimitsOfKinematicFunctions}.
\subsection{Factorization into Soft and Collinear Functions} 
\label{Gen_sec_SoftAndCollinearFunctions}
Having studied how the matrix element and the phase space factorize, we now turn to the factorization of the cross section into soft and collinear functions. We remind the reader that our working assumption, motivated in Section~\ref{sec_MixedIRLimits}, is that in the region 
\begin{equation}
\begin{aligned}
    \{k_i\}&=I_1\cup I_2\cup S \cup\bigcup_{j=3}^{n_J+2}F_j\\
    I_i&=\{k_{i_1},k_{i_2},\dots\},\quad F_i=\{k_{i_1},k_{i_2},\dots\}\quad , S=\{k_{S_1},k_{S_2},\dots\}\, ,
\end{aligned}
\end{equation} the all-order partonic squared matrix element $\lvert M_{b_1b_2;\Af_n}(x_1P_1, x_2P_2, \{k_i\}, p_F)\rvert^2$ approximates as
\begin{equation}
  \label{Gen_eq_MEfactorization}
    \begin{aligned}
        &\lvert M_{b_1b_2\Af_n}(x_1P_1, x_2P_2, \{k_i\}, p_F)\rvert^2=\Tr\biggl(\Hb_\A(\tilde{x}_1P_1, \tilde{x}_2P_2, \{p_i\}, \tilde{p}_F)\\
        &\times \bcalJJ_{\Af_S}(S, n_1, n_2, \{n_i\})\bcalPhat_{a_1b_1;\Af_1}(\tilde{x}_1P_1, I_1)\bcalPhat_{a_2b_2;\Af_2}(\tilde{x}_2P_2, I_2) \prod_{i=3}^{n_J+2}\calPhat_{\Af_i}(\tilde{F}_i)
         \biggr)
  \end{aligned}
\end{equation}
to leading power, independent of the specific choices of scaling dimensions $a,b$ and $c$. In \eqref{Gen_eq_MEfactorization} we defined the \emph{bare hard function}
\begin{equation}
\Hb_{\A}(\tilde{x}_1P_1, \tilde{x}_2P_2, \{p_i\}, \tilde{p}_F)=\ket{M_{\A}(\tilde{x}_1P_1, \tilde{x}_2P_2, \{p_i\}, \tilde{p}_F)}\bra{M_{\A}(\tilde{x}_1P_1, \tilde{x}_2P_2, \{p_i\}, \tilde{p}_F)}\, ,
\end{equation}
where $\ket{M_{\A}}$ denotes the all-order matrix element in color and spin space. 

 For convenience, we also define the set of boosted and rescaled momenta 
\begin{equation}
  \tilde{F}_i=\{\tilde{k}_{i_1},\tilde{k}_{i_2},\dots\}\, .
\end{equation}

Our next step is to group all contributions from the phase space and the matrix element with the same scaling. This leads to the factorization formula for the bare\footnote{Bare in the sense that this does not contain the mass factorization counter term yet.} partonic cross section with a cut on the resolution variable $\qfcut$
\begin{equation}
  \label{gen:eq_FactorizationFormulaFully}
  \begin{aligned}
    &\rd x_1\rd x_2\hat{\sigma}_{b_1b_2}(\qfcut)=\int \frac{\rd \tilde{x}_1\, \rd \tilde{x}_2}{2Q^2}\frac{\rd z_1\rd z_2}{z_1 z_2}\prod_{i=3}^{n_J+2} [\rd p_i]\, \rd \tilde{p}_F \deltabar^{\, d}\biggl[ \sum_{i=3}^{n_J+2} p_i+\tilde{p}_F-q\biggr] \rd \tilde{\Pi}_F \F(\{p_i\}, F)\\
    &\times \sum_\A
   \Tr \Biggl\{ \frac{1}{S_\A} \Hb_\A(\tilde{x}_1P_1, \tilde{x}_2P_2, \{p_i\}, \tilde{p}_F) \Biggl[{\sum_{n=n_J}^\infty}\sum_{\pf_n\in\mathfrak{P}_n}\int \rd\Bb_{a_1b_1}(I_1, z_1)\rd\Bb_{a_2b_2}(I_2, z_2)\\
   &\times\rd\Sb(S, n_1, n_2, \{n_i\})
    \prod_{i=1}^{n_j}\rd\J_{a_i}(\tilde{F}_i)\theta\left(\qfcut-\qft_{\pf_n}(\{k\},p_F)\right)\Biggr] \Biggr\} \, ,
  \end{aligned}
\end{equation}
where we sum over the possible multiplicity of massless QCD final-state partons $n$. Here we also introduced a symmetry factor $S_\A$ to account for identical particles in the Born phase space. $\mathfrak{P}_n$ denotes the set of all possible partitions of the set of momenta $\{k_1, \dots, k_n\}$ into the sectors $I_1, I_2, F_i$ and $S$, and $\pf_n$ denotes a specific choice of such a partition, i.e., a region. Partitions that can be made equal by relabeling the momenta are considered the same, i.e., only one of the equivalent permutations appears in the sum. The functions $\rd\Bb$, $\rd\Sb$, and $\rd\J$ are the fully differential radiative bare beam, soft, and jet functions, respectively, and they are defined below. $\qft_{\pf_n}$ denotes the leading-power approximation of the resolution variable in the region specified by $\pf_n$. A detailed explanation on how to take these limits is provided in Appendix~\ref{sec_LimitsOfKinematicFunctions}, and we provided several examples in Section~\ref{Gen_sec_PCExamples}. We point out that while the differential jet function $\rd\J_{a_i}$ does only depend on the boosted and rescaled momenta $\tilde{F}_i$ and is therefore independent of the recoil $\krec$, the resolution variable $\qft_{\pf_n}$ generally depends on the actual momenta $F_i$ and thus one needs to carefully treat the recoil by expressing the momenta  $F_i$ with the $\tilde{F}_i$ and $k_\mathrm{rec}$ through the relations summarized in \eqref{Gen_eq_LP-relations-for-ktilde-LC}. 

Remember that \eqref{gen:eq_FactorizationFormulaFully} was obtained through the method of regions. We fully expanded the cross section in all relevant regions. In particular, this means that there is no double counting in \eqref{gen:eq_FactorizationFormulaFully}. If we pick one specific region in \eqref{gen:eq_FactorizationFormulaFully}, specified by a partition $\pf_n$, and we expand the corresponding integrand in another region $\pf_n^\prime$, we would obtain a scaleless integral that vanishes in dimensional regularization. However, we will later see that \eqref{gen:eq_FactorizationFormulaFully} is sometimes ill-defined due to the presence of \emph{rapidity divergences}. When regularizing these divergences, we have to be careful to avoid double counting.

The full expression inside the square brackets in \eqref{gen:eq_FactorizationFormulaFully} can be understood as a complicated convolution of the functions $\rd\Bb$, $\rd\Sb$, and $\rd\J$. We will see in \secref{Gen_sec_VariableFactorization} that for specific functional forms of $\qft_{\pf_n}$ the convolution can be simplified significantly. The functions $\rd\Bb$, $\rd\Sb$ and $\rd\J$ are defined as 
\begin{equation}
  \label{Gen_eq_RadiativeFunctionsDifferential}
  \begin{aligned}
    \rd\Bb_{a_1b_1}(I_1, z_1)&=\sum_{\Af_1}\frac{1}{S_{\Af_1}}z_1\rd^d k_{I_1}\rd \Pi_{I_1}\delta\!\left( \bar{z}_1-\frac{k_{I_1}\cdot P_2}{x_1P_1\cdot P_2} \right)\!\bcalPhat_{a_1b_1;\Af_1}(I_i, z_i)\\
    \rd\Sb(S, n_1, n_2, \{n_i\})&=\sum_{\Af_S}\frac{1}{S_{S}}\rd^d k_{S}\rd \Pi_{S}\bcalJJ_{\Af_S}(S, n_1, n_2, \{n_i\})\\
    \rd\J_{a_i}(\tilde{F}_i)&=\sum_{\Af_i}\frac{1}{S_{\Af_i}}2(2\pi)^{d-1}\prod_{j\in F_i}\dktj\delta(1-\sum_j\tilde{z}_j)\delta^{d-2}\left( \tilde{k}_{j, \perp} \right) \calPhat_{\Af_i}(\tilde{F}_i)\, ,
  \end{aligned}
\end{equation}
where the factors $\frac{1}{S_{\Af_1}},\frac{1}{S_{S}}$ and $\frac{1}{S_{\Af_i}}$ are symmetry factors accounting for identical particles. Note that we included the phase space measure for the soft and collinear emissions in the definition of the radiative functions. It will be interesting to compare our definition of the differential radiative jet functions to operator-based definitions of amplitude-level jet functions, for example, defined in~\cite{Bonocore:2015esa, Bonocore:2016awd, Magnea:2018ebr, Beneke:2019oqx, Liu:2020ydl, Liu:2021mac} or to the \emph{collinear transition probability} defined, for example, in~\cite{Agarwal:2021ais}. The differential beam function defined in this way is similar to the one defined in~\cite{Catani:2022sgr}, except that the beam functions in~\cite{Catani:2022sgr} are more inclusive. 

We point out that all our radiative functions and the hard function $\Hb$ still contain poles in $\epsilon$. Some of these poles cancel within single ingredients. For example, the single-emission soft function $ \rd\Sb(\{k_1\}, n_1, n_2, \{n_i\})$ contains explicit poles in $\epsilon$ at order $\alpha_S^2$ due to virtual gluons becoming soft or collinear to the emitted gluon $k_1$. The double-emission soft function $\rd\Sb(\{k_1, k_2\}, n_1, n_2, \{n_i\})$ at order $\alpha_S^2$ does not contain any explicit poles in the squared soft current. However, when the two emitted gluons become collinear, or if one of them becomes much softer than the other, the matrix elements become divergent, and they will also generate $\epsilon$-poles once integrated, which will cancel with the aforementioned poles in the single emission soft function. This cancellation is ensured by the IR-safety of the resolution variable $\qf$. In general, all loops in the ingredients listed in \eqref{Gen_eq_RadiativeFunctionsDifferential} can generate poles; some will be cancelled by analogous contributions with additional emissions. At least in principle, one could expand the relevant emission phase space as a distribution to make this cancellation already explicit at the differential level. However, we will leave this approach to future work. If $\qft_{\pf_n}$ simplifies sufficiently, one can define more inclusive radiative functions where these poles are manifestly cancelled. 

The collinear approximation of the matrix element introduced in \secref{sec_CollinearLimits} also produces spurious poles in the anti-collinear direction. On top of that the soft approximations studied in \secref{sec_SoftLimits} are also singular when emitted partons become collinear to hard legs. These types of singularities generally only cancel in \eqref{gen:eq_FactorizationFormulaFully} once different ingredients in the square brackets in \eqref{gen:eq_FactorizationFormulaFully} are combined. In cases where the resolution variable $\qf$ is insensitive to these spurious singularities, it turns out that the contribution in the square brackets in \eqref{gen:eq_FactorizationFormulaFully} is ill-defined in dimensional regularization. This is the problem of the aforementioned rapidity divergences, which will be discussed in detail in \secref{Gen_sec_RapidityDivergences}.

Even once all the above-mentioned singularities in the differential beam, soft, and jet functions have canceled, the full contribution in the square brackets in \eqref{gen:eq_FactorizationFormulaFully} can still contain some explicit poles and additionally generates $\epsilon$-poles after all integrals have been performed. These poles are the ones that are associated with the fully unresolved configurations, i.e., the ones that are indistinguishable from the Born-like configuration by any detector. These poles will then cancel with the explicit poles in the hard function $\Hb$, except for some $z_i$-dependent poles in the beam functions $\Bb$, which will cancel against corresponding poles in the bare parton distribution functions $f_{b_i}(x_i)$ when we integrate over the parton momentum fractions to obtain the full hadronic cross section
\begin{equation}
  \sigma(\qfcut)=\int \rd x_1\rd x_2 \hat{\sigma}_{b_1b_2}(\qfcut) f_{b_1}(x_1)f_{b_2}(x_2)\, .
\end{equation}
The total hadronic cross section in the limit $Q\gg\qfcut\gg \Lambda_{\mytext{QCD}}$ can now be written in terms of finite ingredients as
\begin{equation}
  \label{eq_GeneralSlicingCoeffs}
\begin{aligned}
    \sigma_{\qfcut>\qf}\left( \mu^2, \mu_F^2 \right)&= \!\sum_{\A, b_1, b_2}\int_0^1\rd \tilde{x}_1 \int_0^1 \rd \tilde{x}_2 \int_{\tilde{x}_1}^1\frac{\rd z_1}{z_1} \int_{\tilde{x}_2}^1\frac{\rd z_2}{z_2}f_{b_1}\left( \frac{\tilde{x}_1}{z_1}, \mu_F^2 \right)f_{b_2}\left( \frac{\tilde{x}_2}{z_2} , \mu_F^2 \right) \int\frac{\rd \Pi_B}{2Q^2}\\
    &\hspace{-2.6cm}\times \Tr\Bigg[ \frac{1}{S_{\A}} \Hb_\A(\tilde{x}_1P_1, \tilde{x}_2P_2, \{p_i\};\mu^2)\BDelta\Bigg]+\mathcal{O}\left(\left(  \frac{\Lambda_{\mytext{QCD}}}{Q} \right)^{p_\Lambda} \right)+\mathcal{O}\left( \left( \frac{\qfcut}{Q} \right)^p \right)\, ,
\end{aligned}
\end{equation} 
where $\rd\Pi_B$ is the Lorentz invariant Born phase, and we defined the renormalized hard function
\begin{equation}
  \begin{aligned}
    \Hb_\A(\tilde{x}_1P_1, \tilde{x}_2P_2, \{p_i\};\mu^2)&=\ket{\amp_{\A}(\tilde{x}_1P_1, \tilde{x}_2P_2, \{p_i\};\mu^2)}\bra{\amp_{\A}(\tilde{x}_1P_1, \tilde{x}_2P_2, \{p_i\};\mu^2)}\\
    &=\lim _{\epsilon \rightarrow 0} \boldsymbol{Z}^{-1}(\epsilon, \{p_i\}, \mu)\Hb_\A(\tilde{x}_1P_1, \tilde{x}_2P_2, \{p_i\}) \left( \boldsymbol{Z}^{-1}(\epsilon, \{p_i\}, \mu) \right)^\dagger\, ,
  \end{aligned}
\end{equation}
and the renormalized PDFs as explained in Appendices~\ref{sec_IRStructureLoops} and \ref{sec_MassFactorizationCounterTerms}.
Using \eqref{gen:eq_FactorizationFormulaFully}, we can extract the formula
\begin{equation}
\begin{aligned}
    &\BDelta=\lim_{\epsilon\to 0} \left( \boldsymbol{Z}(\epsilon, \{p_i\}, \mu) \right)^\dagger \\
      &\Biggl[{\sum_{n=n_J}^\infty}\sum_{\pf_n\in \mathfrak{P}_n }\int \left[ \rd\Bb_{a_1b^\prime_1}(I_1)\otimes\left( \Gamma^{-1}(\mu_F^2) \right)_{b^\prime_1 b_1} \right](z_1)\left[ \rd\Bb_{a_2b^\prime_2}(I_2)\otimes\left( \Gamma^{-1}(\mu_F^2) \right)_{b^\prime_2 b_2} \right](z_2)\\
   &\times\rd\Sb(S, n_1, n_2, \{n_i\})
    \prod_{i=1}^{n_j}\rd\J_{a_i}(\tilde{F}_i)\theta(\qfcut-\qft_{\pf_n}(\{k_i\},p_F))\Biggr]\left( \boldsymbol{Z}(\epsilon, \{p_i\}, \mu) \right)\, ,
\end{aligned}\end{equation}
where $\left( \Gamma^{-1}(\mu_F^2) \right)_{b^\prime_1 b_1}$ is the $\MSbar$ mass factorization counter term defined in \eqref{eq_GammaOperator}. 
The $\BDelta$ function can be expanded to fixed order in perturbation theory as
\begin{equation}
  \BDelta=\sum_{j=0}^\infty\sum_{k=0}^{2j} \BDelta^{jk}_{\A b_1 b_2}\!\!\left(z_1, z_2, \frac{\mu_F^2}{Q^2}, \frac{\mu^2}{Q^2}\right)\!\!
      \left( \frac{\as(\mu)}{\pi} \right)^j\!\!\log^k\!\!\left( \frac{\qfcut}{Q} \right)\, .
\end{equation}
The coefficients $\BDelta^{jk}_{\A b_1 b_2}$ are the ingredients required to define a slicing method based on the resolution variable.

\subsection{Factorization Properties of Resolution Variables}
\label{Gen_sec_VariableFactorization}
Even though the factorization formula \eqref{gen:eq_FactorizationFormulaFully} is completely general, it is not particularly practical and often overcomplicates matters. For example, the differential jet function $\rd\J$ is an operator in spin space. However, it turns out that the spin correlations often drop out when performing the convolution in the square brackets of \eqref{gen:eq_FactorizationFormulaFully}. Another simplification that often occurs is that the approximations of the resolution variable, $\qft_{\pf_n}$, do not depend on the individual momenta in a sector, for instance for $\qf=q_T$, $\qft_{\pf_n}$ only depends on the total transverse momenta of all initial-state collinear and soft partons respectively. This means that one can integrate out the individual momenta in the sectors $I_1$, $I_2$, and $S$, only fixing the total momenta, which leads to much simpler expressions.

In this section, we will show in detail how the specific form of $\qft_{\pf_n}$ allows one to simplify the square brackets in \eqref{gen:eq_FactorizationFormulaFully} by the example of $q_T$ and $\tau_n$. At the end of the section, we will show that if $\qft$ takes another specific form, which we will introduce, one can write the convolution in the square brackets in \eqref{gen:eq_FactorizationFormulaFully} as a product of cumulant beam, soft, and jet functions. 

\paragraph{\boldmath Factorization of $q_T$\unboldmath}
To improve readability, we will use bold symbols to denote transverse momenta in this section. The absolute value $q_T$ of the total transverse momentum $\mathbf{q_T}$ of the resolved system can be used as a resolution variable if there are no resolved jets. We only need to consider real emissions in the sectors $I_1$, $I_2$, and $S$. A big advantage of $q_T$ is that $\qf$ is already homogeneous in the power counting $\lambda$ for every region, and we do not need to make further approximations. Thus, we can write 
\begin{equation}
\qft_{\pf_n}(\{k\}, p_F)=\lvert \mathbf{q_t}\rvert=\lvert -\mathbf{k_{I_1, t}}-\mathbf{k_{I_2, t}}-\mathbf{k_{S, t}}\rvert\, ,
\end{equation}
where $\mathbf{k_{I_1, t}}$, $\mathbf{k_{I_2, t}}$ and $\mathbf{k_{S, t}}$ are the transverse parts of the respective total momenta in the three sectors. The analysis in the previous section also applies if instead of calculating the cumulant cross section for small $q_T$ we would have calculated the differential cross section in $\mathbf{q_T}$---we only have to replace $\theta(\qfcut-\qf)$ with a delta function $\delta^{d-2}(\mathbf{q_T}+\mathbf{k_{I_1, t}}+\mathbf{k_{I_2, t}}+\mathbf{k_{S, t}})$. 
We can then rewrite the square brackets in \eqref{gen:eq_FactorizationFormulaFully} as 
\begin{equation}
  \label{Gen_eq_qtFactorizationFormula}
  \begin{aligned}
    &\Biggl[\sum_{n=n_J}^\infty\sum_{\mathfrak{P}_n}\int\rd\Bb_{a_1b_1}(I_1, z_1)\rd\Bb_{a_2b_2}(I_2, z_2)\rd\Sb(S, n_1, n_2, \{n_i\})\\
     &\times\delta^{d-2}(\mathbf{q_T}+\mathbf{k_{I_1, t}}+\mathbf{k_{I_2, t}}+\mathbf{k_{S, t}})\Biggr]\\
     &=\int\rd^{d-2}\mathbf{k_{I_{1}, t}}\int\rd^{d-2}\mathbf{k_{I_{2}, t}}\int\rd^{d-2}\mathbf{k_{S, t}}\Bb_{a_1b_1}(\mathbf{k_{I_1, t}}, z_1)\Bb_{a_1b_1}(\mathbf{k_{I_2, t}}, z_2)\Sb(\mathbf{k_{S, t}}, n_1, n_2, \{N_i\})\\
     &\times \delta^{d-2}(\mathbf{q_T}+\mathbf{k_{I_1, t}}+\mathbf{k_{I_2, t}}+\mathbf{k_{S, t}})\, ,
  \end{aligned}
\end{equation}
where we defined the transverse-momentum-space beam and soft functions as
\begin{equation}
  \label{eq_TMDqtSpace}
  \begin{aligned}
    &\Bb_{a_1b_1}(\mathbf{k_{t}}, z_1)=\sum_{n=0}^\infty\int  \rd\Bb_{a_1b_1}(I_1, z_1) \delta^{d-2}(\mathbf{k_t}-\mathbf{k_{I_1, t}})\\
    &=\delta^{d-2}(\mathbf{k_t})\delta(1-z_1)\delta_{a_1b_1}\bbbone +\\
    &+z_1\sum_{n=1}^\infty \sum_{\Af_1}\frac{1}{S_{\Af_1}}\int \rd^d k_{I_1}\rd \Pi_{I_1}\delta\!\left( \bar{z}_1-\frac{k_{I_1}\cdot P_2}{x_1P_1\cdot P_2} \right) \delta^{d-2}(\mathbf{k_t}-\mathbf{k_{I_1, t}})\!\bcalPhat_{a_1b_1;\Af_1}(I_1, z_1)\\
    &\Sb(\mathbf{k_{t}}, n_1, n_2, \{N_i\})=\sum_{n=0}^\infty\int \rd\Sb(S, n_1, n_2, \{N_i\})\delta^{d-2}(\mathbf{k_t}-\mathbf{k_{S, t}})\\
    &=\delta^{d-2}(\mathbf{k_t})\bbbone +\\
    &+\sum_{n=1}^\infty \sum_{\Af_S}\frac{1}{S_{S}}\int \rd^d k_{S}\, \rd \Pi_{S}\bcalJJ_{\Af_S}(S, n_1, n_2, \{N_i\})\delta^{d-2}(\mathbf{k_t}-\mathbf{k_{S, t}})\, .
  \end{aligned}
\end{equation}
We accounted for possible heavy quarks in the resolved final state by including some massive direction vectors (four-velocities) $\{N_i\}$ in the squared soft currents. The $\mathbf{k_{t}}$ dependent beam function, sometimes also called \emph{space-like (SL) transverse-momentum dependent (TMD) collinear function}, is formally equivalent to the one defined in~\cite{Catani:2022sgr}, where it is called $\bF$. However, we point out that both the soft function and beam function in \eqref{eq_TMDqtSpace} still contain rapidity divergences that need to be regularized. This will be discussed in more detail in Section~\ref{Gen_sec_RapidityDivergences}. The beam functions are also closely related to the (bare) matching coefficients $I_{i \leftarrow j}\left(z, x_T^2, \epsilon\right)$ appearing in the SCET literature on $q_T$-resummation (c.f.~\cite{Becher:2010tm}).

One can further simplify the convolution in \eqref{Gen_eq_qtFactorizationFormula} by going over to Fourier space, which in this case is the so-called \emph{impact parameter space} or \emph{b-space}. We define the Fourier transform of the transverse momentum space beam and soft functions as
\begin{equation}
  \begin{aligned}
    \tilde{\Bb}_{a_1b_1}(\bb, z_1)&=\int \rd^{d-2}\mathbf{k_{t}}e^{-i \bb\cdot \mathbf{k_t}}\Bb_{a_1b_1}(\mathbf{k_{t}}, z_1)\\
    \tilde{\Sb}(\bb, n_1, n_2, \{n_i\})&=\int \rd^{d-2}\mathbf{k_{t}}e^{-i \bb\cdot \mathbf{k_t}}\Sb(\mathbf{k_{t}}, n_1, n_2, \{N_i\})\, .
  \end{aligned}
\end{equation}
The Fourier transformed convolution in \eqref{Gen_eq_qtFactorizationFormula} now becomes a product
\begin{equation}
  \label{Gen_eq_qtFactorizationFormulaFourier}
  \begin{aligned}
    &\int \rd^{d-2}\mathbf{q_{T}}e^{-i \bb\cdot \mathbf{q_T}}\int\rd^{d-2}\mathbf{k_{I_{1}, t}}\int\rd^{d-2}\mathbf{k_{I_{2}, t}}\int\rd^{d-2}\mathbf{k_{S, t}}\Bb_{a_1b_1}(\mathbf{k_{I_1, t}}, z_1)\Bb_{a_2b_2}(\mathbf{k_{I_2, t}}, z_2)\\
    &\times\Sb(\mathbf{k_{S, t}}) \delta^{d-2}(\mathbf{q_T}+\mathbf{k_{I_1, t}}+\mathbf{k_{I_2, t}}+\mathbf{k_{S, t}})=\tilde{\Bb}_{a_1b_1}(\bb, z_1)\tilde{\Bb}(\bb, z_2)_{a_2b_2} \tilde{\Sb}(\bb, n_1, n_2, \{n_i\})\, .
  \end{aligned}
\end{equation}
This factorized result is particularly useful if one wants to resum the cross section to all orders in perturbation theory. The reason is that one can subtract all the remaining $\epsilon$-poles in the Fourier-transformed beam and soft functions with multiplicative factors without spoiling the factorization.
\noindent\paragraph{\boldmath Factorization of $\tau_n$\unboldmath}
The $n$-jettiness resolution variable, $\tau_n$, was defined originally in~\cite{Stewart:2010tn}. To match the convention in this paper, where resolution variables have mass dimension one, we define $\mathcal{T}_n=Q\tau_n$. As usual, $Q$ is an IR-safe function of the momenta that reduces to the center of mass energy at the Born level. Jettiness then approximates to
\begin{equation}
  \mathcal{T}_n(\{k\})Q\sim t_{I_1}+t_{I_2}+\sum_{i=3}^{n_J+2} s_i+\sum_{k\in S}\min_{i}(2p_i\cdot k)\, ,
\end{equation}
where we defined the total invariant mass in sector $F_i$ as $s_i$ and the transverse virtualities of the initial-state emissions
\begin{equation}
  t_{I_i}=2\tilde{x}_1P_i\cdot k_{I_i}=\frac{z_i}{1-z_i}\left( k_{I_i}^2+\lvert k_{I_i, t}^2 \rvert \right)\, .
\end{equation}
Again, we do not need the individual momenta of every parton to perform the convolution in \eqref{gen:eq_FactorizationFormulaFully}, and one can define much more inclusive beam, jet, and soft functions. Defining 
\begin{equation}
  \begin{aligned}
    \Bb_{a_1b_1}(t, z_1)&=\sum_{n=0}^\infty\int  \rd\Bb_{a_1b_1}(I_1, z_1) \delta(t-t_{I_1})\\
    &=\delta(t)\delta(1-z_1)\delta_{a_1b_1}\bbbone +\\
    &+z_1\sum_{n=1}^\infty\sum_{\Af_1}\frac{1}{S_{\Af_1}}\int \rd^d k_{I_1}\rd \Pi_{I_1}\delta\!\left( \bar{z}_1-\frac{k_{I_1}\cdot P_2}{x_1P_1\cdot P_2} \right) \delta(t-t_{I_1})\!\bcalPhat_{a_1b_1;\Af_1}(I_1, z_i)
    \end{aligned}
\end{equation}
\begin{equation}
  \begin{aligned}
    \J_{a_i}(t)&=\sum_{n=1}^\infty\int  \rd\J_{a_i}(F_i) \delta(t-s_i)\\
    &=\delta(t)\bbbone +\\
    &+\sum_{n=2}^\infty \sum_{\Af_i} \frac{1}{S_{\Af_i}}2(2\pi)^{d-1}\prod_{j\in F_i}\dktj\delta(1-\sum_j\tilde{z}_j)\delta^{d-2}\left( \tilde{k}_{j, \perp} \right) \delta(t-\tilde{s}_i)\!\calPhat_{\Af_i}(\tilde{F}_i)
  \end{aligned}
\end{equation}
\begin{equation}
  \begin{aligned}
    \Sb(t, n_1, n_2, \{n_i\})&=\sum_{n=0}^\infty\int\rd\Sb(S, n_1, n_2, \{N_i\})\delta \!\left( t-\sum_{k\in S}\min_{i}(2p_i\cdot k) \right)\\
    &=\delta(t)\bbbone +\\
    &+\sum_{n=1}^\infty\sum_{\Af_S}\frac{1}{S_{S}}\int \rd^d k_{S}\, \rd \Pi_{S}\bcalJJ_{\Af_S}(S, n_1, n_2, \{N_i\})\delta \!\left( t-\sum_{k\in S}\min_{i}(2p_i\cdot k) \right)
  \end{aligned}
\end{equation}
one can rewrite the convolution in \eqref{gen:eq_FactorizationFormulaFully} as
\begin{align}
    &\Biggl[\sum_{n=n_J}^\infty\sum_{\mathfrak{P}_n}\int\rd\Bb_{a_1b_1}(I_1, z_1)\rd\Bb_{a_2b_2}(I_2, z_2)\rd\Sb(S, n_1, n_2, \{n_i\})\\
    &\times \prod_{i=1}^{n_j}\rd\J_{a_i}(\tilde{F}_i)\theta(\qfcut Q-t_{I_1}+t_{I_2}+\sum_{i=3}^{n_J+2} s_i+\sum_{k\in S}\min_{i}(2p_i\cdot k))\Biggr]\\
    &=\int \rd t_{I_1}\Bb_{a_1b_1}(t_{I_1}, z_1)\int \rd t_{I_2}\Bb_{a_2b_2}(t_{I_2}, z_2)\int\prod_{i=3}^{n_J+2}\rd s_i \J_{a_i}(s_i)\int \rd t_S\\
    &\times \Sb(t_S, n_1, n_2, \{n_i\})\theta\!\left( \qfcut Q -t_{I_1}-t_{I_2}-\sum_{i=3}^{n_J+2}s_i-t_S \right)\, .
\end{align}
This convolution could also be written as a product in Laplace space.
\paragraph{Factorization into Cumulant Functions}
A class of resolution variables that is particularly useful for slicing methods is characterized by the leading-power behavior
\begin{equation}
  \label{Gen_eq_MaximumFactorization}
  \qft_{\pf_n}(\{k\})=\max\left( \qft_{C_1}(I_1), \qft_{C_1}(I_2), \qft_{C_3}(\tilde{F_3}), \dots, \qft_S(S)\right)\, .
\end{equation}
On the right-hand side, we have used the notation $C_{1(2)}$, to refer to the region where the momenta in $I_{1(2)}$ are all collinear to the proton $P_1$ and all other sectors are trivial, i.e., the other initial-state collinear sector and the soft sector are empty and each final-state collinear sector only contains one parton. Similarly, $C_3$ refers to the region where the momenta $\tilde{F}_3$ are all collinear to the jet $p_3$ and all other sectors are trivial, and $S$ refers to the region where the momenta $S$ are all soft, and all other sectors are trivial (c.f. Appendix~\ref{sec_ShorthandNotationRegions} for more details on this notation). In \eqref{Gen_eq_MaximumFactorization}, we have suppressed a dependence on the fixed hard configuration $\{p_i\}\cup \{\tilde{p}_F\}$ on both sides. 
Only the boosted and rescaled momenta $\tilde{F}_i$ appear on the right-hand side of \eqref{Gen_eq_MaximumFactorization}. We will show in \secref{sec_ktness_approximations} that one can modify the original definition of $\ktness$~\cite{Buonocore:2022mle} such that \eqref{Gen_eq_MaximumFactorization} is satisfied. 

One can then define the bare cumulant radiative functions
\begin{align}
  \label{Gen_eq_CumulantRadiativeFunctions}
    &\hat{\Bb}_{a_1b_1}(\qfcut, z_1)=\sum_{n=0}^\infty\int  \rd\Bb_{a_1b_1}(I_1, z_1)\theta\!\left( \qfcut -  \qft_{C_1}(I_1)\right)\notag\\
    &\hspace{0.5cm}=\theta(\qfcut)\delta(1-z_1)\delta_{a_1b_1}\bbbone +\notag\\
    &\hspace{0.5cm}+z_1\sum_{n=1}^\infty\sum_{\Af_1}\frac{1}{S_{\Af_1}}\int \rd^d k_{I_1}\rd \Pi_{I_1}\delta\!\left( \bar{z}_1-\frac{k_{I_1}\cdot P_2}{x_1P_1\cdot P_2} \right) \theta\!\left( \qfcut -  \qft_{C_1}(I_1)\right)\!\bcalPhat_{a_1b_1;\Af_1}(I_1, z_i)\notag\\
    &\hat{\J}_{a_i}(\qfcut)=\sum_{n=1}^\infty\int  \rd\J(\tilde{F}_i) \theta\!\left( \qfcut -  \qft_{C_i}(\tilde{F}_i)\right)\notag\\
    &\hspace{0.5cm}=\theta(\qfcut)\bbbone +\notag\\
    &\hspace{0.5cm}+\sum_{n=2}^\infty\sum_{\Af_i}\frac{1}{S_{\Af_i}}2(2\pi)^{d-1}\prod_{j\in F_i}\dktj\delta(1-\sum_j\tilde{z}_j)\delta^{d-2}\left( \tilde{k}_{j, \perp} \right) \calPhat_{\Af_i}(\tilde{F}_i)\theta\!\left( \qfcut -  \qft_{C_i}(\tilde{F}_i)\right)\notag\\
    &\hat{\Sb}(\qfcut, n_1, n_2, \{n_i\})=\sum_{n=0}^\infty\int\rd\Sb(S, n_1, n_2, \{N_i\})\theta\!\left( \qfcut -  \qft_S(S)\right)\notag\\
    &\hspace{0.5cm}=\theta(\qfcut)\bbbone +\notag\\
    &\hspace{0.5cm}+\sum_{n=1}^\infty\sum_{\Af_S}\frac{1}{S_{S}}\int \rd^d k_{S}\, \rd \Pi_{S}\bcalJJ_{\Af_S}(\{ k\}_S, n_1, n_2, \{N_i\})\theta\!\left( \qfcut -  \qft_S(S)\right)
\end{align}
in terms of which the convolution in \eqref{gen:eq_FactorizationFormulaFully} can be written as
\begin{equation}
  \label{Gen_eq_CumulantFactorizationFormula}
  \begin{aligned}
    &\Biggl[\sum_{n=n_J}^\infty\sum_{\mathfrak{P}_n}\int \rd\Bb_{a_1b_1}(I_1, z_1)\rd\Bb_{a_1b_1}(I_2, z_2)\rd\Sb(S, n_1, n_2, \{n_i\})\\
    &\times \prod_{i=1}^{n_j}\rd\J_{a_i}(\tilde{F}_i)\theta\!\left(\qfcut -\max\left( \qft_{C_1}(I_1), \qft_{C_1}(I_2), \qft_{C_3}(\tilde{F_3}), \dots, \qft_S(S)\right)\right)\Biggr]\\
    &=\hat{\Bb}_{a_1b_1}(\qfcut, z_1)\hat{\Bb}_{a_2b_2}(\qfcut, z_2)\prod_{i=3}^{n_J+2} \hat{\J}_{a_i}(\qfcut)\hat{\Sb}(\qfcut)\, ,
  \end{aligned}
\end{equation}
where we made use of the identity
\begin{equation}
  \theta(\qfcut-\max\{a_1, a_2, \dots, a_n\})=\theta(\qfcut-a_1)\theta(\qfcut-a_2)\dots\theta(\qfcut-a_n)\, .
\end{equation}
Thus, we see that variables of this type naturally factorize in cumulant space, which is the natural space to consider when studying slicing methods.

For future reference, we here define the power expansion of the bare cumulant functions as
\begin{equation}
  \label{Gen_eq_BareCumulantExpansion}
  \begin{aligned}
    \hat{\Bb}_{a_1b_1}(\qfcut, z_1)&=\theta(\qfcut)\delta(1-z_1)\delta_{a_1b_1}\bbbone+\sum_{n=1}^\infty\left( \frac{\alpha_0}{\pi}\right)^n\hat{\Bb}_{a_1b_1}^{(n)}(\qfcut, z_1)\\
    \hat{\J}_{a_i}(\qfcut)&=\theta(\qfcut)\bbbone +\sum_{n=1}^\infty\left( \frac{\alpha_0}{\pi}\right)^n\hat{\J}_{a_i}^{(n)}(\qfcut)\\ 
    \hat{\Sb}(\qfcut, n_1, n_2, \{n_i\})&=\theta(\qfcut)\bbbone+\sum_{n=1}^\infty\left( \frac{\alpha_0}{\pi} \right)^n \hat{\Sb}^{(n)}(\qfcut, n_1, n_2, \{n_i\})\, .
  \end{aligned}
\end{equation} In Appendix~\ref{App_NLOJetFunctions}, we calculate the NLO cumulant jet functions for a quite general set of resolution variables.

\section{\texorpdfstring{\boldmath Rapidity Divergences and the $z_N$-Prescription\unboldmath}{Rapidity Divergences and the zN-Prescription}}
\label{Gen_sec_RapidityDivergences}
This section discusses rapidity divergences, an issue we have alluded to repeatedly in the preceding sections. To illustrate the problem, let us perform a sample calculation. We will try to calculate the NLO cumulant quark jet function for $\ktness$  (c.f. Section~\ref{sec_ktness_definition}). $\ktness$ has the scaling dimensions $a=b=c=1$.
The leading-power approximation of the resolution variable for the case where we have two particles $\{k_1, k_2\}$ collinear to each other (and along, say, jet $p_i$) reads
\begin{equation}
  \qft_{C_i}(\{\tilde{k}_1,\tilde{k}_2\})=\lvert \tilde{k}_{1, \perp}\rvert=\lvert \tilde{k}_{2, \perp}\rvert\, ,
\end{equation}
where we point out that $\tilde{k}_{1(2)}=k_{1(2)}$ exactly because there are no other emissions in the soft or initial-state collinear sectors that could generate a recoil.

Inserting the tree-level splitting kernel defined in Appendix~\ref{sec_APkernels}, we find
\begin{equation}
  \label{Gen_eq_RapidityDivergenceExample}
  \begin{aligned}
    \hat{\J}_q^{(1)}&= \mu^{2\epsilon} \frac{e^{\epsilon \gamma_E}}{\Gamma(1-\epsilon)}4\!\!\int\frac{\rd^d \tilde{k}_1\delta_+(\tilde{k}_1^2)\rd^d \tilde{k}_2\delta_+(\tilde{k}_2^2)}{\Omega_{d-2}}\delta(1-\sum_j\tilde{z}_j)\delta^{d-2}( \tilde{k}_{j, \perp} ) \frac{\hat{P}_{q\to gq}}{2\tilde{k}_1\cdot\tilde{k}_2}\theta\!( \qfcut- \lvert \tilde{k}_{1, \perp}\rvert)\\
    &= \mu^{2\epsilon} \frac{e^{\epsilon \gamma_E}}{\Gamma(1-\epsilon)}\int_0^1 \rd z \int \frac{\rd^{d-2}\tilde{k}_\perp}{\Omega_{d-2}}\frac{\hat{P}_{q\to gq}}{\lvert \tilde{k}_{\perp} \rvert^2}\theta\!\left( \qfcut- \lvert \tilde{k}_{\perp}\rvert\right)\\
    &=-\left( \frac{\mu^2}{\qfcut^2} \right)^\epsilon\frac{1}{2\epsilon} \frac{e^{\epsilon \gamma_E}}{\Gamma(1-\epsilon)}\int_0^1 \rd z \hat{P}_{q\to gq}\, ,
  \end{aligned}
\end{equation}
where we solved the delta functions and rewrote the integral with simpler variables in the second line, and we performed the integration over the transverse momentum to get to the third line. We also defined the volume of the unit sphere
\begin{equation}
  \Omega_d=\frac{2\pi^{\frac{d}{2}}}{\Gamma(\frac{d}{2})}\, . 
\end{equation}Now it becomes obvious that we have a problem---the splitting kernel becomes singular when the gluon becomes soft $\hat{P}_{q\to gq} \sim \frac{2C_F}{z}$, and the integral becomes divergent. This would not have happened if the resolution variable depended on $z$ in the limit $z \to 0$. For example, if we had $\qft_{C_i} = \frac{\lvert \tilde{k}_{1, \perp} \rvert^2}{z_1 z_2 Q}$ (as would be the case for $Q \tau_n$), the transverse momentum integral in \eqref{Gen_eq_RapidityDivergenceExample} would lead to
\begin{equation}
  \hat{\J}_q^{(1)} = -\left( \frac{\mu^2}{Q \qfcut} \right)^\epsilon \frac{1}{2\epsilon} \frac{e^{\epsilon \gamma_E}}{\Gamma(1-\epsilon)} \int_0^1 \rd z \, z^{-\epsilon} (1-z)^{-\epsilon} \hat{P}_{q\to gq} \, .
\end{equation}
 We see that the integral is now finite. In fact, dimensional regularization comes to the rescue, turning the $z\to 0$ divergence into a pole in $\epsilon$. Dimensional regularization will always be sufficient to make jet functions finite unless $b=c$ ($Q\tau_n$ has the scaling dimensions $a=b=\frac{1}{2}, c=1$).

This section aims to better understand the origin of rapidity divergences, and to choose a method for dealing with them. 
\subsection{Origin of Rapidity Divergences}
The origin of the rapidity divergences encountered above can be traced back to the matrix element. When defining the collinear limit along jet $p_i$, we defined a lightlike reference vector $n_i$ along the jet direction. We also had to specify a reference vector $\bar{n}_i$ to define the anti-collinear directions, or in other words, to define what we mean by a longitudinal momentum fraction and a transverse momentum. In this subsection, we will drop the index $i$ and always consider the same jet direction. Then, the collinear momentum fraction of a parton $k$ in the collinear splitting is
\begin{equation}
  z=\frac{\tilde{k}\cdot \bar{n}}{p\cdot \bar{n}}\sim\frac{k\cdot \bar{n}}{p\cdot \bar{n}}\, .
\end{equation}
Let us again focus on the example of a quark splitting into a gluon with momentum $k$ and another quark with momentum $l$. Then, the endpoint $z\to 0$ of the collinear approximation of the matrix element behaves as
\begin{equation}
  \label{Gen_eq_RapidityDivergenceExampleMatrixElement}
  \lvert M\rvert^2\sim \frac{C_F}{k\cdot l z}=C_F \frac{p\cdot \bar{n}}{k\cdot l \, k\cdot \bar{n}}\, .
\end{equation}
By expanding the matrix element in the collinear limit, we have introduced an extra collinear singularity along the direction of the auxiliary vector $\bar{n}$. To see why this divergence is called a rapidity divergence, we integrate \eqref{Gen_eq_RapidityDivergenceExampleMatrixElement} over the collinear emission phase space parametrized in terms of the gluon rapidity $\eta$. We assume that the resolution variable is parametrized as 
\begin{equation}
  \qft_{C_i}\sim Q \left( \frac{ k_\perp}{Q} \right)^{\frac{1}{b}}z^{\frac{1}{c}-\frac{1}{b}}=Q \left( \frac{Q}{p_+} \right)^{\frac{1}{c}-\frac{1}{b}}
  \left( \frac{ k_\perp}{Q} \right)^{\frac{1}{c}}e^{-\left( \frac{1}{b}-\frac{1}{c} \right)\eta}\, ,
\end{equation}
where $p_+$ is the plus component of the hard momentum.
The relevant integral then becomes 
\begin{equation}
  \label{Gen_eq_RapidityDivergenceRapidityFirst}
  \begin{aligned}
    \int \rd \Pi_{\mytext{coll}} \lvert M\rvert^2 \theta\!\left( \qfcut-  \qft_{C_i}\right)&\sim \int_0^1\frac{\rd z}{z}\int_0^\infty \rd  k_\perp  k_\perp^{-1-2\epsilon}\theta\!\left( \qfcut-Q \left( \frac{\lvert k_\perp\rvert}{Q} \right)^{\frac{1}{b}}z^{\frac{1}{c}-\frac{1}{b}} \right)\\
    &\hspace{-2cm}=\int_0^\infty \rd  k_\perp  k_\perp^{-1-2\epsilon} \int_{-\infty}^{\log\left( \frac{p_+}{k_\perp} \right)}\rd \eta \, \theta\!\left( \qfcut-Q \left( \frac{Q}{p_+} \right)^{\frac{1}{c}-\frac{1}{b}}
    \left( \frac{ k_\perp}{Q} \right)^{\frac{1}{c}}e^{-\left( \frac{1}{b}-\frac{1}{c} \right)\eta}\right)\\
    &\hspace{-2cm}=-\frac{c}{c-b}\int_0^{k_\perp^{\mathrm{max}}} \rd  k_\perp  k_\perp^{-1-2\epsilon}\log\left( \frac{k_\perp}{k_\perp^{\mathrm{max}}} \right)=\frac{1}{4\epsilon^2}\frac{c}{c-b}\left( \frac{\qfcut}{Q} \right)^{-2\epsilon b}Q^{-2\epsilon}\, ,
  \end{aligned}
\end{equation}
where we introduced $k_\perp^{\mathrm{max}}=Q\left( \frac{\qfcut }{Q} \right)^b$, and we assumed $c\neq b$ to get to the last line. The main insight from this calculation comes from the second line. We see that there is an upper boundary in the rapidity integral, which stems from the fact that when the jet splits into a gluon and a quark, the ``plus'' components of the momenta are conserved in the collinear limit (c.f., the $\delta$-function containing the momentum fractions $z_i$ in \eqref{Gen_eq_RadiativeFunctionsDifferential}). However, in the collinear limit, the ``minus'' components of the momenta are not conserved; thus, the gluon rapidity can become arbitrarily negative from pure phase space considerations. From the second line in \eqref{Gen_eq_RapidityDivergenceRapidityFirst}, we see that the cut on the resolution variable $\qf$ introduces a lower cutoff
\begin{equation}
  \eta_{\mathrm{min}} = -\frac{c}{c-b} \log\! \left(\frac{k_\perp^{\mathrm{max}}}{k_\perp} \right) + \log\! \left( \frac{p_+}{Q} \right)\, ,
\end{equation}
which tends to $-\infty$ in the limit $c \to b$\footnote{In this example, we assume $b\geq c$ to simplify the discussion.}. We conclude that the rapidity divergence in the collinear region is, in fact, due to gluons with rapidities that tend to $-\infty$, i.e., anti-collinear gluons. The rapidity divergence is regulated by the resolution variable $\qf$ if $b\neq c$. Note again that rapidity divergences arose due to the expansion of the collinear phase space. They are thus also a consequence of the phase space expansion performed in the method of regions---rapidity divergences never appear in an exact QCD calculation.

The divergence at very negative rapidities is unphysical and, in the full result, cancels once contributions from different regions are combined. However, one might wonder whether the rapidity divergence in the collinear limit ought to be regulated by $\qfcut$. The purpose of a resolution variable is to split the phase space into unresolved/IR and resolved parts. Now, $\qfcut$ also plays the extra role of putting a cut on negative rapidities. This cut affects the poles of the jet, soft and beam functions, making them dependent on the resolution variable. Alternatively, it can be useful to regulate the rapidity divergences with an extra regulator even for resolution variables that satisfy $b\neq c$. A similar observation was made in~\cite{Bauer:2020npd}.
\subsection{Rapidity Regulators}
This section will give an overview of rapidity regulators used in the literature. All of these regulators can be used directly within our framework. At the end of the section, we will expand on a new scheme that we will call the $z_N$-prescription, which was first introduced in~\cite{Catani:2022sgr}.

General discussions on rapidity divergences, as well as the related rapidity renormalization group, can be found in~\cite{Collins:1992tv, BenekeFA, Manohar_2007, Collins:2008ht, Becher:2010tm, Chiu:2011qc, Chiu:2012ir, vladimirov2017structure}.
Rapidity regularization schemes that have been used in the literature include hard cutoffs~\cite{Balitsky_1996, Jalilian_Marian_1998, Kovchegov:1999yj, Manohar_2007}, tilting Wilson lines off the light cone~\cite{Ji:2004wu,Collins:2011zzd}, the delta regulator~\cite{Chiu_2009}, the $\eta$ regulator~\cite{Chiu:2011qc, Chiu:2012ir}, analytic continuation of propagators~\cite{Beneke_2004, Chiu_2008}, adding an analytic regulator in the emission phase space \cite{Becher_2012}, the exponential regulator~\cite{li2016exponential} and the recently proposed pure rapidity regulator~\cite{Ebert:2018gsn, Vita:2020ckn} which is similar to the $\eta$ regulator but simplifies the treatment of subleading power corrections.

All the above regulators introduce a mass scale, typically called $\nu$, and a small parameter, say $\eta$. In some cases, $\nu$ can simultaneously play the role of the small parameter. At the integrand level, the rapidity regulator vanishes for $\eta\to 0$. However, the limit and the integral do not commute---the integrals will typically feature poles in $\eta$ or logs in $\eta\nu$. These divergences can be subtracted, for instance, in \msbar, and the resulting renormalized radiative functions will satisfy a renormalization group equation called the \emph{rapidity renormalization group}~\cite{Chiu:2011qc}.

While there is a lot that can be said about the different regulators mentioned above, we stress one property of rapidity regulators in particular: Many rapidity regulators spoil the power counting performed in the method of regions, and they thus introduce spurious regions that must be subtracted to avoid over-counting. Suppose, for example, that we had regulated our integral in \eqref{Gen_eq_RapidityDivergenceExample} as
\begin{equation}
  \label{Gen_eq_RapidityDivergenceExpReg}
  \begin{aligned}
    \hat{\J}_q^{(1)}&= \mu^{2\epsilon} \frac{e^{\epsilon \gamma_E}}{\Gamma(1-\epsilon)}4\int\frac{\rd^d \tilde{k}_1\delta_+(\tilde{k}_1^2)\rd^d \tilde{k}_2\delta_+(\tilde{k}_2^2)}{\Omega_{d-2}}e^{-\left( \frac{\eta}{\nu} e^{-\gamma_E}\left( n+\bar{n} \right)\cdot \left(\tilde{k}_1+\tilde{k}_2  \right) \right)}\\&\times\delta\bigg(1-\sum_j\tilde{z}_j\bigg)\delta^{d-2}\left( \tilde{k}_{j, \perp} \right) \frac{\hat{P}_{q\to gq}}{2\tilde{k}_1\cdot\tilde{k}_2}\theta\!\left( \qfcut- \lvert \tilde{k}_{1, \perp}\rvert\right)\, ,
  \end{aligned}
\end{equation}
where we introduced the rapidity regulator $\eta$ and the scale $\nu$.
The idea of the method of regions was to expand fully in all relevant regions. Here, we are calculating a jet function, and thus, the relevant scaling is $(k_+, k_-, k_\perp)\sim (1, \lambda^2, \lambda)Q$ for both $\tilde{k}_1$ and $\tilde{k}_2$. However, 
\begin{equation}
  \left( n+\bar{n} \right)\cdot \left(\tilde{k}_1+\tilde{k}_2  \right) =\tilde{k}_{1, +}+\tilde{k}_{2, +}+\tilde{k}_{1, -}+\tilde{k}_{2, -}
\end{equation}
is not fully expanded in this region since we are retaining the minus components. One crucial element of the method of regions, as discussed in \secref{Gen_sec_MethodOfRegionsResolution}, is that there is no double-counting between different regions because the double-counting contributions are scaleless. However, the introduction of the new scale $\nu$ in the rapidity regulator spoils this argument---\eqref{Gen_eq_RapidityDivergenceExpReg} now contains both contributions in the collinear and the soft region $\tilde{k}_2\sim \lambda Q$. This soft contribution needs to be subtracted to avoid double counting. In the SCET literature, this procedure is called \emph{zero-bin subtraction}. Applying the soft scaling to \eqref{Gen_eq_RapidityDivergenceExpReg}, we can see that the zero bin is 
\begin{equation}
  \begin{aligned}
    \hat{\J}_q^{(1)}\Big\vert_{0-\mytext{bin}}&=\mu^{2\epsilon} \frac{e^{\epsilon \gamma_E}}{\Gamma(1-\epsilon)}2C_F\int\frac{\rd^d \tilde{k}_2\delta_+(\tilde{k}_2^2)}{\Omega_{d-2}}e^{-\left( \frac{\eta}{\nu} e^{-\gamma_E}\left( n+\bar{n} \right)\cdot \tilde{k}_2  \right)} \frac{n\cdot \bar{n}}{n \cdot\tilde{k}_2 \, \bar{n} \cdot\tilde{k}_2}\theta\!\left( \qfcut- \lvert \tilde{k}_{2, \perp}\rvert\right)\\
    &=\left( \frac{\mu^2}{\qfcut^2} \right)^\epsilon\frac{C_F}{\epsilon^2} \frac{e^{\epsilon \gamma_E}}{\Gamma(1-\epsilon)}\left( 1+2\epsilon\log \left( \frac{\qfcut \eta}{\nu} \right) \right)\, ,
  \end{aligned}
\end{equation}
where we dropped higher-order terms in $\eta$.
The full result for the jet function with this regulator reads
\begin{equation}
  \begin{aligned}
    \hat{\J}_q^{(1)}&\sim\mu^{2\epsilon} \frac{e^{\epsilon \gamma_E}}{\Gamma(1-\epsilon)}\int_0^1 \rd z \int \frac{\rd^{d-2}k_\perp}{\Omega_{d-2}}\frac{\exp\left( -\frac{\eta e^{-\gamma_E }k_\perp^2}{\nu p_+ z(1-z)} \right)}{z(1-z)}\frac{z(1-z)\hat{P}_{q\to gq}}{\lvert k_{\perp} \rvert^2}\theta\!\left( \qfcut- \lvert k_{\perp}\rvert\right)\\
    &=C_F\left( \frac{\mu^2}{\qfcut^2} \right)^\epsilon\frac{e^{\epsilon \gamma_E}}{\Gamma(1-\epsilon)}\left( \frac{1}{\epsilon^2} +\frac{3}{4\epsilon}+\frac{1}{4}+\frac{\log\left[ \frac{\qfcut^2 \eta}{p_+\nu} \right]}{\epsilon}\right)\, ,
  \end{aligned}
\end{equation}
and subtracting the zero bin, one finds 
\begin{equation}
  \hat{\J}_q^{(1)}-\hat{\J}_q^{(1)}\Big\vert_{0-\mytext{bin}}=C_F\frac{e^{\epsilon \gamma_E}}{\Gamma(1-\epsilon)}\left( \frac{3}{4\epsilon}+\frac{1}{4}+\frac{\log\left[ \frac{ \nu}{p_+} \right]-\log\left[ \eta \right]}{\epsilon}\right)\, .
\end{equation}
After calculating the other jet functions and the soft function with the same regulator and putting everything together,  the $\log\eta$ term cancels. However, $\log\left(  \frac{\nu}{p_+}\right) $ cannot appear in the soft function (because it only depends on the directions of the jets), and thus the $\log\left(  \frac{\nu}{p_+}\right)$ will not cancel entirely in the final result. This is perhaps surprising because $p_+$ is a hard scale, and the soft and jet functions are expanded around the soft and collinear scales. We thus say that the log of $\frac{\nu}{p_+}$ is anomalous. In SCET, the appearance of this logarithm is called the \emph{collinear anomaly}~\cite{Becher:2010tm}. However, we point out that the appearance of the collinear anomaly is not a feature of SCET itself; it is a consequence of the method of regions that appears whenever two regions overlap and live at the same scale, requiring an additional (rapidity) regulator.

\subsection{\texorpdfstring{\boldmath The $z_N$-Prescription\unboldmath}{The zN-Prescription}}
\label{Gen_sec_zNPrescription}
Most of the regularization schemes mentioned in the last section are defined on beam, jet, and soft functions already written in terms of matrix elements of operators, either in the effective theory or in QCD. Such matrix elements contain lightlike Wilson lines. One can then regularize the rapidity divergences by either shifting the reference direction in the Wilson lines off the light cone or by slightly adapting the momentum space definition of the Wilson lines by adding additional momentum factors to the exponents (the $
\eta$ regulator and the pure rapidity regulator) or by slightly moving the origin of the Wilson lines (the exponential regulator). For the delta regulator, one instead adds additional small mass terms to the effective theory Lagrangian, and the analytic regulator can be implemented by adding extra suppression factors to the phase space. In the spirit of this paper, it is not feasible to define the rapidity regulator by modifying Wilson lines. Instead of writing the radiative functions as matrix elements of operators of QCD or SCET fields, we have defined them in \eqref{Gen_eq_RadiativeFunctionsDifferential} as integrals of soft and collinear functions that appear when approximating the matrix element in the maximally unresolved limits. In particular, this means that the virtual integrals in real-virtual contributions are already performed in \eqref{Gen_eq_RadiativeFunctionsDifferential}, meaning that we cannot a posteriori shift the Wilson lines appearing in the loop integrals off the light cone. 

Instead of defining the regularization scheme by modifying the Wilson lines, we will redefine the collinear matrix elements introduced in \secref{sec_CollinearLimits} such that they do not change in the collinear limit while removing the spurious collinear divergences along the lightlike reference direction $\bar{n}$. We will achieve this by replacing some, but not all, collinear momentum fractions in the splitting kernels with energy fractions. An operator definition, of this procedure will be left to future work.

This paper will focus on rapidity divergences in the jets. However, the regularization procedure we will use has first been introduced in \cite{Catani:2022sgr} to regularize rapidity divergences in beam functions. In a collinear sector defined by the jet momentum $p$ along the lightlike vector $n$ and the anti-collinear lightlike vector $\bar{n}$ (c.f. \secref{Gen_sec_PhaseSpaceFactorization}), the momentum fraction of a collinear emission $\tilde{k}_i$ is defined as\footnote{At leading power the momentum fraction of $\tilde{k}_i$ is the same as the momentum fraction of $k_i$. Thus, we will drop tildes on the $z_i$ variables.}
\begin{equation}
  z_i=\frac{\tilde{k}_i\cdot \bar{n}}{p\cdot \bar{n}}\, .
\end{equation}

To define an energy fraction, one instead uses a massive, i.e., time-like reference vector $N$ and defines\footnote{We point out that $\sum_i z_{N, i}\neq 1$.}
\begin{equation}
  z_{N, i}=\frac{\tilde{k}_i\cdot N}{p\cdot N}\, .
\end{equation}

In any rest frame of $N$, $z_{N, i}$ is the energy ratio of $\tilde{k}_i$ and $p$. We will always assume that the vector $N$ is a linear combination of the respective $n$ and $\bar{n}$ such that the transverse momentum $\tilde{k}_\perp$ with respect to $n$ and $\bar{n}$ is also transverse to $N$, i.e., $\tilde{k}_\perp \cdot N=0$. We will typically use $N=n+\bar{n}\propto q$, where $q$ is the total four-momentum of the hard system, but in principle, we allow for different $N_i$ for every jet $p_i$, as long as $N_i$ is a linear combination of $n_i$ and $\bar{n}_i$. 

Let us study a dummy resolution variable for dijet production at electron-proton colliders at NLO to motivate this procedure. If we are only interested in the NLO corrections, we only need to define the resolution variable on single-emission contributions. The real emission phase space only contains a quark and an anti-quark with momenta $k_1$ and $k_2$ and a gluon with momentum $k$. We can then define the dummy resolution variable as $\lvert \tilde{k}_\perp \rvert$, where $\tilde{k}_\perp$ is the transverse momentum of the gluon with respect to the thrust axis. The dummy variable has the power counting exponents $b=c=1$. Note that this variable is not a well-defined resolution variable because it singles out the gluon, but it still works as a resolution variable for NLO calculations. The cumulant quark jet function for the dummy variable agrees with the $\ktness$ quark jet function we tried to calculate in \eqref{Gen_eq_RapidityDivergenceExample}, but obtained an ill-defined result. To make the quark jet function finite, we now define the $z_N$-prescription splitting kernel $\hat{P}_{N, gq}(z)$ as\footnote{Up to power corrections, we could also have replaced the second $z$ with a $z_N$. Here we chose not to do so, to illustrate that one does not have to replace every $z$. This will be useful for higher multiplicity splittings later on.}
\begin{equation}
  \label{Gen_eq_zNSplittingKernelgqLO}
  \hat{P}_{N, q\to gq}(z)\equiv 2C_F \left(\frac{1}{z_N} -1 + \frac{1-\epsilon}{2}z \right)\, , 
\end{equation}
which leads to the jet function in the $z_N$-prescription
\begin{equation}
  \label{Gen_eq_NLOznJetFunction}
  \hat{\J}_{N, q}^{(1)}\equiv \mu^{2\epsilon} \frac{e^{\epsilon \gamma_E}}{\Gamma(1-\epsilon)}\int_0^1 \rd z \frac{\rd^{d-2}\tilde{k}_\perp}{\Omega_{d-2}}\frac{\hat{P}_{N, q\to gq}(z)}{\lvert \tilde{k}_{\perp} \rvert^2}\theta\!\left( \qfcut- \lvert \tilde{k}_{\perp}\rvert\right)\, .
\end{equation}
To evaluate this integral we rewrite 
\begin{equation}
  \label{eq_zNDefinition}
  z_N=\frac{\tilde{k}_1\cdot N}{p\cdot N}=z+\frac{N^2 \lvert \tilde{k}_{\perp} \rvert^2}{4(p\cdot N)^2}\frac{1}{z}\, , 
\end{equation}
where $p$ is the hard momentum of the quark jet splitting into $k_1$ and $k$.
In the pure collinear limit, $\tilde{k}_\perp \sim \lambda Q$ and $z\sim 1$, $z_N$ and $z$ agree up to power corrections of order $\lambda^2$. However, in the soft region, $\tilde{k}_\perp \sim \lambda Q$ and $z\sim \lambda$, the difference between $z$ and $z_N$ is of the same order as $z$ itself. We also notice that the soft region picks up a dependence on the hard scale $\frac{4(p\cdot N)^2}{N^2}$, which implies that the soft zero bin is not scaleless, and we will need to deal with it to avoid over counting. The integral in \eqref{Gen_eq_NLOznJetFunction} can be calculated by using the identity
\begin{equation}
  \begin{aligned}
    \frac{1}{z_N}&= \left( \frac{1}{z+\frac{N^2 \lvert \tilde{k}_{\perp} \rvert^2}{4(p\cdot N)^2}\frac{1}{z}} \right)_++\frac{1}{2}\log\left( \frac{1+\frac{N^2 \lvert \tilde{k}_{\perp} \rvert^2}{4(p\cdot N)^2}}{\frac{N^2 \lvert \tilde{k}_{\perp} \rvert^2}{4(p\cdot N)^2}} \right)\delta(z)\\
    &=\left( \frac{1}{z} \right)_+-\frac{1}{2}\log\left(\frac{N^2 \lvert \tilde{k}_{\perp} \rvert^2}{4(p\cdot N)^2}  \right)\delta(z)+\mathcal{O}(\lambda)\, ,
  \end{aligned}
\end{equation}
where we used that the plus prescription in the first term on the right-hand side removes the pole in the soft limit, thus allowing us to expand the parentheses in the collinear limit, leading to the second line. Performing the trivial integrals in \eqref{Gen_eq_NLOznJetFunction} we find
\begin{equation}
  \hat{\J}_{N, q}^{(1)}=\left( \frac{\mu^2}{\qfcut^2} \right)^\epsilon\frac{e^{\epsilon \gamma_E}}{\Gamma(1-\epsilon)}C_F\left(\frac{1}{2\epsilon}\left[ \frac{1}{\epsilon}+\log\left( \frac{N^2\qfcut^2}{4(p\cdot N)^2} \right) \right]+\frac{ 3}{4\epsilon} +\frac{1}{4} \right) + \mathcal{O}(\lambda)\, .
\end{equation}
We point out that the introduction of the new scale $N^2$  led to the ``anomalous'' appearance of the large logarithm $\log\left( \frac{N^2\qfcut^2}{4(p\cdot N)^2} \right)$. This logarithm can be given a natural interpretation by parametrizing the vector $N$ as
\begin{equation}
  N=\sqrt{N^2}\left( e^\eta \frac{n}{2}+ e^{-\eta} \frac{\bar{n}}{2}\right)\, .
\end{equation}
Then the large logarithm can be rewritten as
\begin{equation}
  \label{eq_log_withrapidity}
  \log\left( \frac{N^2\qfcut^2}{4(p\cdot N)^2} \right)=2\log\left( \frac{\qfcut Q e^{\eta}}{2 p\cdot q} \right)=2\log\left( \frac{\qfcut}{2E_{J}} \right)+2\eta\, ,
\end{equation}
where $E_{J}$ is the energy of the jet $p$ in the rest frame of $q$. This means that the large logarithm could be eliminated from the jet function with the choice $\eta \sim -\log\left( \frac{\qfcut}{2E_{J}} \right)$. In this paper, we are mainly concerned with fixed-order predictions. However, if we were interested in resumming logarithms of $\qfcut/Q$ to all orders, we would calculate the jet function with $\eta \sim -\log\left( \frac{\qfcut}{2E_{J}} \right)$ and set up a renormalization group equation in $\eta$ to resum the logarithms of $ \frac{N^2\qfcut^2}{4(p\cdot N)^2} $. The term \emph{rapidity renormalization group} is therefore particularly appropriate. The remaining large logarithms could then be resummed using standard renormalization group equations in $\mu$.

Continuing with our fixed-order discussion, we analyze the zero-bin contribution. We first calculate the soft limit of the modified collinear approximation of the matrix element. We find  
\begin{equation}
  \label{Gen_eq_SoftlimitNLOznMatrixelement}
  \frac{\lvert M^{(0)}_{gq\dots}(k, k_1, \dots)\rvert^2}{\lvert M^{(0)}_{q\dots}(p, \dots)\rvert^2}\xrightarrow[z_N]{k_1 \parallel k}  8\pi \as^u\frac{\hat{P}_{N, q\to gq}(z)}{2\tilde{k}_1\cdot \tilde{k}}\xrightarrow{\tilde{k}\to 0}  4\pi \as^u J^{(0)}_{N}(p, \tilde{k})\, ,
\end{equation}
where the first arrow indicates that we take the collinear limit and then apply the $z_N$-prescription, and the second arrow denotes the soft gluon limit (while the quark stays collinear along the jet direction $p$). Here, we defined a new eikonal factor for single gluon emission
\begin{equation}
  J_{N}(p_i, k)\equiv \sum_{n=0}^\infty \left( \frac{\alpha_0}{\pi} \right)^n J_{N}^{(n)}(p_i, k)\, ,
\end{equation}
which, to the lowest order, is given by
\begin{equation}
  J_{N}^{(0)}(p_i, k)=2C_i \frac{p_i\cdot N}{\left( p_i\cdot k \right) \left( N\cdot k \right)}\, ,
\end{equation}
where $C_i$ is the quadratic Casimir of the hard leg $i$. The result in \eqref{Gen_eq_SoftlimitNLOznMatrixelement} looks similar to the soft approximation of the tree-level matrix element, which for dijet production is given by 
\begin{equation}
  \frac{\lvert M_{gq\bar{q}\dots}(k,k_1, k_2 \dots)\rvert^2}{\lvert M_{q\bar{q}}(p_3,p_4 \dots)\rvert^2}\xrightarrow{k\to 0} 8\pi \as^u C_F\frac{p_3\cdot p_4}{\left( p_3\cdot k \right) \left( p_4\cdot k \right)}=4\pi \as^u \bJJ^{(0)}(k)\, .
\end{equation}
In fact, for dijet production, our collinear reference vectors are $p_3\propto n_3=\bar{n}_4$ and $p_4\propto n_4=\bar{n}_3$. Thus, for both jets, the possible choices of $N$ are linear combinations of $p_3$ and $p_4$. Curiously, if one chooses the same $N$ for both jets, we find 
\begin{equation}
  J_{N}^{(0)}(p_3, k)+ J_{N}^{(0)}(p_4, k)=\bJJ^{(0)}(k)\, .
\end{equation}
The zero bin in \eqref{Gen_eq_NLOznJetFunction} for the jet with momentum $p_3$ is given by 
\begin{equation}
  \label{Gen_eq_ZeroBinNLOznJetFunction}
  \begin{aligned}
    \hat{\J}_{N, q}^{(1)}(p_3)\Big\vert_{0-\mytext{bin}}&=\mu^{2\epsilon} \frac{e^{\epsilon \gamma_E}}{\Gamma(1-\epsilon)}\int \frac{\rd^dk\delta_+(k^2)}{\Omega_{d-2}}J_{N}^{(0)}(p_3, k)\theta\!\left( \qfcut- \lvert k_{\perp}\rvert\right)\, ,
  \end{aligned}
\end{equation}
where we relabeled the boosted and rescaled momentum $\tilde{k}$ as $k$ to better match the soft function. 
The zero bin is not well-defined in this case because $J^{(0)}_N(p_3, k)$ has an unregularized collinear singularity if $k$ becomes parallel to $p_3$. However, we note that the soft function 
\begin{equation}
  \hat{\Sb}^{(1)}=\mu^{2\epsilon} \frac{e^{\epsilon \gamma_E}}{\Gamma(1-\epsilon)}\int \frac{\rd^dk\delta_+(k^2)}{\Omega_{d-2}}\bJJ^{(0)}(k)\theta\!\left( \qfcut- \lvert k_{\perp}\rvert\right)
\end{equation}
is also ill-defined because it has two collinear singularities that are not regularized by dimensional regularization. Putting together the soft function and the zero bins, we find 
\begin{equation}
  \hat{\Sb}^{(1)}-\hat{\J}_{N, q}^{(1)}(p_3)\Big\vert_{0-\mytext{bin}}-\hat{\J}_{N, q}^{(1)}(p_4)\Big\vert_{0-\mytext{bin}}=0\, .
\end{equation}
The vanishing of the soft function minus the zero bins, the \emph{soft subtracted function}, is a special case. In general, this is only true if there is only one emitting dipole in the hard function, and if the resolution variable is the same in the soft limit and subsequent collinear and soft limits. However, we can use this simple example to motivate how to define the $z_N$-prescription at higher orders to get the simplest possible soft subtracted function. 

The above derivation of the soft subtracted function and the regularized jet function above was not rigorous. In particular, the jet function at intermediate steps and the zero bins were not well-defined due to the rapidity divergences. For a rigorous treatment of the $z_N$-prescription, we note that all ingredients in the general factorization formula in \eqref{gen:eq_FactorizationFormulaFully} are well-defined as long as the resolution variable $\qf$ is not of the $\SCETII$ type, i.e., as long as $a\neq c$ and $b\neq c$. In this case, rapidity regularization is not necessary, however, we can still choose to use the $z_N$-prescription to define rapidity regularized beam and jet functions and subtract all relevant zero bins, which are now also well-defined. One can then group the zero bins with the soft function to define a subtracted soft function. At this point, all ingredients in the factorization formula are free of rapidity divergences and one can smoothly extend it to the case where the resolution variable is of the $\SCETII$ type. 

In the following, we will discuss the $z_N$-prescription at higher orders. While the $z_N$-prescription can be applied for any resolution variable, we will in the following focus on variables that satisfy a factorization formula at the cumulant level of the form in \eqref{Gen_eq_CumulantFactorizationFormula}. The procedure straightforwardly generalizes to other variables, and Section~\ref{chap_FactorizationFormulaForKtNess} also discusses versions of $\ktness$ that only approximately factorize at the cumulant level.  We modify the collinear splitting kernels entering the beam and jet functions by replacing some longitudinal momentum fractions with energy fractions to get rid of rapidity divergences\footnote{We point out that modifying the collinear matrix elements will, in general, also introduce subleading power corrections in $\lambda$. In this paper, we will always drop such power corrections in the radiative functions $\hat{\Bb}_N, \hat{\J}_N(\qfcut)$ and $\hat{\Sb}_{\mytext{sub}}(\qfcut)$, but they should be kept if one is interested in subleading powers.}. Assuming that the cumulant factorization is not spoiled by the $z_N$-prescription, one finds a new factorization formula 
\begin{equation}
  \label{Gen_eq_CumulantFactorizationFormulaZN}
\begin{aligned}
    \hat{\Bb}_{a_1b_1}(\qfcut, z_1)\hat{\Bb}_{a_2b_2}(\qfcut, z_2)\prod_{i=3}^{n_J+2} \hat{\J}_{a_i}(\qfcut)\hat{\Sb}(\qfcut)&=\hat{\Bb}_{N, a_1b_1}(\qfcut, z_1)\hat{\Bb}_{N, a_2b_2}(\qfcut, z_2)\\
    &\quad\times\prod_{i=3}^{n_J+2} \hat{\J}_{N, a_i}(\qfcut)\hat{\Sb}_{\mytext{sub}}(\qfcut)\, ,
\end{aligned}\end{equation}
where we defined
\begin{equation}
  \label{Gen_eq_CumulantRadiativeFunctionsN}
  \begin{aligned}
    \hat{\Bb}_{N, a_1b_1}(\qfcut, z_1)&=\theta(\qfcut)\delta(1-z_1)\delta_{a_1b_1}\bbbone \\
    &\hspace{-2.2cm}+z_1\sum_{n=1}^\infty\sum_{\Af_1}\frac{1}{S_{\Af_1}}\int \rd^d k_{I_1}\rd \Pi_{I_1}\delta\!\left( \bar{z}_1-\frac{k_{I_1}\cdot P_2}{x_1P_1\cdot P_2} \right) \theta\!\left( \qfcut -  \qft_{C_1}(I_1)\right)\!\bcalPhat_{N, a_1b_1;\Af_1}(I_1, z_1)\\
    \hat{\J}_{N, a_i}(\qfcut)&=\theta(\qfcut)\bbbone \\
    &\hspace{-2.2cm}+\sum_{n=2}^\infty \sum_{\Af_i}\frac{1}{S_{\Af_i}}2(2\pi)^{d-1}\!\prod_{j\in F_i}\dktj\delta\bigg(1-\sum_j\tilde{z}_j\bigg)\delta^{d-2}\!\bigg( \sum_j\tilde{k}_{j, \perp} \!\!\bigg) \calPhat_{N, \Af_i}(\tilde{F}_i)\theta\!\Big( \qfcut -  \qft_{C_i}(\tilde{F}_i)\Big) .
  \end{aligned}
\end{equation}
Some modified collinear matrix elements $\calPhat_{N, \Af_i}$ entering jet functions will be given below.
The soft subtracted function $\hat{\Sb}_{\mytext{sub}}(\qfcut)$ collects all soft contributions as well as the zero bins. We point out that the idea of grouping zero bins with the soft function to construct a subtracted soft function is not new. A similar function was, for example, constructed for $q_T$-resummation in heavy quark production~\cite{Catani:2014qha, Catani:2023tby}, or in \cite{Buonocore:2022mle}. The zero bins of the jet functions are defined by expanding $\hat{\J}_{N}(\qfcut)$ to leading power in $\lambda$ in all non-vanishing regions where not all momenta in the phase space have the collinear scaling $k\sim (1, \lambda^{2a}, \lambda^a)Q$, and summing over all these regions. The zero bins of the beam functions are defined analogously. We define the perturbative expansions of the $z_N$-prescription radiative functions in cumulant space as
\begin{equation}
  \label{Gen_eq_BareCumulantExpansion_zN}
  \begin{aligned}
    \hat{\Bb}_{N, a_1b_1}(\qfcut, z_1)&=\theta(\qfcut)\delta(1-z_1)\delta_{a_1b_1}\bbbone+\sum_{n=1}^\infty\left( \frac{\alpha_0}{\pi} \right)^n\hat{\Bb}_{N, a_1b_1}^{(n)}(z_1)\\
    \hat{\J}_{N, a_i}(\qfcut)&=\theta(\qfcut)\bbbone +\sum_{n=1}^\infty\left( \frac{\alpha_0}{\pi} \right)^n\hat{\J}_{N, a_i}^{(n)}\\
    \hat{\Sb}_{\mytext{sub}}(\qfcut, n_1, n_2, \{n_i\})&=\theta(\qfcut)\bbbone+\sum_{n=1}^\infty\left( \frac{\alpha_0}{\pi} \right)^n \hat{\Sb}_{\mytext{sub}}^{(n)}(N, n_1, n_2, \{n_i\})\, .
  \end{aligned}
\end{equation}

To define the modified collinear matrix elements $\bcalPhat_{N, a_1b_1;\Af_1}$ and $\calPhat_{N, \Af_i}$, we use the following guiding principle: \emph{We want to define the $z_N$-prescription such that the soft subtracted function vanishes if we only have one emitting dipole (either in the initial- or in the final-state) and if the resolution variable is sufficiently simple, as is the case for example for our dummy variable $\lvert \tilde{k}_\perp \rvert$ studied above. 
}

In this paper, we will only discuss the modified splitting kernels for jet functions, and at NNLO, we only give the modifications needed for quark jet functions. Modifications to the beam functions and gluon jet functions will not be discussed in this paper, but we will provide them in future work.  
\paragraph{NLO}
We already defined the $z_N$-prescription for the quark splitting matrix element in \eqref{Gen_eq_zNSplittingKernelgqLO}. Here we repeat it for completeness's sake and add the $z_N$-prescription for the gluon splitting matrix elements.

The NLO $z_N$-prescription consists of the replacements
\begin{equation}
\begin{aligned}
     \hat{P}^{ss^{\prime}}_{N, q\to gq}(z_1)&\equiv\delta_{s s^{\prime}}  2C_F \left(\frac{1}{z_{N, 1}} -1 + \frac{1-\epsilon}{2}z_1 \right)\\
  \hat{P}_{N, g\to g g}^{\mu \nu}\left(z_1, k_{\perp} ; \epsilon\right)&\equiv 2 C_A\!\!\left[g^{\mu \nu}\!\!\left(\!2-\frac{1}{z_{N, 1}}-\frac{1}{z_{N, 2}}\right)-2(1-\epsilon) z_1z_2 \frac{k_{\perp}^\mu k_{\perp}^\nu}{k_{\perp}^2}\right]\, .
  \end{aligned}
\end{equation} 
The $g\to q\bar{q}$ splitting kernel remains unmodified, and the identities 
\begin{equation}
  \label{Gen_eq_SymmetrySplittingKernels}
  \hat{P}^{ss^{\prime}}_{N, q\to gq}(z_1)= \hat{P}^{ss^{\prime}}_{N, \bar{q}\to g\bar{q}}(z_1)= \hat{P}^{ss^{\prime}}_{N, q\to qg}(z_1)\Bigg\rvert_{1\leftrightarrow 2}= \hat{P}^{ss^{\prime}}_{N, q\to\bar{q}g}(z_1)\Bigg\rvert_{1\leftrightarrow 2}
\end{equation}
still hold, which defines the $z_N$-prescription for all remaining splitting kernels. Note that while $z_1+z_2=1$, the respective $z_N$ are not constrained in this way. For example, one cannot not replace $\frac{1}{z_{N, 2}}$ with 
$\frac{1}{1-z_{N, 1}}$ in the $g\to gg$ splitting kernel. Instead, $z_{N,2}$ is obtained from \eqref{eq_zNDefinition} through the replacement $z\to z_2=1-z_1$. We calculate the result for the NLO $z_N$-prescription cumulant jet function at NLO for very general resolution variables in Appendix~\ref{App_NLOJetFunctions}.

The zero bins of the NLO quark and gluon jet functions are both given by \eqref{Gen_eq_ZeroBinNLOznJetFunction}; one simply used the respective Casimir factors $C_F$ and $C_A$ in the definition of $J_N^{(0)}(p_i, k)$. The soft subtracted function up to order $\alpha_s$ is found to be
\begin{equation}
  \label{Gen_eq_SsubZnNLO}
  \begin{aligned}
    \hat{\Sb}_{\mytext{sub}}^{(1)}=&\mu^{2\epsilon} \frac{e^{\epsilon \gamma_E}}{\Gamma(1-\epsilon)}\int \frac{\rd^dk\delta_+(k^2)}{\Omega_{d-2}}\Bigg[ \bJJ^{(0)}(k)\theta\!\left( \qfcut-\qft_S\right)\\
    &\quad-\sum_{\alpha=1}^{n_J+2}J^{(0)}_{N_\alpha}(p_\alpha, k)\theta\!\left( \qfcut-\qft_{C_\alpha, S}\right)\Bigg]\, ,
  \end{aligned}
\end{equation}
where we accounted for the possibility of using a different time-like reference vector $N_\alpha$ for each collinear direction.
 The expression $\qft_{C_\alpha, S}$ refers to the approximation of the resolution variable that one obtains by first expanding $\qf$ in the limit where $k$ is collinear to the hard leg with momentum $p_\alpha$ and then expanding the result in the soft gluon limit (c.f. Appendix~\ref{sec_ShorthandNotationRegions}). We point out that the definition \eqref{Gen_eq_SsubZnNLO} is almost the same as the wide-angle soft function derived in \cite{Buonocore:2023rdw} with very different methods. 

\paragraph{NNLO---RV terms}
The $z_N$-prescription for the $1\to 2$ splitting kernels at one loop is defined as follows. We do not change $P_{g\to q\bar{q}}^{(1)}$, and we leave $ \hat{P}^{(1)}_{a\to a_1a_2, H}(z)$ (c.f. Appendix~\ref{sec_APkernels}) unchanged for all splittings. For the $q\to gq$ splitting, we define
\begin{equation}
\hat{P}^{(1)}_{N, q\to gq, C}(z_1, z_2)\equiv  \hat{P}^{(1)}_{q\to gq, C}(z_1, z_2) + C_A C_F \frac{ z_1^{-\epsilon -1} \Gamma (1-\epsilon ) \Gamma
  (1+\epsilon)}{\epsilon ^2}
  \left(1 - \frac{z_1}{z_{N, 1}} \right)\, .
\end{equation}
For the $g\to gg$ splitting, we replace
\begin{equation}
  \begin{aligned}
    \hat{P}^{(1)}_{N, g\to gg, C}(z_1, z_2)\equiv  \hat{P}^{(1)}_{g\to gg, C}(z_1, z_2) &+C_A C_A \frac{ z_1^{-\epsilon -1} \Gamma (1-\epsilon ) \Gamma
    (1+\epsilon)}{\epsilon ^2}
    \left(1 - \frac{z_1}{z_{N, 1}} \right)\\
    &+ C_A C_A \frac{ z_2^{-\epsilon -1} \Gamma (1-\epsilon ) \Gamma
    (1+\epsilon)}{\epsilon ^2}
    \left(1 - \frac{z_2}{z_{N, 2}} \right)\, .
  \end{aligned}
\end{equation}
The remaining splitting kernels are defined by the analog of the relations \eqref{Gen_eq_SymmetrySplittingKernels}.

The full splitting kernels in the $z_N$-prescription are then given by 
\begin{equation}
  \hat{P}^{(1)}_{N, a\to a_1a_2}(z_1, z_2)\equiv \cos(\pi \epsilon) c_P \left(  \hat{P}^{(1)}_{a\to a_1a_2, H}(z_1, z_2)+ \hat{P}^{(1)}_{N, a\to a_1a_2, C}(z_1, z_2) \right)\, .
\end{equation}

Replacing both the leading and next-to-leading order splitting kernels with their respective $z_N$-prescription counterparts, we can look at the soft limit of the collinear approximation. We can now generalize \eqref{Gen_eq_SoftlimitNLOznMatrixelement} to the next order in $\as$. Neglecting azimuthal correlations, we find 
\begin{equation}
  \begin{aligned}
  \lvert M_{ga_2, \dots}(k_1, k_2, \dots)\rvert^2&\xrightarrow[z_N]{k_1 \parallel k_2} 8\pi \as^u\lvert M_a(p_1, \dots)\rvert^2\frac{\hat{P}_{N, a\to ga_2}(z_1, z_2)}{2\tilde{k}_1\cdot \tilde{k}_2}\\&\xrightarrow{k_2\to 0} 
 4\pi  \as^u \lvert M_{a, \dots}(p_1, \dots)\rvert^2J_{N}(p_3, \tilde{k}_2)\, .
  \end{aligned}
\end{equation}
The one-loop coefficient of $J_{N}(p_3, k_2)$ reads
\begin{equation}
  J_{N}^{(1)}(p_i, k)=- C_A c_S \left(\mu^2\frac{n_i\cdot \bar{n}_i}{2n_i\cdot k \bar{n}_i\cdot k} \right)^\epsilon \frac{\cos(\pi \epsilon)}{2\epsilon^2} J_{N}^{(0)}(p_i, k)\, ,
\end{equation}
where $c_S$ was defined in \eqref{eq_cs_definition}. One easily checks that 
\begin{equation}
  \label{Gen_eq_JN1Sum}
  J_{N}^{(1)}(p_3, k)+J_{N}^{(1)}(p_4, k)= \bJJ^{(1)}(k)
\end{equation}
in the case of dijet production, if we take the same $N$ for both jets. 

As in the tree-level case, there would have been many possible ways to define the $z_N$-prescription that are equivalent to leading power. However, one still needs to be careful. For example, replacing every $z_i$ with a $z_{N, i}$ would lead to a different zero bin with less convenient properties, i.e., \eqref{Gen_eq_JN1Sum} would not hold.

Finally, we note that one can in general define the two-loop coefficient of the soft function due to single soft gluon emission as
\begin{equation}
  \label{Gen_eq_SsubZnSingleGluonRV}
  \begin{aligned}
    \hat{\Sb}^{(2)}_{g, \mytext{sub}}=&\mu^{2\epsilon} \frac{e^{\epsilon \gamma_E}}{\Gamma(1-\epsilon)}\int \frac{\rd^dk\delta_+(k^2)}{\Omega_{d-2}}\Bigg[\bJJ^{(1)}(k)\theta\!\left( \qfcut-\qft_S\right)\\
    &\quad-\sum_{\alpha=1}^{n_J+2}J_{N_\alpha}^{(1)}(p_\alpha, k)\theta\!\left( \qfcut-\qft_{C_\alpha, S}\right)\Bigg]\, ,
  \end{aligned}
\end{equation}
where the definition of $\qft_{C_\alpha, S}$ was given below \eqref{Gen_eq_SsubZnNLO}, and we again allowed the possibility to use different reference vectors $N_\alpha$ for each collinear direction. Note that in general, the one-loop eikonal kernel $\bJJ^{(1)}(k)$ contains a tripole term (c.f. \eqref{eq_bJJ1gOneLoop}).

\paragraph{NNLO---RR terms}
Here we only list the modified splitting kernels for quark splitting processes. We only need to modify the splittings of type $q\to \bar{q}_1^{\prime} q_2^{\prime} q_3$ and $q\to g_1 g_2 q_3$ as the others do not feature any rapidity divergences. The $z_N$-prescription for the $q\to \bar{q}_1^{\prime} q_2^{\prime} q_3$ splitting reads
\begin{equation}
  \begin{aligned}
    \left\langle\hat{P}_{N, q\to\bar{q}_1^{\prime} q_2^{\prime} q_3}\right\rangle &= \frac{1}{2}C_FT_R\frac{s_{123}}{s_{12}}\Biggl[
      \frac{1}{z_{1, N}+z_{2, N}}\left(-\frac{(z_1+z_2)t_{12, 3}^2}{s_{12}s_{123}} + 4z_3 + (z_1-z_2)^2\right) \\
      &\quad+ (1-2\epsilon) \left(z_1+z_2-\frac{s_{12}}{s_{123}} \right)\Biggr]\, .
  \end{aligned}
\end{equation} 
We point out that only the combination $z_1+z_2$, i.e., the longitudinal momentum fraction of the gluon that splits into the $q\bar{q}$ pair, was replaced with an energy fraction. It turns out that this modification only introduces zero bins if both $q$ and $\bar{q}$ are in the soft sector. In this limit, the tree-level matrix element behaves as
\begin{equation}
  \begin{aligned}
    \lvert M_{q^\prime_1 \bar{q}^\prime_2 q_3\dots}(k_1, k_2, k_3, \dots)\rvert^2&\xrightarrow[z_N]{k_1 \parallel k_2\parallel k_3} \left( \frac{8 \pi \asbare}{s_{123}} \right)^2 \left\langle\hat{P}_{N, q\to\bar{q}_1^{\prime} q_2^{\prime} q_3}\right\rangle  \lvert M_{q\dots}(p, \dots)\rvert^2\\
    &\xrightarrow{k_1\sim k_2\to 0}\left( 4\pi \asbare \right)^2 J_{N, q\bar{q}}^{(0)}(p, \tilde{k}_1, \tilde{k}_2)\lvert M_{q\dots}(p, \dots)\rvert^2\, ,
  \end{aligned}
\end{equation}
where we defined
\begin{equation}
    J_{N, q\bar{q}}^{(0)}(p_i, k_1, k_2)\equiv J_{N}^{(0)}(p_i, k_1+k_2)T_R\mathcal{S}_{q\bar{q}}(n_i, \bar{n}_i;k_1, k_2)\, ,
\end{equation}
where $\mathcal{S}_{q\bar{q}}$ is the function we defined in \eqref{eq_Sqqbar}.
If there is only a single back-to-back emitting dipole one finds 
\begin{equation}
  J_{N, q\bar{q}}^{(0)}(p_3, k_1, k_2)+J_{N, q\bar{q}}^{(0)}(p_4, k_1, k_2)= \bJJ_{q\bar{q}}^{(0)}(k_1, k_2)\, ,
\end{equation}
just as we wanted. One can then define the soft subtracted function due to soft quark pair emission as
\begin{equation}
  \label{Gen_eq_SsubZnqqbar}
  \begin{aligned}
    \hat{\Sb}^{(2)}_{q\bar{q}, \mytext{sub}}=&\mu^{4\epsilon} \left( \frac{e^{\epsilon \gamma_E}}{\Gamma(1-\epsilon)} \right)^2\int \frac{\rd^dk_1\delta_+(k_1^2)}{\Omega_{d-2}}\int \frac{\rd^dk_2\delta_+(k_2^2)}{\Omega_{d-2}}\Bigg[\bJJ_{q\bar{q}}^{(0)}(k_1, k_2)\theta\!\left( \qfcut-\qft_{S}\right)\\
    &\quad-\sum_{\alpha=1}^{n_J+2}J_{N_\alpha, q\bar{q}}^{(0)}(p_\alpha, k_1, k_2)\theta\!\left( \qfcut-\qft_{C_\alpha, S}\right)\Bigg]\, ,
  \end{aligned}
\end{equation}
where $\qft_{S} = \qft_{S}(k_1, k_2)  $   is the limit of the resolution variable if we only have two soft emissions, and $\qft_{C_\alpha, S}(k_1, k_2)$ is obtained by first expanding the resolution variable in the limit where $k_1$ and $k_2$ are collinear to the hard leg with momentum $p_\alpha$ and then in the limit where $k_1$ and $k_2$ are soft (c.f. Appendix~\ref{sec_ShorthandNotationRegions}). We allowed the possibility of using different reference vectors $N_\alpha$ for each collinear direction in \ref{Gen_eq_SsubZnqqbar}.

For the $q\to ggq$ splittings, we separately discuss the modification to the Abelian and non-Abelian parts of the collinear matrix elements. The $z_N$-prescription for the non-Abelian part reads
\begin{equation}
  \begin{aligned}
    &\left\langle\hat{P}_{N, q\to g_1 g_2 q_3}^{(\mathrm{nab})}\right\rangle = \left\langle\hat{P}_{q\to g_1 g_2 q_3}^{(\mathrm{nab})}\right\rangle-\left( 1-\frac{z_1+z_2}{z_{1, N}+z_{2, N}} \right)\Biggl[ \frac{(1-\epsilon ) (z_1 s_{23}-z_2 s_{13})^2}{(z_1+z_2)^2 s_{12}^2}  \\
    &+\frac{(z_1+2 z_2) s_{123}^2}{z_2 (z_1+z_2) s_{12} s_{13}}-\frac{s_{123}^2}{2 z_1 z_2 s_{13} s_{23}}+\frac{(z_1-3 z_2) s_{123}}{z_2 (z_1+z_2) s_{12}}-\frac{s_{123}}{z_1 (z_1+z_2) s_{13}} +\left( 1\leftrightarrow 2 \right)\Biggr]\, .
  \end{aligned}
\end{equation}
Again, one only needs to replace the combination $z_1+z_2$ by $z_{1,N}+z_{2,N}$, which implies that leading-power zero bins arise only in the region where both gluons become soft simultaneously. In this region, the modified matrix element behaves as
\begin{equation}
  \begin{aligned}
    \lvert M_{g_1 g_2 q_3\dots}(k_1, k_2, k_3, \dots)\rvert^2&\xrightarrow[z_N, C_F C_A]{k_1 \parallel k_2\parallel k_3} \left( \frac{8 \pi \asbare}{s_{123}} \right)^2C_F C_A\left\langle\hat{P}_{N, q\to g_1 g_2 q_3}^{(\mathrm{nab})}\right\rangle \lvert M_{q\dots}(p, \dots)\rvert^2\\
    &\xrightarrow{k_1\sim k_2\to 0}\left( 4\pi \asbare \right)^2 C_F C_AJ_{N, gg}^{(0), (\mathrm{nab})}(p, \tilde{k}_1, \tilde{k}_2)\lvert M_{q\dots}(p, \dots)\rvert^2\, ,
  \end{aligned}
\end{equation} 
where in the first limit, we take the (azimuthally averaged) triple-collinear approximation, only keep the contribution proportional to $C_F C_A$, and go over to the $z_N$-prescription. We defined the eikonal kernel
\begin{equation}
    C_F C_A J_{N, gg}^{(0), (\mathrm{nab})}(p_i, k_1, k_2)\equiv J_{N}^{(0)}(p_i, k_1+k_2)C_A\mathcal{S}_{gg}\left( n_i, \bar{n}_i;k_1, k_2 \right)\, ,
\end{equation}
where $\mathcal{S}_{gg}$ is the function we defined in \eqref{eq_Sgg}.
We again note that if there is only a single back-to-back emitting dipole one finds
\begin{equation}
  J_{N, gg}^{(0), (\mathrm{nab})}(p_3, k_1, k_2)+J_{N, gg}^{(0), (\mathrm{nab})}(p_4, k_1, k_2)= \bJJ_{gg}^{(0), (\mathrm{nab})}(k_1, k_2)
\end{equation}
if one uses the same $N$ for both jets. Non-Abelian emission of two soft gluons then contributes as
\begin{equation}
  \label{Gen_eq_SsubZngg_nonAbelian}
  \begin{aligned}
    \hat{\Sb}^{(2), (\mathrm{nab})}_{gg, \mytext{sub}}=&\mu^{4\epsilon} \left( \frac{e^{\epsilon \gamma_E}}{\Gamma(1-\epsilon)} \right)^2\frac{1}{2}\int \frac{\rd^dk_1\delta_+(k_1^2)}{\Omega_{d-2}}\int \frac{\rd^dk_2\delta_+(k_2^2)}{\Omega_{d-2}}\Bigg[\bJJ_{gg}^{(0), (\mathrm{nab})}(k_1, k_2)\theta\!\left( \qfcut-\qft_{S}\right)\\
    &\quad-\sum_{\alpha=1}^{n_J+2}J_{N_\alpha, gg}^{(0), (\mathrm{nab})}(p_\alpha, k_1, k_2)\theta\!\left( \qfcut-\qft_{C_\alpha, S}\right)\Bigg]
  \end{aligned}
\end{equation}
to the two-loop soft function. We allowed the possibility of using different reference vectors $N_\alpha$ for each collinear direction in \ref{Gen_eq_SsubZngg_nonAbelian}.

Finally, we come to the Abelian part of the $q\to ggq$ splitting. The splitting kernel can be split into a strongly ordered and a non-strongly ordered part. The non-strongly ordered part does not contain any soft endpoints, and thus it does not need to be modified. For the strongly ordered part, we want the $z_N$-modified splitting kernels to be still written as a product of two modified $1\to 2$ splitting kernels, similar to the unmodified splitting kernels. This naturally leads us to the modification
\begin{equation}
  \label{Gen_eq_zNggqAbelian}
  \begin{aligned}
    &\left\langle\hat{P}_{N, q\to g_1 g_2 q_3}^{(\mathrm{ab})}\right\rangle = \left\langle\hat{R}_{q\to g_1 g_2 q_3}^{(\mathrm{ab})}\right\rangle+\Biggl(\frac{s_{123}\hat{P}_{N, q\to gq}(z_1)\hat{P}_{N, q\to gq}\left(\frac{z_2}{1-z_1}\right)}{s_{23}}+(1\leftrightarrow 2)\Biggr)\, , 
  \end{aligned}
\end{equation}
where
\begin{equation}
  \hat{P}_{N, q\to gq}\left(\frac{z_2}{1-z_1}\right)\equiv  2C_F \left(\frac{1-z_1}{z_{N, 2}} -1 + \frac{1-\epsilon}{2}\frac{z_2}{1-z_1} \right)\, ,
\end{equation}
and $ \left\langle\hat{R}_{q\to g_1 g_2 q_3}^{(\mathrm{ab})}\right\rangle$ is defined in \eqref{eq_Rab}.
Again, there are many choices for the $z_N$-prescription that are equivalent to leading power. The double-soft limit of the modified matrix element reads
\begin{equation}
  \begin{aligned}
    \lvert M_{g_1 g_2 q_3\dots}(k_1, k_2, k_3, \dots)\rvert^2&\xrightarrow[z_N, C_F^2]{k_1 \parallel k_2\parallel k_3} \left( \frac{8 \pi \asbare}{s_{123}} \right)^2C_F^2\left\langle\hat{P}_{N, q\to g_1 g_2 q_3}^{(\mathrm{ab})}\right\rangle \lvert M_{q\dots}(p, \dots)\rvert^2\\
    &\xrightarrow{k_1\sim k_2\to 0}\left( 4\pi \asbare \right)^2 J_{N, gg}^{(0), (\mathrm{ab})}(p, \tilde{k}_1, \tilde{k}_2)\lvert M_{q\dots}(p, \dots)\rvert^2\, ,
  \end{aligned}
\end{equation} 
where we defined
\begin{equation}
  J_{N, gg}^{(0), (\mathrm{ab})}(p_i, k_1, k_2)\equiv 4C^2_i \frac{p_i\cdot N}{\left( p_i\cdot k_1 \right) \left( N\cdot k_1 \right)}\frac{p_i\cdot N}{\left( p_i\cdot k_2 \right) \left( N\cdot k_2 \right)} =J_{N}^{(0)}(p_i, k_1)J_{N}^{(0)}(p_i, k_2)\, .
\end{equation}
Even for dijet production, we have 
\begin{equation}
  J_{N, gg}^{(0), (\mathrm{ab})}(p_3, k_1, k_2)+J_{N, gg}^{(0), (\mathrm{ab})}(p_4, k_1, k_2)\neq \bJJ_{gg}^{(0), (\mathrm{ab})}(k_1, k_2)\, .
\end{equation}
Thus, we must ask ourselves whether we have chosen the ideal $z_N$-prescription. 

However, there are three types of zero bins that we have been forgetting so far. The first type is additional zero bins stemming from our manipulation of the triple-collinear limit. The second type stems from zero bins in double-collinear limits, i.e., limits where we have two separate pairs of partons that are collinear along two different jets. The third type is zero bins introduced in the limit where one parton is soft, and two others are collinear.

To understand the first type, note that in \eqref{Gen_eq_zNggqAbelian} we did not only replace the combination $z_1+z_2$ with $z_{1, N}+z_{2, N}$, but we also replaced individual momentum fractions. This has an important consequence---the $z_N$-prescription not only introduces zero bins in the double-soft limit but also in the collinear-soft limits, for example, when $k_1$ and $k_3$ are collinear along $p_i$ and $k_2$ is soft. In this limit, the modified triple-collinear matrix element reads
\begin{equation}
  \label{Gen_eq_CollinearSoftLimitOfPABggq}
  \begin{aligned}
    \lvert M_{g_1 g_2 q_3 \dots}(k_1, k_2, k_3, \dots) \rvert^2 &\xrightarrow[z_N, C_F C_A]{k_1 \parallel k_2 \parallel k_3} \left( \frac{8 \pi \asbare}{s_{123}} \right)^2 C_F^2 \left\langle \hat{P}_{N, q \to g_1 g_2 q_3}^{(\mathrm{ab})} \right\rangle \lvert M_{q \dots}(p, \dots) \rvert^2 \\
    &\hspace{-2cm}\xrightarrow{k_1 \parallel k_3, k_2 \to 0} \left( \frac{8 \pi \asbare}{s_{13}} \right) \hat{P}_{q \to g_1 q_3}(z_1, z_3) (4\pi \asbare) J_N^{(0)}(p, \tilde{k}_2) \lvert M_{q \dots}(p, \dots) \rvert^2\, ,
  \end{aligned}
\end{equation} 
which introduces a zero bin in the region $F_i = \{k_1, k_3\}$, $S = \{k_2\}$, which we need to subtract. However, by taking the strict collinear limit we again end up with a splitting kernel $\hat{P}_{g_1q_3}(z_1, z_3)$ only containing momentum fractions. This introduces new rapidity divergences, and it also leads to contributions that are not of the form anticipated on the right-hand side of  \eqref{Gen_eq_CumulantFactorizationFormulaZN}. To avoid this problem, we must reintroduce the $z_N$-prescription in the second line in \eqref{Gen_eq_CollinearSoftLimitOfPABggq}. This leads to a new zero bin in the region where both $k_1$ and $k_2$ are soft. We can summarize this as 
\begin{equation}
  \begin{aligned}
    \lvert M_{g_1 g_2 q_3\dots}(k_1, k_2, k_3, \dots)\rvert^2&\xrightarrow[z_N, C_F C_A]{k_1 \parallel k_2\parallel k_3} \left( \frac{8 \pi \asbare}{s_{123}} \right)^2C_F^2\left\langle\hat{P}_{N, q\to g_1 g_2 q_3}^{(\mathrm{ab})}\right\rangle \lvert M_{q\dots}(p, \dots)\rvert^2\\
    &\hspace{-2cm}\xrightarrow[z_N]{k_1 \parallel k_3, k_2\to 0}\left( \frac{8 \pi \asbare}{s_{13}} \right)\hat{P}_{N, q\to g_1 q_3}(z_1, z_3)(4\pi\asbare )J_N^{(0)}(p, \tilde{k}_2)\lvert M_{q\dots}(p, \dots)\rvert^2\\
    &\hspace{-2cm}\xrightarrow{k_1 \sim k_2 \to 0}(4\pi\asbare )J_{N}^{(0)}(p_i, \tilde{k}_1)(4\pi\asbare )J_{N}^{(0)}(p_i, \tilde{k}_2)\lvert M_{q\dots}(p, \dots)\rvert^2\, ,
  \end{aligned}
\end{equation} 
where the second line is a zero bin and thus will enter with a minus sign in the full result, while the third line is a zero bin of a zero bin, and thus will enter with a plus sign in the full result. We will call zero bins of zero bins \emph{double-zero bins} in what follows.

The second type of zero bins is introduced in the limit where two pairs of partons are collinear along two different jets. In this limit, the modified matrix element behaves as
\begin{equation}
  \begin{aligned}
    \lvert M_{g_1 q_3, g_2 q_4\dots}(k_1, k_3;k_2, k_4, \dots)\rvert^2&\xrightarrow[z_N]{k_1 \parallel k_3;k_2\parallel k_4} \left( 8 \pi \asbare \right)^2 \frac{\hat{P}_{Nq_i\to g_1q_3}(z_1, z_3)}{s_{13}}\frac{\hat{P}_{N, q_j\to g_2q_4}(z_2, z_4)}{s_{24}}\\&\quad\times\lvert M_{q_i, q_j\dots}(p_i, p_j, \dots)\rvert^2\, ,
  \end{aligned}
\end{equation}
where the quark jet with momentum $p_i$ splits into $k_1$ and $k_3$, while the quark jet with momentum $p_j$ splits into $k_2$ and $k_4$. $z_1$ and $z_3=1-z_1$ are momentum fractions with respect to $p_i$, while $z_2$ and $z_4=1-z_2$ are momentum fractions with respect to $p_j$. The $z_N$-prescription introduces zero bins both in the double-soft and the collinear-soft region. Again, we want to write the collinear-soft zero bins with a $z_N$-regularized splitting kernel. This introduces another double-zero bin in the double-soft region. For brevity, we no longer write out all the limits.

Finally, the third region where zero bins are introduced is the limit where one gluon is soft, and two partons become collinear. In this region, the $z_N$ modified matrix element behaves as
\begin{equation}
  \begin{aligned}
    \lvert M_{g_1 g_2 q_3\dots}(k_1, k_2, k_3, \dots)\rvert^2&\xrightarrow[z_N]{k_1 \parallel k_3, k_2\to 0} \left( \frac{8 \pi \asbare}{s_{13}} \right)\hat{P}_{N, q\to g_1 q_3}(z_1, z_3) \\
    &\times (4\pi\asbare )\braket{ M_{q\dots}(p, \dots)| \bJJ^{(0)}(\tilde{k}_2) |M_{q\dots}(p, \dots) }\, .
  \end{aligned}
\end{equation} 
This region only introduces a zero bin in the double-soft region.

We can combine all contributions living in the region $F_i=\{k_1, k_2\}$, $S=\{k\}$, including zero bins. The combined contribution reads
\begin{equation}
  \label{Gen_eq_PNSsubZN}
  \begin{aligned}
    &\left( \mu^{2\epsilon} \frac{e^{\epsilon \gamma_E}}{\Gamma(1-\epsilon)} \right)^{2}\int_0^1 \rd z_1 \int\frac{\rd^{d-2}\tilde{k}_{\perp,1}}{\Omega_{d-2}}\frac{\hat{P}_{N_i, q\to gq}(z_1)}{\lvert \tilde{k}_{\perp,1} \rvert^2}\Biggl[\\
    & \theta\!\left( \qfcut-\qft_{C_i}(\tilde{k}_1, \tilde{k}_2)\right) \int \frac{\rd^dk\delta_+(k^2)}{\Omega_{d-2}}\bJJ^{(0)}(k)\theta\!\left( \qfcut-\qft_S(k)\right)\\
    &-\theta\!\left( \qfcut-\qft_{C_i}(\tilde{k}_1, \tilde{k}_2)\right)\! \int \frac{\rd^dk\delta_+(k^2)}{\Omega_{d-2}}\sum_{\alpha\neq i}^{n_J+2}J^{(0)}_{N_\alpha}(p_\alpha, k)\theta\!\left( \qfcut-\qft_{C_\alpha, S}(k)\right)\\
    &-\int \frac{\rd^dk\delta_+(k^2)}{\Omega_{d-2}}J^{(0)}_{N_i}(p_i, k)\theta\!\left( \qfcut-\qft_{C_i, C_iS}(\{\tilde{k}_1, \tilde{k}_2\}, \{k\})\right)\Biggr]\\
    &=\hat{\J}_{N_i, q}^{(1)}\hat{\Sb}_{\mytext{sub}}^{(1)}\,.
  \end{aligned}
\end{equation}
 We allowed for the possibility to use different time-like reference vectors $N_\alpha$ for each collinear sector. In \eqref{Gen_eq_PNSsubZN}, the second line comes from the original region $F_i=\{k_1, k_2\}$, $S=\{k\}$ after modifying the matrix element, the third line comes from zero bins of the double-collinear regions where $F_i=\{k_1, k_2\}$, $F_j=\{k, k_3\}$, $S=\emptyset$, and the fourth line is a zero bin of the triple collinear region where $F_i=\{k_1, k_2, k\}$. To get to the last line, we assumed that $\qft_{C_i, C_iS}$---the approximation of the resolution variable where one first takes the triple-collinear and then the collinear-soft limit---satisfies\footnote{Assuming that the various limits commute, this is a straightforward consequence of the factorization property \eqref{Gen_eq_MaximumFactorization}. Then we could write 
\begin{equation}
  \begin{aligned}
    \qft_{C_i, C_iS}&=\qft_{C_iS, C_i}=\max\left( \qft_{C_i}(\tilde{k}_1, \tilde{k}_2), \qft_{S}(k)\right)\bigg\rvert_{C_i}\\
    &=\max\left( \qft_{C_i}(\tilde{k}_1, \tilde{k}_2), \qft_{S, C_i}(k)\right)=\max\left( \qft_{C_i}(\tilde{k}_1, \tilde{k}_2), \qft_{C_i, S}(k)\right)\, .
  \end{aligned}
\end{equation}
}
\begin{equation}
  \qft_{C_i, C_iS}(\{\tilde{k}_1, \tilde{k}_2\}, \{k\})=\max\left( \qft_{C_i}(\tilde{k}_1, \tilde{k}_2), \qft_{C_i, S}(k)\right)\, .
\end{equation}

Based on the calculation in \eqref{Gen_eq_PNSsubZN}, we have demonstrated that the assumption used in equation \eqref{Gen_eq_CumulantFactorizationFormulaZN}—--that the $z_N$-prescription does not spoil the cumulant factorization—--is reasonable for a large set of variables, at least up to NNLO. We are not sure whether it is even possible for the $z_N$-prescription to spoil cumulant space factorization for a resolution variable satisfying \eqref{Gen_eq_MaximumFactorization}.

Finally, we can combine the double-soft region and all zero and double-zero bins to write the contribution due to Abelian emissions of two soft gluons to the two-loop soft function. We find 
\begin{equation}
  \label{Gen_eq_SsubZngg_Abelian}
  \begin{aligned}
    \hat{\Sb}^{(2), (\mathrm{ab})}_{gg, \mytext{sub}}=&\mu^{4\epsilon} \left( \frac{e^{\epsilon \gamma_E}}{\Gamma(1-\epsilon)} \right)^2\frac{1}{2}\int \frac{\rd^dk_1\delta_+(k_1^2)}{\Omega_{d-2}}\int \frac{\rd^dk_2\delta_+(k_2^2)}{\Omega_{d-2}}\Bigg[\bJJ_{gg}^{(0), (\mathrm{ ab})}(k_1, k_2)\Theta_{S}\\
    &\quad-\sum_{\alpha}J_{N_\alpha}^{(0)}(p_\alpha, k_1)J_{N_\alpha}^{(0)}(p_\alpha, k_2)\left( \Theta_{C_\alpha , S} -\Theta_{C_\alpha, S C_\alpha, S}-\Theta_{C_\alpha, C_\alpha S, S}\right)\\
    &-\sum_{\alpha\neq \beta}J_{N_\alpha}^{(0)}(p_\alpha, k_1)J_{N_\beta}^{(0)}(p_\beta, k_2)\left( \Theta_{C_\alpha C_\beta, S}- \Theta_{C_\alpha C_\beta, C_\alpha S, S}- \Theta_{C_\alpha C_\beta, S C_\beta, S}\right) \\
    &-\sum_\alpha \left( J_{N_\alpha}^{(0)}(p_\alpha, k_1)\bJJ^{(0)}(k_2)\Theta_{C_\alpha S, S}  +\left( k_1\leftrightarrow k_2 \right) \right) \Bigg]\, ,
  \end{aligned}
\end{equation}
where we used a shorthand notation for the $\theta$-functions, for instance 
\begin{equation}
  \Theta_{C_\alpha C_\beta, S C_\beta, S}=\theta\!\left( \qfcut-\qft_{C_\alpha C_\beta, S C_\beta, S}\right)\, ,
\end{equation}
where $\qft_{C_\alpha C_\beta, S C_\beta, S}$ is the approximation of $\qf$ where one first expands in the double-collinear limit, where $k_1$ is collinear along the jet $p_\alpha$ and $k_2$ is collinear along the jet $p_\beta$, then in the soft-collinear limit, where $k_1$ is soft and $k_2$ is still collinear along the jet $p_\beta$, and finally in the double-soft limit (c.f. Appendix~\ref{sec_ShorthandNotationRegions}). The $\Theta_{S}$ term in \eqref{Gen_eq_SsubZngg_Abelian} stems from the double-soft limit, while $\theta$-functions with one ``,'' stem from zero bins, and $\Theta$-functions with two ``,'' stem from double-zero bins. In \eqref{Gen_eq_SsubZngg_Abelian}, we allowed for the choice of different reference vectors $N_\alpha$ in each sector. If we only have two colored hard legs, we choose the reference vectors $N_\alpha$ equal, and additionally, all $\theta$-functions agree with $\Theta_{S}$, then $\hat{\Sb}^{(2), (\mathrm{ab})}_{gg, \mytext{sub}}=0$ in accordance with our guiding principle for constructing the $z_N$-prescription. 

Another simplification in $\hat{\Sb}^{(2), (\mathrm{ab})}_{gg, \mytext{sub}}$ occurs if the variable respects the \emph{non-Abelian exponentiation theorem}~\cite{Gatheral:1983cz, Frenkel:1984pz}. This is the case if each $\theta$-function in \eqref{Gen_eq_SsubZngg_Abelian} can be written as a product of two $\theta$-functions that only depend on $k_1$ and $k_2$, respectively. To be precise, if the variable satisfies
\begin{equation}
  \begin{aligned}
    \Theta_{S}&=\Theta_{S}(k_1)\Theta_{S}(k_2)\\
    \Theta_{C_\alpha , S} &=\Theta_{C_\alpha, S C_\alpha, S}=\Theta_{C_\alpha, C_\alpha S, S}=\Theta_{C_\alpha, S}(k_1)\Theta_{C_\alpha, S}(k_2)\\
    \Theta_{C_\alpha C_\beta, S}&= \Theta_{C_\alpha C_\beta, C_\alpha S, S}= \Theta_{C_\alpha C_\beta, S C_\beta, S}=\Theta_{C_\alpha, S}(k_1)\Theta_{C_\beta, S}(k_2)\\
    \Theta_{C_\alpha S, S}&=\Theta_{C_\alpha, S}(k_1)\Theta_{S}(k_2)  \quad\Theta_{SC_\alpha, S}=\Theta_{C_\alpha, S}(k_2)\Theta_{S}(k_1)\, ,
  \end{aligned}
\end{equation}
then also the soft subtracted function factorizes as 
\begin{equation}
  \hat{\Sb}^{(2), (\mathrm{ab})}_{gg, \mytext{sub}}=\frac{1}{2}\left( \hat{\Sb}^{(1)}_{\mytext{sub}} \right)^2\, .
\end{equation}
We do not know whether the $z_N$-prescription can spoil non-Abelian exponentiation for a resolution variable satisfying $ \Theta_{S}=\Theta_{S}(k_1)\Theta_{S}(k_2)$.

The full two-loop subtracted function is given by
\begin{equation}
  \hat{\Sb}^{(2)}_{\mytext{sub}}=\hat{\Sb}^{(2)}_{g, \mytext{sub}}+n_f \hat{\Sb}^{(2)}_{q\bar{q}, \mytext{sub}}+\hat{\Sb}^{(2), (\mathrm{ab})}_{gg, \mytext{sub}}+\hat{\Sb}^{(2), (\mathrm{nab})}_{gg, \mytext{sub}}\, .
\end{equation}

We point out that even though we have not yet defined the $z_N$ modified initial-state collinear splitting kernels, nor have we defined the $z_N$-modified gluon splitting kernels for triple-collinear splittings, the soft subtracted function formulas \eqref{Gen_eq_SsubZnSingleGluonRV}, \eqref{Gen_eq_SsubZnqqbar}, \eqref{Gen_eq_SsubZngg_nonAbelian}, \eqref{Gen_eq_PNSsubZN} and \eqref{Gen_eq_SsubZngg_Abelian} are already fully general. This means that the $z_N$-prescription for the beam functions and the gluon jet functions can be defined in such a way that these results are reproduced. This will be the subject of future work.

\section{\texorpdfstring{\boldmath The Case of ${\ktness}$ \unboldmath}{The Case of ktness}}
\label{chap_FactorizationFormulaForKtNess}
This section applies the techniques introduced in Sections~\ref{Gen_sec_FactorizationForResolutionVariables} and \ref{Gen_sec_RapidityDivergences} to the $\ktness$ resolution variable. Section~\ref{sec_ktness_definition} reviews the definition of $\ktness$, while Section~\ref{sec_ktness_approximations} examines how its different variants behave in various limits and the resulting factorization formulas.

\subsection{\boldmath The \texorpdfstring{$\ktness$}{kTness} Variable\unboldmath}
\label{sec_ktness_definition}

The resolution variable $n$-$\ktness$ is an $n-$jet resolution variable that is recursively IR-safe and continuously global. A definition of $\ktness$ was first provided in Ref.~\cite{Buonocore:2022mle}. Here, we give a slightly more general definition, and we will point out some choices that can be made when defining the resolution variable. 

The definition of $n$-$\ktness$ on a phase space configuration with $n+k$ partons is based on running an exclusive $k_t$-type jet clustering algorithm~\cite{Catani:1993hr, Ellis:1993tq} down to the stage where only $n+1$ pseudojets remain. We define the distance between two pseudojets and the distance between a pseudojet and the beam as
\begin{equation}
  \label{eq_ktn_distance}
  \begin{aligned}
    d_{ij}&\equiv M_{ij}\frac{R_{ij}}{D}\\
    d_{iB}&\equiv k_{Ti}\, ,
  \end{aligned}
\end{equation}
where $d_{ij}$ is the distance between pseudojets $i$ and $j$, $d_{iB}$ is the distance between the pseudojet $i$ and the beam, and we used the notation $k_{Ti}=\lvert k_{ti} \rvert$. We have split the distance $d_{ij}$ into a product of a term $M_{ij}$, which has the dimensions of a mass, and a dimensionless \emph{angular separation} $R_{ij}$. $D$ is a dimensionless parameter that can be used to adjust the size of the jets. Possible choices for $M_{ij}$ and $R_{ij}$ are  
\begin{equation}
  \label{eq_ktn_distances}
\begin{aligned}
  M_{ij}&=\min(k_{Ti}, k_{Tj}) \quad M_{ij}=\frac{k_{Ti}k_{Tj}}{k_{Ti} + k_{Tj}}\quad M_{ij}=\frac{k_{Ti}k_{Tj}}{\lvert k_{ti} + k_{tj}\rvert}\\
  R_{ij}^2&=\Delta y_{ij}^2+\Delta \phi_{ij}^2\quad R_{ij}^2=2 (\cosh\Delta y_{ij}-\cos\Delta \phi_{ij}) \quad R_{ij}^2=-\left( \frac{k_i}{k_{iT}}-\frac{k_j}{k_{jT}} \right)^2\, .
\end{aligned}  
\end{equation}
For lepton collider processes, we instead use 
\begin{equation}
  \label{eq_ktn_distances_epem}
  \begin{aligned}
    M_{ij}&=\min(E_i, E_j) \quad M_{ij}=\frac{E_i E_j}{ E_i + E_j}\\
  R_{ij}^2&=2(1-\cos\theta_{ij})\, ,
  \end{aligned}
\end{equation}
and one does not need to consider the distances with the beams $d_{iB}$ when performing the clustering; this also means that $D=1$ can always be assumed for lepton colliders without loss of generality.

To define $n$-$\ktness$ for a configuration with $n+k$ final-state massless QCD partons, we employ the following recursive procedure:
\begin{enumerate}[start=0]
  \item  If $k=0$, set $\ktness=0$ and stop. Otherwise, initialize the \emph{recoil momentum} as $k_{\mathrm{{rec}}}=0$. Add all massless QCD partons to the list of pseudojets.
  \item  Compute the minimum among all distances $\{d_{ij}\}$ and $\{d_{iB}\}$. If
  there are at least $ n+2$ pseudojets, continue with step 2. If only $n+1$ pseudojets remain, proceed to step 3.
  \item If the minimum is $d_{iB}$, remove pseudojet $i$ from the list and update the recoil momentum as $k_{\mathrm{{rec }}} \rightarrow \sigma k_{\mathrm{{rec }}}+k_i$. If instead the minimum is $d_{ij}$, remove pseudojets $i$ and $j$ and replace them by a new pseudojet with momentum $k_{ij}$. Return to step 1.
  \item If the minimum is $d_{iB}$, update the recoil momentum as $k_{\mathrm{rec}}\to \sigma k_{\mathrm{rec}}+k_i$ and set $\ktness=k_{T,\mathrm{rec}}$. If instead the minimum is $d_{ij}$, set $\ktness=d_{ij}$.
\end{enumerate}
Besides the choice of distance measure $d_{ij}$, two further ingredients enter the definition of $\ktness$. The first is the recombination scheme used in step 2. The default, also adopted in~\cite{Buonocore:2022mle}, is the $E$-scheme,
\begin{equation}
  k_{ij} = k_i + k_j \, .
\end{equation}
As an alternative, we also consider the winner-take-all (WTA) scheme~\cite{Bertolini:2013iqa},\footnote{Note, that we define $k_{T ij}$ by the magnitude of the transverse momentum vector $k_{t i}+k_{t j}$, which differs from the original definition in~\cite{Bertolini:2013iqa}, where $k_{T ij}$ is defined as the sum of the magnitudes $k_{T i}+k_{T j}$.}
\begin{equation}
  \label{eq_wta-scheme}
  k_{ij}=k_{T ij}\hat{n}_{ij}, \quad k_{T ij}=\lvert k_{t i}+k_{t j}\rvert, \quad \hat{n}_{ij}= \begin{cases}\frac{k_i}{k_{T i}} & \mathrm { if }\, k_{T i}>k_{T j} \\ \frac{k_j}{k_{T j}} & \mathrm { if }\, k_{T j}>k_{T i}\, .\end{cases}
\end{equation}
At lepton colliders, one instead defines the WTA scheme exactly as in \eqref{eq_wta-scheme} but with $E_i$ and $E_j$ instead of $k_{Ti}$ and $k_{Tj}$. The WTA scheme has the advantage that it leaves jets massless. We will also see that the WTA scheme simplifies the construction of a factorization formula for $\ktness$ because the jet directions are not influenced by soft recoil. The second choice is how to treat the recoil. The recoil can either be summed over iterations, or one can instead drop the recoil from earlier steps. For the first choice, we would set $\sigma=1$ in steps two and three, while for the second choice, we would set $\sigma=0$. The advantage of the first choice is that initial state collinear radiation is treated the same as in $q_T$-resummation. One can therefore reuse the well-known beam functions of $q_T$-resummation to a certain degree. The advantage of the second choice is that the factorization formula looks simpler. 

We regard $\ktness$ as a natural generalization of $q_T$. It scales like a transverse momentum whenever radiated partons are soft and collinear to a beam or a jet. $\ktness$ is not the first $n$-jet resolution variable with this property. In~\cite{Bertolini:2017efs}, several generalizations of $n$-jettiness were introduced, some of which also scale like transverse momenta in all soft-collinear limits. To our knowledge, however, $\ktness$ is the first such variable that has been explicitly developed and tested for generic $n$-jet processes at NLO. A detailed study of its application as a slicing variable to compute NNLO corrections to dijet production at lepton colliders and Higgs decays to a (massless) $b\bar{b}$-pair is in preparation~\cite{inprep2}.

\subsection{\texorpdfstring{Factorization Formulas for $\ktness$}{Factorization Formulas for ktness}}\label{sec_ktness_approximations}
The definition of $\ktness$ is based on the inter-particle distances $d_{ij}$ defined in \eqref{eq_ktn_distance}. Thus, we need to understand in detail how $d_{ij}$ approximates when the pseudojet momenta $p_i$ and $p_j$ are combinations of momenta from the sectors $I_1, I_2, S, F_3, \dots $. We already showed in \secref{Gen_sec_PhaseSpaceFactorization} how one can approximate the distances $d_{ij}$ for $y_{n(n+1)}$, which is a version of $n$-$\ktness$ at lepton colliders, and we found that the appropriate power counting parameters are $a=b=c=1$. We will extend the results of \secref{Gen_sec_PhaseSpaceFactorization} to include more general distances, and we will also include initial-state collinear particles in our analysis. The approximations of the distances for the case where the particles have been combined with the $E-$scheme will be presented, accompanied by an explanation of the changes required in the WTA scheme.

\begin{table}[b]
  
  \centering
  \begin{tabular}{c|ccccc}
    $d_{ij}$& $I_1$ & $I_2$& $F_k$ & $F_l$ & $S$\\
    \hline
    $I_1$ & $d_{ij}$ & $\sim Q$ & $\sim Q$ & $\sim Q$& $\sim Q$\\    
    $I_2$ & $\sim Q$ & $d_{ij} $ & $\sim Q$& $\sim Q$& $\sim Q$\\
    $F_k$ & $\sim Q$ & $\sim Q$ & $d_{c, ij}$ & $\sim Q$ & $d_{s, ij}$\\
    $F_l$ & $\sim Q$ & $\sim Q$ & $\sim Q$ & $d_{c, ij}$ & $d_{s, ij}$\\
    $S$ & $\sim Q$ & $\sim Q$ & $d_{s, ji}$ & $d_{s, ji}$ & $d_{ij}$  \\
  \end{tabular}
  \caption{Approximation of the inter-particle distances $d_{ij}$ where the pseudojets $i$ and $j$ come from any two sectors.}
  \label{Fac:tab_ktn-scalings}

\end{table}

The first observation when considering $\ktness$ for hadron colliders is that an initial-state collinear particle, e.g., $k_i\in I_1$, can never cluster with a particle from another sector. The reason is that the rapidity of $k_i$ is of order $-\log(\lambda)$ while the rapidities of final-state collinear and soft partons are of order one, and the rapidities of particles in $I_2$ are even of order $+\log(\lambda)$. Therefore, $k_i$ will always cluster with the beam or with another particle in $I_1$ before it could cluster with a particle in another sector. A second observation is that distances between partons in different final-state collinear sectors are of order $Q$; therefore, they never cluster. The only remaining distances of order $\lambda Q$ are those between particles in the same sector, those between final-state collinear and soft particles, and beam distances for soft or initial-state collinear particles. These findings are summarized in Table~\ref{Fac:tab_ktn-scalings}. In the table, we have also indicated the distances for which no approximation can be made. For instance, the distance between two soft particles cannot be further approximated. The table also contains the approximated distances $d_{c, ij}$ and $d_{s, ij}$, which we now define. For the pseudojets in the collinear sector, say $F_k$, we also allow the possibility that they have previously been clustered with soft particles. We thus write $k_i = k^c_i + k^s_i$, where $k^c_i$ is the momentum sum of the collinear constituents from sector $F_k$ and $k^s_i$ is the momentum sum of all soft constituents from sector $S$. One can then write
\begin{equation}
  \label{Fac:eq_dijc}
  \begin{aligned}
    d_{c, ij}^2 &= -\frac{M_{ij}^2(z_i, z_j)}{D^2} \left( \frac{k^c_{i, \perp} + k^s_{i, \perp}}{z^c_i} - \frac{k^c_{j, \perp} + k^s_{j, \perp}}{z^c_j} \right)^2\\
    &\sim -\frac{M_{ij}^2(\tilde{z}_i, \tilde{z}_j)}{D^2} \left( \frac{\tilde{k}^c_{i, \perp} + k^s_{i, \perp}}{\tilde{z}^c_i} - \frac{\tilde{k}^c_{j, \perp} + k^s_{j, \perp}}{\tilde{z}^c_j} \right)^2,
  \end{aligned}
\end{equation}
where we indicate that the approximation is the same, at leading power in $\lambda$, whether one uses the boosted and rescaled momenta $\tilde{k}^c_i$, which are centered around the jet direction, or the true momenta $k^c_i$. Note that $k_{i(j)}^s$ has no boosted and rescaled counterpart because it lives in the soft sector. In \eqref{Fac:eq_dijc}, we introduced the shorthand notation 
\begin{equation}
  \label{eq_MijofzDefinition}
  M_{ij}(z_i, z_j)\equiv \begin{cases}
    \frac{M_{ij}(z_i p_k, z_j p_k)}{p_{Tk}} & \text{ for hadron colliders}\\
    \frac{M_{ij}(z_i p_k, z_j p_k)}{p_k^0} & \text{ for lepton colliders}
  \end{cases}\, ,
\end{equation} where $p_k$ is the momentum of the Born-level parent associated with the collinear sector $F_k$ (c.f. \eqref{Momentum_With_Homogeneous_Scaling}). The expression for $M_{ij}(z_i, z_j)$ for hadron (lepton) colliders can also be obtained from Eqs.~\eqref{eq_ktn_distances} and \eqref{eq_ktn_distances_epem} by replacing all transverse momenta (energies) with momentum fractions. We point out that the transverse momenta and momentum fractions entering $d_{c, ij}$ are defined with respect to the corresponding reference vectors, say $n_k$ and $\bar{n}_k$ in sector $F_k$. 

Note that in the WTA recombination scheme, the recoil due to soft radiation can be neglected in $d_{c, ij}$, meaning that we can calculate the distance between collinear particles in sector $F_k$ without information about the soft sector. However, $d_{c, ij}$ also complicates matters slightly because the total transverse momentum $k_{i(j), \perp}^c$ in \eqref{Fac:eq_dijc} needs to be replaced with the respective combined momentum obtained with the WTA scheme. We denote the momentum obtained from clustering $k_i$ and $k_j$ in any recombination scheme by $k_{ij}$. In the WTA scheme, if $k_i$ and $k_j$ are collinear particles, we get
\begin{equation}
  z_{ij} = z_i + z_j\, , \quad  k_{\perp ij} = \begin{cases}
    \frac{z_i + z_j}{z_i} k_{\perp i} & \text{ if } z_i > z_j\\
    \frac{z_i + z_j}{z_j} k_{\perp j} & \text{ else}\, ,
  \end{cases}
\end{equation}
 while if $k_i$ is in a final-state collinear sector and $k_j$ is in the soft sector, we find
\begin{equation}
  z_{ij} = z_i\, , \quad  k_{\perp ij} = k_{\perp i}
\end{equation}
to leading power. If both $k_i$ and $k_j$ are in the soft sector or the same IS collinear sector, one cannot make any approximation when combining the two particles in the WTA scheme.

The leading-power expansion for the distance between a soft pseudojet $k_i$ and a collinear particle $k_j$ in the sector $F_k$ reads
\begin{equation}
  \label{Fac:eq_dijs}
  d_{s, ij}=\lvert k_{ti}\rvert \frac{R(k_i, p_k)}{D}\, ,
\end{equation}
where $R(k_i, p_k)$ is the angular distance between the soft particle $k_i$ and the hard parent momentum $p_k$. For lepton colliders, one replaces $k_{ti}\to E_i$ in \eqref{Fac:eq_dijs}. 

We will now go through all possible regions that appear in NNLO calculations with $n$-$\ktness$ slicing, and we will analyze whether $\ktness$ factorizes as described in \eqref{Gen_eq_MaximumFactorization} for different definitions, and if not, which terms spoil this factorization. In the following, it can be helpful to reference Appendix~\ref{sec_LimitsOfKinematicFunctions} for technical details on how to take the limits and Appendix~\ref{sec_ShorthandNotationRegions} for the shorthand notation we use for the various regions. 

\paragraph{Two initial-state collinear emissions in opposite sectors: $I_1=\{k_1\}, I_2=\{k_2\}, \tilde{F}_i=\{p_i\}, S=\emptyset$}
Since the momenta $k_1$ and $k_2$ are in opposite initial-state collinear sectors, they cannot cluster with each other and the following formulas are independent of the recombination scheme. For the standard recoil definition of $\ktness$ ($\sigma=1$) one finds 
\begin{equation}
  \label{Fac:eq_qftI1I2}
  \qft_{\mathfrak{p}_{n_J+2}}(\{k\}) \equiv\qft_{C_1C_2}(\{k_1\},\{k_2\})= \lvert q_T \rvert = \lvert k_{1, t}+k_{2, t} \rvert
\end{equation}
while for the definition without recoil ($\sigma=0$), we would find 
\begin{equation}
  \label{Fac:eq_qftI1I1_norec}
  \qft_{C_1C_2}(\{k_1\},\{k_2\}) = \max(\qft_{C_1}(k_1), \qft_{C_2}(k_2))=\max(\lvert k_{1, t}\rvert, \lvert k_{2, t}\rvert)\, .
\end{equation}
For the recoil definition of $\ktness$, it would thus make sense to define the beam functions in impact parameter space. In contrast, for the no-recoil definition, one would define the beam functions directly in cumulant space. Note that $\qft$ also has homogeneous scaling in the soft sectors. Thus, the approximation of $\qf$ going into the zero bins introduced by the $z_N$-prescription will have the same form, i.e.,
\begin{equation}
  \begin{aligned}
    \qft_{C_1C_2}(\{k_1\},\{k_2\})&=\qft_{C_1C_2,C_1S}(\{k_1\},\{k_2\})=\qft_{C_1C_2,SC_2}(\{k_1\},\{k_2\})=\qft_{C_1C_2,S}(k_1,k_2)\\
    &=\qft_{C_1C_2,C_1S,S}(k_1,k_2)=\qft_{C_1C_2,SC_2,S}(k_1,k_2)\,.
  \end{aligned}
\end{equation}
\paragraph{Two initial-state collinear emissions in the same sector: $I_1=\{k_1, k_2\}, I_2=\emptyset, \tilde{F}_i=\{p_i\}, S=\emptyset$}

For the $\sigma=1$ definitions of $\ktness$, we find
\begin{equation}
  \label{Fac:eq_qftI1I1}
  \qft_{\mathfrak{p}_{n_J+2}}(\{k\}) \equiv \qft_{C_1}(k_1,k_2)= \lvert q_T \rvert = \lvert k_{1, t}+k_{2, t} \rvert\,,
\end{equation}
while for the definitions with $\sigma=0$, one needs to perform the clustering. The clustering distance cannot be approximated. We find
\begin{equation}
  \label{Fac:eq_qftI1I1_withrec}
  \qft_{C_1}(k_1,k_2)= \max \left( \lvert k_{1, t}\rvert, \lvert k_{2, t}\rvert \right) + \theta\!\left( R_{12}< D \right)\left[ \lvert k_{1, t}+k_{2, t} \rvert - \max \left( \lvert k_{1, t}\rvert, \lvert k_{2, t}\rvert \right)\right]\, .
\end{equation}
 In the case where we do not consider the beam recoil, $\qft$ is not homogeneous in all zero bins, and one needs to perform the expansions
\begin{equation}\label{Fac:eq_qftI1I1ZeroBins_norec}
  \begin{aligned}
    \qft_{C_1, S}(k_1, k_2)&=\qft_{C_1}(k_1, k_2)\\
    \qft_{C_1, C_1S}(\{k_1\}, \{k_2\})&=\qft_{C_1, SC_1}(\{k_1\}, \{k_2\})=\max \left( \lvert k_{1, t}\rvert, \lvert k_{2, t}\rvert \right)
  \end{aligned}
\end{equation}
in the zero bins and 
\begin{equation}
  \label{Fac:eq_qftI1I1DoubleZeroBins_norec}
    \qft_{C_1, C_1S, S}(k_1, k_2)=\qft_{C_1, SC_1, S}(k_1, k_2)=\max \left( \lvert k_{1, t}\rvert, \lvert k_{2, t}\rvert \right)
\end{equation}
in the double-zero bins. For the collinear-soft zero bins, we again used the fact that $R_{12}\sim -\log(\lambda)\gg D$ if one of the two particles is collinear and the other is soft.
\paragraph{One initial-state collinear and one final-state collinear emission: $I_1=\{k_1\}, I_2=\emptyset, \dots, F_i=\{k_2, k_3\}, \dots, S=\emptyset$} 
Initial-state collinear and final-state collinear emissions cannot cluster together, thus $\qft$ simplifies to
\begin{equation}
  \label{Fac:eq_qftI1Fi}
  \begin{aligned}
    \qft_{\mathfrak{p}_{n_J+2}}(\{k\})&=\qft_{C_1C_i}(\{k_1\}, \{\tilde{k}_2, \tilde{k}_3\})=\max\left(\lvert k_{t, 1}\rvert, d_{c, 23}   \right)=\max\left( \qft_{C_1}(k_1),\qft_{C_i}\left( \tilde{k}_2,\tilde{k}_3 \right) \right)
  \end{aligned}
\end{equation}
for all $\ktness$ definitions introduced in \secref{sec_ktness_definition}. 
The zero bins are 
\begin{equation}
  \label{Fac:eq_qftI1FiZeroBins}
  \begin{aligned}
   \qft_{C_1C_i, SC_i}(\{k_1\}, \{\tilde{k}_2, \tilde{k}_3\})&=\qft_{C_1C_i}(\{k_1\}, \{\tilde{k}_2, \tilde{k}_3\})=\max\left(\lvert k_{t, 1}\rvert, d_{c, 23}   \right)\\
   \qft_{C_1C_i, C_1S}(\{k_1\}, \{k_3\})&=\max\left(\lvert k_{t, 1}\rvert, \frac{\lvert k_{\perp, 3}\rvert}{D}   \right)
   \, ,
  \end{aligned}
\end{equation}
where $k_{\perp, 3}$ is defined with respect to the reference vectors $n_i$ and $\bar{n}_i$.
For the double-zero bins, one gets 
\begin{equation}
  \label{Fac:eq_qftI1FiZeroZeroBins}
  \begin{aligned}
    \qft_{C_1C_i, C_1S, S}(k_1, k_3)= \qft_{C_1C_i, SC_i, S}(k_1, k_3)=\max\left(\lvert k_{t, 1}\rvert, \frac{\lvert k_{\perp, 3}\rvert}{D}   \right)\, .
  \end{aligned}
\end{equation}
\paragraph{Two final-state collinear emissions in two different sectors: $I_1=\emptyset, I_2=\emptyset, \dots, F_i=\{k_1, k_2\}, \dots, F_j=\{k_3, k_4\}, \dots, S=\emptyset$} 
In all $\ktness$ definitions introduced in \secref{sec_ktness_definition} $\ktness$ simplifies to 
\begin{equation}
  \qft_{\mathfrak{p}_{n_J+2}}(\{k\})=\qft_{C_iC_j}(\{\tilde{k}_1, \tilde{k}_2\}, \{\tilde{k}_3, \tilde{k}_4\})=\max\left( d_{c, 12}, d_{c, 34} \right)\, ,
\end{equation}
where we again point out that the transverse momenta and momentum fractions in $d_{c, 12}$ and $d_{c, 34}$ have to be understood with respect to the corresponding reference vectors $n_i$ and $\bar{n}_i$ in sector $F_i$, and $n_j$ and $\bar{n}_j$ in sector $F_j$.
For the zero bins, one finds
\begin{equation}
  \begin{aligned}
    \qft_{C_iC_j, C_iS}(\{\tilde{k}_1, \tilde{k}_2\},\{ k_4\})&=\max\left( d_{c, 12}, \frac{\lvert k_{4, \perp}\rvert}{D}  \right)\\
    \qft_{C_iC_j, SC_j}(\{k_2\}, \{\tilde{k}_3, \tilde{k}_4\})&=\max\left( \frac{\lvert k_{2, \perp}\rvert}{D}, d_{c, 34}  \right)\\
    \qft_{C_iC_j, S}(k_2, k_4)&=\max\left( \frac{\lvert k_{2, \perp}\rvert}{D}, \frac{\lvert k_{4, \perp}\rvert}{D} \right)\, ,
  \end{aligned}
\end{equation}
where we assumed that the particles that go soft carry the momenta $k_2$ and $k_4$, respectively. Again, the transverse momenta have to be understood with respect to the corresponding reference vectors.

The approximations for the double-zero bins are also given as 
\begin{equation}
  \label{Fac:eq_qftFiFjZeroZeroBins}
  \qft_{C_iC_j, SC_j, S}(k_2, k_4)=\qft_{C_iC_j, C_iS, S}(k_2, k_4)=\max\left( \frac{\lvert k_{2, \perp}\rvert}{D}, \frac{\lvert k_{4, \perp }\rvert}{D} \right)\, .
\end{equation}

\paragraph{Triple-collinear limit: $I_1=\emptyset, I_2=\emptyset, \dots, F_i=\{k_1, k_2, k_3\}, \dots, S=\emptyset$}
In the triple-collinear region, $\ktness$ approximates as 
\begin{equation}
  \label{Fac:eq_qftCCC}
  \qft_{\mathfrak{p}_{n_J+2}}(\{k\})=\qft_{C_i}(\{\tilde{k}_1, \tilde{k}_2, \tilde{k}_3\})=\theta\!\left( \min\left( d_{c, 13}, d_{c, 23} \right)-d_{c, 12} \right)d_{c, (1+2)3}+(\mathrm{perm.})\, ,
\end{equation}
where, on the right-hand side, one also needs to sum over the permutations of the momenta and $d_{c, (1+2)3}$ is the collinearly approximated distance between the pseudojet $k_3$ and the pseudojet obtained by combining $k_1$ with $k_2$ in the chosen recombination scheme. The approximations needed for the collinear-soft zero bin are
\begin{equation}
  \label{Fac:eq_qftCCC_zerobinCS}
  \begin{aligned}
    \qft_{\tilde{C}_i, \tilde{C}_iS}(\{\tilde{k}_1, \tilde{k}_2\}, \{k_3\})&=\theta\!\left(\frac{ \lvert k_{3, \perp}\rvert}{D}-d_{c, 12} \right)\frac{\lvert k_{3, \perp}\rvert}{D}\\
    &+\theta\!\left( d_{c, 12} -\frac{\lvert k_{3, \perp}\rvert}{D}\right)\left[ \theta(\tilde{d}_{c, 13}-\tilde{d}_{c, 23} )d_{c, (2+3)1} +\left( 1\leftrightarrow 2 \right)\right]\\
    &=\max \left(d_{c, 12}, \frac{\lvert k_{3, \perp}\rvert}{D}  \right)\\
    &+\theta\!\left( d_{c, 12} -\frac{\lvert k_{3, \perp}\rvert}{D}\right)\left[ \theta(\tilde{d}_{c, 13}-\tilde{d}_{c, 23} )\left( d_{c, (2+3)1} -d_{c, 12} \right)+\left( 1\leftrightarrow 2 \right)\right]\, ,
  \end{aligned}
\end{equation}
where now $d_{c, (1+3)2}$ has to be understood as in \eqref{Fac:eq_dijc} with $k_{i, \perp}^s\to k_{3, \perp}$. Note that the term in the last line spoils the maximum factorization, and it would vanish in the WTA scheme, where $d_{c, (2+3)1} -d_{c, 12}=0$. The tildes in $\theta(\tilde{d}_{c, 13}-\tilde{d}_{c, 23} )$ and the corresponding $1\leftrightarrow2$ term indicate that, since both $d_{c,13}$ and $d_{c,23}$ expand to $\tfrac{k_{\perp,3}}{D}$ in the collinear-soft limit, the distances must be expanded to next order. The corresponding expansion depends on the exact definition of the distance $d_{ij}$, and they are generally quite subtle.  We give explicit formulas for $\tilde{d}_{c, ij}$ for the case of $y_{23}$ in Appendix~\ref{App_Approximationsy23}. A future publication will provide the formulas for $\ktness$ with hadron-collider distances.

The approximation for the soft-soft zero bin reads 
\begin{equation}
  \label{Fac:eq_qftCiCiCi_zerobinSS}
  \begin{aligned}
    \qft_{C_i, S}(k_2, k_3)&=\theta(\frac{1}{D}\min(\lvert k_{2, \perp}\rvert, \lvert k_{3\perp}\rvert)-d_{c, 23})\frac{\lvert k_{\perp, 23}  \rvert}{D}\\
    &+\theta(d_{c, 23}-\frac{1}{D}\min(\lvert k_{2, \perp}\rvert, \lvert k_{3\perp}\rvert)) \frac{1}{D}\max(\lvert k_{2, \perp}\rvert, \lvert k_{3\perp}\rvert)\, ,
  \end{aligned}
\end{equation}
from which we already see that the Abelian part of the $\ktness$ soft subtracted function will not satisfy non-Abelian exponentiation due to the region of phase space where the two emissions cluster together. The approximation for the double-zero bin is 
\begin{equation}
  \label{Fac:eq_qftCiCiCi_zerozerobins}
  \begin{aligned}
    \qft_{C_i, C_iS, S}(k_2, k_3)&= \frac{1}{D}\max(\lvert k_{2, \perp}\rvert, \lvert k_{3\perp}\rvert)\\
    &+\theta\!\left( \lvert k_{2, \perp}\rvert- \lvert k_{3, \perp} \rvert\right)  \theta(\tilde{\tilde{d}}_{c, 13}-\tilde{\tilde{d}}_{c, 23} )\frac{1}{D} \left( \lvert k_{23, \perp}\rvert - \lvert k_{2, \perp}\rvert \right)\\
    \qft_{C_i, SC_i, S}(k_2, k_3)&= \frac{1}{D}\max(\lvert k_{2, \perp}\rvert, \lvert k_{3\perp}\rvert)\\
    &+\theta\!\left( \lvert k_{3, \perp}\rvert- \lvert k_{2, \perp} \rvert\right)  \theta(\tilde{\tilde{d}}_{c, 12}-\tilde{\tilde{d}}_{c, 23} )\frac{1}{D} \left( \lvert k_{23, \perp}\rvert - \lvert k_{3, \perp}\rvert \right)\, ,
  \end{aligned}
\end{equation}
where the second and fourth lines would vanish in the WTA scheme. Here we put double tildes in $\theta(\tilde{\tilde{d}}_{c, 13}-\tilde{\tilde{d}}_{c, 23} )$ and the corresponding terms with $(2\leftrightarrow 3)$. It indicates that one needs to expand $\theta(\tilde{d}_{c, 13}-\tilde{d}_{c, 23} )$ in the double-soft limit while keeping the leading, non-vanishing term. We give explicit expressions for $\tilde{\tilde{d}}_{c, ij}$ for the case of $y_{23}$ in Appendix~\ref{App_Approximationsy23}.
\paragraph{Initial-state collinear and soft sector: $I_1=\{k_1\}, I_2=\emptyset, \tilde{F}_i=\{p_i\}, S=\{k_2\}$}
We again point out that the soft particle $k_2$ cannot cluster with $k_1$ due to the large rapidity difference. Thus, the following results are independent of the recombination scheme. However, $k_2$ can still either be closer to the beam or one of the jets. For the $\ktness$ definition with recoil ($\sigma=1$), the $\ktness$ approximation reads
\begin{equation}
  \label{Fac:eq_qftI_1S}
  \begin{aligned}
    \qft_{\mathfrak{p}_{n_J+2}}(\{k\})&=\qft_{C_1S}(\{k_1\},\{ k_2\})=\max\left( \lvert k_{1, t} \rvert, \qft_S(k_2) \right)\\
    &\quad+\theta\!\left( \min_i\left( R(p_i, k_2)\right)-D \right)\left[ \lvert k_{1, t}+k_{2, t}\rvert- \max\left( \lvert k_{1, t} \rvert, \qft_S(k_2) \right)\right]\\
    &=\lvert k_{1, t}+k_{2, t}\rvert\\
    &\quad +\theta\!\left( D- \min_i\left( R(p_i, k_2)\right) \right)\left[\max\left( \lvert k_{1, t} \rvert, \qft_S(k_2) \right)- \lvert k_{1, t}+k_{2, t}\rvert\right]\, ,
  \end{aligned}
\end{equation}
where we used the function $\qft_S(k_2)$ to denote the value of $\ktness$ one would get if $k_2$ was the only emission, i.e.,
\begin{equation}
  \label{Fac:eq_qftS}
  \qft_S(k_2)\equiv  \lvert k_{t, 2}\rvert \min\left(\{1, \frac{R(p_i, k_2)}{D}\}_i \ \right)\, .
\end{equation}
The second line in \eqref{Fac:eq_qftI_1S} would vanish if we defined $\ktness$ without using the beam recoil, i.e., with $\sigma=0$. We wrote \eqref{Fac:eq_qftI_1S} in two equivalent ways---the first two lines show how $\ktness$ almost factorizes in cumulant space, while the third and fourth lines show how $\ktness$ almost factorizes in impact parameter space. 

The $z_N$-prescription adds a zero bin in the double-soft region. However, $\qft_{C_1S}(k_1, k_2)$ is already homogeneous in this region and thus
\begin{equation}
  \qft_{C_1S, S}(k_1, k_2)=\qft_{C_1S}(\{k_1\},\{ k_2\})\, .
\end{equation}
The expressions for the region where $k_1$ is instead in $I_2$ are the same.

\paragraph{Final-state collinear and soft sector: $I_1=\emptyset, I_2=\emptyset, F_i=\{k_1, k_2\}, S=\{k_3\}$}
The approximation of $\ktness$ reads 
\begin{equation}
  \label{Fac:eq_qftFCS}
  \begin{aligned}
    & \qft_{\mathfrak{p}_{n_J+2}}(\{k\})=\qft_{C_iS}\left( \{\tilde{k}_1, \tilde{k}_2\}, \{k_3\} \right)\sim \max\left(d_{c, 12}, \qft_S(k_3)\right)\\
    &\quad +\Theta_i(k_3)\theta\!\left( d_{c, 12}-k_{t, 3}\frac{R(p_i, k_3)}{D} \right)\left[ \theta\!\left( \tilde{d}_{13}-\tilde{d}_{23} \right)\left( d_{c, 1(2+3)}-d_{c, 12} \right)+ \left( 1\leftrightarrow 2 \right)\right]\, ,
  \end{aligned}
\end{equation}
where, for $d_{c, 1(2+3)}$, on uses \eqref{Fac:eq_dijc} with $k_{j, \perp}^s\to k_{3, \perp}$. The notation $\Theta_{i}(k_3)$ means that $k_3$ is closest to jet $i$, i.e., $\Theta_{i}(k_3)=1$ if $\qft_S(k_3)=k_{t, 3}\frac{R(p_i, k_3)}{D}$ and $\Theta_{i}(k_3)=0$ otherwise. Here we also used the notation $\tilde{d}_{13}$ and $\tilde{d}_{23}$ which is the distance between the respective particles expanded in the limit where momentum $k_{1(2)}$ has the scaling $k_{1(2)} \in F_i$ and $k_3$ has soft scaling expanded to the first subleading power. Note that at leading power in $\lambda$ both $d_{13}$ and $d_{23}$ approximate to $k_{t, 3}R(p_i, k_3)/D$ which is why one needs the next power. In practice, the calculation of $\tilde{d}_{13}$ and $\tilde{d}_{23}$ is subtle and depends on the details of the chosen $d_{ij}$ function. An explicit example for $\tilde{d}_{ij}$ is given for $y_{23}$ in Appendix~\ref{App_Approximationsy23}. Note again that all these difficulties can be avoided by using the WTA scheme, where the second line in \eqref{Fac:eq_qftFCS} would vanish, because $ d_{c, 1(2+3)}-d_{c, 12}=0$ would hold to leading power.

The $z_N$-prescription again introduces a zero bin in the double-soft sector. The approximation for $\ktness$ in the zero bin reads 
\begin{equation}\label{Fac:eq_qftFiSZerobin}
  \begin{aligned}
    \qft_{C_iS, S}(k_2, k_3)&=\max\left(\frac{k_{\perp, 2}}{D}, \qft_S(k_3)\right)\\
    &+\Theta_i(k_3)\theta\!\left( \frac{k_{\perp, 2}}{D}-k_{t, 3}\frac{R(p_i, k_3)}{D} \right)\left[ \theta\!\left(\tilde{ \tilde{d}}_{13}-\tilde{\tilde{d}}_{23} \right)\frac{1}{D} \left( \lvert k_{23, \perp}\rvert - \lvert k_{2, \perp}\rvert \right)\right]\, ,
  \end{aligned}
\end{equation}
where the double-tilde notation in $ \theta\!\left(\tilde{ \tilde{d}}_{13}-\tilde{\tilde{d}}_{23} \right)$ indicates that one needs to expand $ \tilde{d}_{13}-\tilde{d}_{23} $ in the double-soft limit, keeping only the leading non-vanishing term. The second line in \eqref{Fac:eq_qftFiSZerobin} would vanish in the WTA scheme. 
\paragraph{Double-soft region: $I_1=\emptyset, I_2=\emptyset, \tilde{F}_i=\{p_i\}, S=\{k_1, k_2\}$}
In the double-soft region, $\ktness$ simplifies to
\begin{equation}
  \label{Fac:eq_qftSS}
  \begin{aligned}
    \qft_{\mathfrak{p}_{n_J+2}}(\{k\}) &= \qft_{S}(k_1,k_2)=\max\left( \qft_S(k_1), \qft_S(k_2) \right)\\
    &\quad + \theta\!\left( \min\left(  \qft_S(k_1), \qft_S(k_2)\right)-d_{12}\right)\left[\qft_S(k_{12})-\max\left( \qft_S(k_1), \qft_S(k_2) \right) \right]\\
    &\quad+\theta\!\left(d_{12}- \min\left(  \qft_S(k_1), \qft_S(k_2)\right)\right)\theta\!\left( \min(\{R(p_i, k_1), R(p_i, k_2)\}_i)-D \right)\\
    &\quad\times\left[\lvert k_{t, 1}+k_{t, 2}\rvert -\max\left(\lvert k_{t, 1}\rvert  , \lvert k_{t, 2}\rvert\right) \right]\, .
  \end{aligned}
\end{equation}
Here, we defined $\qft_S(k_{12})$, which is the approximation one would find for $\ktness$ in the double-soft region if $k_1$ and $k_2$ were clustered together, and no other emissions were present, i.e., 
\begin{equation}
  \qft_S(k_{12})=\lvert k_{t, 12}\rvert \min\left(\{1, \frac{R(p_i, k_{12})}{D}\}_i \ \right)\, .
\end{equation}
In \eqref{Fac:eq_qftSS}, the last two lines are only present if we define $\ktness$ using the recoil collected during the clustering ($\sigma=1$). The second line in \eqref{Fac:eq_qftSS}, on the other hand, is always present in any $\ktness$ definition, and it collects the clustering effects that spoil non-Abelian exponentiation in the soft sector.

We have now analyzed the approximation for $\ktness$ in all sectors relevant for NNLO calculations in $n$-$\ktness$ slicing. We notice that $\ktness$ factorizes in cumulant space, i.e., satisfies \eqref{Gen_eq_MaximumFactorization}, except in the regions where both soft and either initial-state or final-state collinear radiation is present. The terms that spoil the factorization are due to two effects---either maximum factorization is spoiled because we use the recoil collected during the clustering in the definition of $\ktness$ ($\sigma=1$), or, because soft particles cluster with final-state collinear particles. In the following, we will call the first problem the \emph{beam recoil problem} and the second problem the \emph{soft recoil problem}. If we use a version that does not collect the beam recoil ($\sigma=0$) and a recoil-free recombination scheme (like the WTA scheme), then both problems are avoided. In this case, our analysis shows that \eqref{Gen_eq_MaximumFactorization} holds to NNLO. When extending the analysis for $\ktness$ with no recoil and WTA scheme to higher orders, a big part of the analysis still runs through. In particular, the distances between collinear particles are always unchanged, even if we cluster them together with soft particles, and distances between soft and collinear particles can always be approximated with the distance between the soft particle and the respective jet. This means that we can define $\ktness$ separately on the particles in each of the sectors $I_1, I_2, \dots$, ignoring the other emissions. At leading power in the WTA scheme, the clustering distances $\underbar{d}\equiv{d_1,\dots,d_k}$ encountered along the clustering sequence are nondecreasing, i.e. $\max(\underbar{d})=d_k$. Consequently, $\ktness$ with WTA recombination and no beam recoil ($\sigma=0$) factorizes in cumulant space to all orders, and the factorization formulas \eqref{Gen_eq_CumulantFactorizationFormula} and \eqref{Gen_eq_CumulantFactorizationFormulaZN} hold to all orders (under the simplified assumption \eqref{eq_GeneralSoftCollinearFactorization}, i.e. neglecting Glauber effects). This definition of $\ktness$ is therefore the most convenient for extensions to $\mytext{N}^3\mytext{LO}$ or for all-order resummation.

In this paper, we are mostly interested in $\ktness$ as a slicing variable for NNLO calculation, and thus we do not necessarily require an all-order factorization formula. Instead, we present two approximate $\ktness$ factorization formulas at NNLO that hold for the more complicated definitions of $\ktness$ with recoil or E-scheme. 

If we use a $\ktness$ definition that does not collect the beam recoil ($\sigma=0$) but uses $E$-scheme recombination, the partonic cumulant $\ktness$ cross section approximates to
\begin{equation}
  \label{Fac:eq_ktnessFactorization}
  \begin{aligned}
    &\rd x_1\rd x_2\hat{\sigma}_{b_1b_2}(\qfcut)=\int \frac{\rd \tilde{x}_1\, \rd \tilde{x}_2}{2Q^2}\frac{\rd z_1\rd z_2}{z_1 z_2}\prod_{i=3}^{n_J+2} [\rd p_i]\, \rd \tilde{p}_F \deltabar^{\, d}\biggl[ \sum_{i=3}^{n_J+2} p_i+\tilde{p}_F-q\biggr]\, \rd \tilde{\Pi}_F \F(\{p_i\})\\
    &\times \sum_\A
   \Biggl \langle \Hb_\A(\tilde{x}_1P_1, \tilde{x}_2P_2, \{p_i\}, p_F) \biggl[\dots\biggr] \Biggr \rangle\, ,
  \end{aligned}
\end{equation}
where 
\begin{equation}\label{Fac:eq_ktnessFactorizationNoBeamRecoil}
  \begin{aligned}
    \biggl[\dots\biggr]&=\hat{\Bb}_{N, a_1b_1}(\qfcut, z_1)\hat{\Bb}_{N, a_2b_2}(\qfcut, z_2)\\
    &\quad\times\prod_{i=3}^{n_J+2} \hat{\J}_{N, a_i}(\qfcut)\hat{\Sb}_{\mytext{sub}}(\qfcut) + \left( \frac{\as}{\pi} \right)^2\sum_{i=3}^{n_J+2} \JS_{\hspace{-0.05cm}N, i}(\qfcut)+\mathcal{O}(\as^3)\, .
  \end{aligned}
\end{equation}
Here, the cumulant beam functions are defined through \eqref{Gen_eq_CumulantRadiativeFunctionsN} using the $\ktness$ approximations $\qft_{C_{1(2)}}(k_1)=\lvert k_{t, 1}\rvert $ for the single-emission pieces and \eqref{Fac:eq_qftI1I1_norec} for the double-emission piece. The cumulant jet functions are defined through \eqref{Gen_eq_CumulantRadiativeFunctionsN} with $\qf_{C_i}(\tilde{k}_1,\tilde{k}_2)=d_{c, 12}$ for the single-emission terms, and \eqref{Fac:eq_qftCCC} for the triple-collinear terms. We also defined the terms breaking the cumulant factorization
\begin{equation}
  \label{Fac:eq_JSNi_norec}
  \begin{aligned}
    \JS_{\hspace{-0.05cm}N, i}\!(\qfcut)&\equiv \sum_{\Af_i}\mu^{4\epsilon} \left( \frac{e^{\epsilon \gamma_E}}{\Gamma(1-\epsilon)} \right)^2\int_0^1 \rd \tilde{z}_1 \frac{\rd^{d-2}\tilde{k}_{\perp, 1}}{\Omega_{d-2}}\frac{\hat{P}_{N, \Af_i}(\tilde{z}_1)}{\lvert \tilde{k}_{\perp, 1} \rvert^2} \int \frac{\rd^dk_3\delta_+(k_3^2)}{\Omega_{d-2}}\\
    &\hspace{-1.2cm}\times \Biggl\{\bJJ^{(0)}(k_3)\left[ \theta\!\left(\qfcut - \qft_{C_iS}\left( \{\tilde{k}_1, \tilde{k}_2\}, \{k_3\} \right)\right)-\theta(\qfcut -  \max\left(d_{c, 12}, \qft_S(k_3)\right)) \right]\\
    &\hspace{-1.2cm}-J^{(0)}_{N}(p_i, k_3)\left[ \theta\!\left(\qfcut - \qft_{C_i, C_iS}\left( \{\tilde{k}_1, \tilde{k}_2\},\{ k_3\} \right)\right)-\theta\!\left(\qfcut -  \max \left(d_{c, 12}, \frac{\lvert k_{3, \perp}\rvert}{D}  \right)\right) \right]\Biggr\}\, ,
  \end{aligned}
\end{equation}
where $\qft_{C_iS}$ was given in \eqref{Fac:eq_qftFCS}, $\qft_{C_i, C_iS}$  in \eqref{Fac:eq_qftCCC_zerobinCS} and $\tilde{k}_1$ and $\tilde{k}_2$ are two momenta collinear to jet $p_i$ while $k_3$ is soft. The sum over $\Af_i$ only contains the $qg$ final state for quark jets and the $q\bar{q}$ and $gg$ final state for gluon jets. $\JS_{\hspace{-0.05cm}N, i}$ can, in principle, be a matrix in spin space and thus feature non-trivial spin correlations for gluon jets. Note that the third line in \eqref{Fac:eq_JSNi_norec} is not summed over the different jets because all zero bins in the collinear-soft region factorize, except the ones coming from the triple collinear region. However, $\JS_{\hspace{-0.05cm}N, i}\!(\qfcut)$ is still free of collinear divergences (i.e. rapidity divergences) because if $k_3$ goes collinear to any hard direction that is not $p_i$, then $\Theta_i(k_3)=0$ and the second line in \eqref{Fac:eq_JSNi_norec} vanishes, while if $k_3$ becomes collinear to $p_i$ then the $\theta$-functions in \eqref{Fac:eq_JSNi_norec} match pairwise and the collinear divergence cancels between $\bJJ^{(0)}(k_3)$ and $J^{(0)}_{N}(p_i, k_3)$. The $\theta$-functions in \eqref{Fac:eq_JSNi_norec} cancel in any NLO type IR limit. For example, if $k_3\to 0$, keeping the other momenta fixed, then $d_{c, (1+3)2}-d_{c, 12}\to 0$ and thus, the $\theta$-funcions in the second line of \eqref{Fac:eq_JSNi_norec} cancel. In fact, $\JS_{\hspace{-0.05cm}N, i}$ has to be free of NLO-type IR singularities because there are no corresponding real-virtual terms violating cumulant factorization that could cancel these poles. Finally, note that $ \JS_{\hspace{-0.05cm}N, i}$ also has no pole in the fully unresolved IR limit $\tilde{k}_{\perp, 1}\sim \lvert k_3\rvert \to 0$ because of the difference of $\theta$-functions. Thus, $ \JS_{\hspace{-0.05cm}N, i}$ is finite and suitable for numerical integration. We again point out that $ \JS_{\hspace{-0.05cm}N, i}\!(\qfcut)=0$ in the WTA scheme.

Finally, the single emission soft functions in \eqref{Fac:eq_ktnessFactorizationNoBeamRecoil} are defined through \eqref{Gen_eq_SsubZnNLO} and \eqref{Gen_eq_SsubZnSingleGluonRV} respectively, where one uses \eqref{Fac:eq_qftS} for $\qft_S$ and 
\begin{equation}
  \qft_{C_\alpha ,S}(k)=\begin{cases}
    \lvert k_t\rvert & \mathrm{if } \, \alpha\in \{1, 2\}\\
    \frac{\lvert k_\perp\rvert}{D} & \mathrm{else } \, .
  \end{cases}
\end{equation} 
For the double-emission soft functions, one uses \eqref{Gen_eq_SsubZnqqbar}, \eqref{Gen_eq_SsubZngg_nonAbelian}, and \eqref{Gen_eq_SsubZngg_Abelian}, and $\qft_{S}$ can be found in \eqref{Fac:eq_qftSS}, dropping the third and fourth line. $ \qft_{C_\alpha , S}$ is found in \eqref{Fac:eq_qftCiCiCi_zerobinSS} for $\alpha>2$ and in \eqref{Fac:eq_qftI1I1ZeroBins_norec}, for $\alpha\leq 2$. $\qft_{C_\alpha, S C_\alpha, S} $ and $\qft_{C_\alpha ,  C_\alpha S, S} $ are found in \eqref{Fac:eq_qftCiCiCi_zerozerobins} for $\alpha>2$ and in \eqref{Fac:eq_qftI1I1DoubleZeroBins_norec} for $\alpha\leq 2$. $\qft_{C_\alpha C_\beta, C_\alpha S, S}$ and $\qft_{C_\alpha C_\beta, SC_\beta, S}$ is found in \eqref{Fac:eq_qftI1I1_norec} if both $\alpha\leq$ and $\beta\leq 2$, in \eqref{Fac:eq_qftI1FiZeroZeroBins}, if either $\alpha>2$ or $\beta>2$, and in \eqref{Fac:eq_qftFiFjZeroZeroBins} if both $\alpha>2$ and $\beta>2$. $\qft_{C_{\alpha}S, S}(k_1, k_2)=\qft_{SC_{\alpha}, S}(k_2, k_1)$ reads $\max\left( \lvert k_{1, t} \rvert, \qft_S(k_2) \right)$, if $\alpha\leq 2$, and \eqref{Fac:eq_qftFiSZerobin} else.

If we instead want to use the original $\ktness$-definition with $E-$scheme and beam recoil introduced in \cite{Buonocore:2022mle}, it is more convenient to put the factorization formula in a slightly different form. For this purpose, we define the function 
\begin{equation}
  \label{Fac:eq_BBfunction}
  \begin{aligned}
    \hat{\BB}_{\hspace{-0.05cm}N, a_1b_1, a_2b_2}(\qfcut, z_1, z_2)&\equiv \int\rd^{d-2}\mathbf{k_{I_{1}, t}}\int\rd^{d-2}\mathbf{k_{I_{2}, t}}\Bb_{N, a_1b_1}(\mathbf{k_{I_1, t}}, z_1)\Bb_{N, a_1b_1}(\mathbf{k_{I_2, t}}, z_2) \\
    &\quad\times\theta(\qfcut-\lvert \mathbf{k_{I_1, t}}+\mathbf{k_{I_2, t}}\rvert)\, ,
  \end{aligned}
\end{equation}
where $\Bb_{N,a_1b_1}(\mathbf{k_{I_1, t}}, z_1)$ is the $\mathbf{q_T}$-resummation beam function in the $z_N$-prescription. For $\mathbf{q_T}$-resummation in color-singlet production, the soft subtracted function is zero, and thus $\hat{\BB}$ fully entails all radiative contributions to the cumulant cross section at leading power. One can thus extract the result for $ \hat{\BB}_{\hspace{-0.05cm}N, a_1b_1, a_2b_2}(\qfcut, z_1, z_2)$ from any $\mathbf{q_T}$-resummation calculation. The approximate factorization formula for the cumulant $\ktness$ cross section is then obtained through the insertion
\begin{equation}\label{Fac:eq_ktnessFactorizationWithBeamRecoil}
  \begin{aligned}
    \biggl[\dots\biggr]&= \hat{\BB}_{\hspace{-0.05cm}N, a_1b_1, a_2b_2}(\qfcut, z_1, z_2)
    \prod_{i=3}^{n_J+2} \hat{\J}_{N, a_i}(\qfcut)\hat{\Sb}_{\mytext{sub}}(\qfcut) \\
    &\quad+ \left( \frac{\as}{\pi} \right)^2\sum_{i=3}^{n_J+2} \JS_{\hspace{-0.05cm}N, i}\!(\qfcut)+ \left( \frac{\as}{\pi} \right)^2 \left( \BS_{\hspace{-0.05cm}N, 1}\!\!(\qfcut)+\BS_{\hspace{-0.05cm}N, 2}\!\!(\qfcut) \right)+\mathcal{O}(\as^3)
  \end{aligned}
\end{equation}
in \eqref{Fac:eq_ktnessFactorization}.
Here $\JS_{\hspace{-0.05cm}N, i}\!(\qfcut)$ is again the function in \eqref{Fac:eq_JSNi_norec} and $\BS_{\hspace{-0.05cm}N, 1(2)}(\qfcut)$ is defined as
\begin{equation}
  \label{eq_BSfunctions}
  \begin{aligned}
    \BS_{\hspace{-0.05cm}N, 1}\!\!(\qfcut)&\equiv \mu^{4\epsilon} \left( \frac{e^{\epsilon \gamma_E}}{\Gamma(1-\epsilon)} \right)^2 \!\!z_1\!\!\int \rd^{d-1}k_1 \delta_+(k_1^2) \delta\!\left( \bar{z}_1-\frac{k_{I_1}\cdot P_2}{x_1P_1\cdot P_2} \right) \frac{\hat{P}_{N, a_1b_1}}{p_1\cdot k_1}\int \frac{\rd^dk_2\delta_+(k_2^2)}{\Omega_{d-2}}\\
    &\hspace{-1cm}\times \Biggl\{\bJJ^{(0)}(k_2)\left[ \theta\!\left(\qfcut - \qft_{C_1S}\left(\{ k_1\}, \{k_2\} \right)\right)-\theta(\qfcut -  \max\left(\lvert k_{t, 1} \rvert, \qft_S(k_2)\right)) \right]\\
    &\hspace{-1cm}-\left( J^{(0)}_{N}(p_1, k_2)+ J^{(0)}_{N}(p_2, k_2)\right)\left[ \theta\!\left(\qfcut - \lvert k_{t, 1} +k_{t, 2}\rvert\right)-\theta\!\left(\qfcut -  \max \left(\lvert k_{t, 1} \rvert, \lvert k_{t, 2}\rvert  \right)\right) \right]\Biggr\}\, ,
  \end{aligned}
\end{equation}
where $\qft_{C_1S}$ can be found in \eqref{Fac:eq_qftI_1S} and $\BS_{\hspace{-0.05cm}N, 2}\!\!(\qfcut)$ is the same expression with $p_1\leftrightarrow p_2$, $P_1\leftrightarrow P_2$, $z_1\leftrightarrow z_2$ and $C_1\leftrightarrow C_2$. One can again show that this function is free of IR divergences, and it is thus suitable for numerical integration. In general, $\BS_{\hspace{-0.05cm}N,1}\!\!(\qfcut)$ is also an operator in spin space if the hard parton with momentum $p_1$ is a gluon.

The single emission soft function is the same as in the previous example. The double-emission soft subtracted function is also the same as for $\ktness$ without recoil except that now $\qft_{S}$ is to be taken from \eqref{Fac:eq_qftSS} without dropping the third and fourth line, $ \qft_{C_\alpha, S}$ is found in \eqref{Fac:eq_qftCiCiCi_zerobinSS} for $\alpha>2$ and in \eqref{Fac:eq_qftI1I1}, for $\alpha\leq 2$. $\qft_{C_\alpha, S C_\alpha, S} $ and $\qft_{C_\alpha ,  C_\alpha S, S} $ are found in \eqref{Fac:eq_qftCiCiCi_zerozerobins} for $\alpha>2$ and in \eqref{Fac:eq_qftI1I1} for $\alpha\leq 2$. $\qft_{C_\alpha C_\beta, C_\alpha S, S}$ and $\qft_{C_\alpha C_\beta, SC_\beta, S}$ can be found in \eqref{Fac:eq_qftI1I2} if both $\alpha\leq$ and $\beta\leq 2$, in \eqref{Fac:eq_qftI1FiZeroZeroBins}, if either $\alpha>2$ or $\beta>2$ and in \eqref{Fac:eq_qftFiFjZeroZeroBins} if both $\alpha>2$ and $\beta>2$. $\qft_{C_{\alpha}S, S}(k_1, k_2)=\qft_{SC_{\alpha}, S}(k_2, k_1)$ can be found in \eqref{Fac:eq_qftI_1S} if $\alpha\leq 2$ and in \eqref{Fac:eq_qftFiSZerobin} else.

\section{Conclusion and Outlook}
\label{sec_Conclusion}
In this paper, we presented a framework to define and calculate perturbative ingredients for arbitrary resolution variables at NNLO and potentially beyond. While the paper builds on well-known results and techniques, like the IR factorization of amplitudes and the method of regions, we addressed several technical problems that arise for complicated resolution variables like $\ktness$. We presented a general way to expand the phase space integrals in the method of regions, correctly treating the recoil in final-state collinear sectors due to the presence of soft and initial-state collinear sectors. We showed how one derives a general factorization formula in terms of fully differential soft, jet, and beam functions, and we provided several examples of how these factorization formulas simplify for specific resolution variables. 

The $z_N$-prescription is a fully general method to regularize rapidity divergences in beam and jet functions. We presented a general treatment of the zero-bin problem arising in the $z_N$-prescription by defining a soft subtracted function that is free of rapidity divergences. Our regularized jet and the soft subtracted functions are well suited for numerical integration. The definitions of the soft subtracted, jet, and beam functions do not distinguish between \SCETI and \SCETII type resolution variables, providing alternative factorization approaches to \SCETI resolution variables like $\tau_n$. While we did not use SCET in this paper, many of the problems solved in our framework are also relevant for SCET calculations. Our treatment of the zero-bin problem in the $z_N$-prescription produced insights that are also relevant for other rapidity regulator choices.

As a first application of our framework, we derived factorization formulas for different definitions of $\ktness$, and we provided a new definition, leveraging the WTA scheme, that factorizes to all orders in cumulant space. A paper detailing the application of this result to the NNLO calculation of dijet production at lepton colliders is in preparation~\cite{inprep2}. A method for the numerical calculation of jet functions in the $z_N$-prescription has been presented recently~\cite{Buonocore:2025ysd}.

There are several directions in which this work can be extended. To complete the framework at NNLO, we will also need to provide the $z_N$-modified splitting kernels for gluon splittings and spacelike splittings. An important future extension is to repeat the analysis of this paper while keeping subleading powers in the phase space. This would allow one to derive formulas for the subleading power contributions to the cross section below the cut written in terms of well-defined derivatives of matrix elements, which could also be expressed through SCET operator matrix elements and Wilson coefficients. Another possible direction would be to use our framework for higher-order calculations and/or resummation. To push in this direction, it would be useful to have an all-order definition of the $z_N$-modified splitting kernels, which would allow us to study the rapidity anomalous dimension of our jet and beam functions. Using this together with our fully differential definition of beam, jet, and soft functions, one can envision a general framework to perform numerical resummation at NNLL$'$ accuracy.

\section*{Acknowledgments}
I thank Luca Buonocore, Prasanna Dhani, Massimiliano Grazzini, Flavio Guadagni, Stefan Kallweit, Luca Rottoli and Chiara Savoini for help and advice during various stages of this project. I also thank Thomas Becher and Sebastian Jaskiewicz for useful discussions. This work was supported by the Swiss National Science Foundation (SNSF)
under contracts 200020$\_$219367 and 200021$\_$219377.

\bibliography{biblio}

\appendix

\section{Splitting kernels}
\label{sec_APkernels}
In the initial-state collinear limit, we use the regularized and spin-polarized LO splitting kernels
\begin{equation}
  \begin{aligned}
  	P_{qq}^{ss'}(z, \hat{k}_t;\epsilon) &= P_{qq}(z;\epsilon)\, \delta^{ss'} \\
  	P_{qg}^{ss'}(z, \hat{k}_t;\epsilon)  &= P_{qg}(z;\epsilon) \, \delta^{ss'}  \\
  	P_{gg}^{\mu\nu}(z, \hat{k}_t;\epsilon)  &= P_{gg}(z;\epsilon)(-g^{\mu\nu}) - 2G_{gg}(z) \biggl( -g^{\mu\nu} - 2(1-\epsilon)\hat{k}_t^{\mu}\hat{k}_t^{\nu}  \biggr)  \\
  	P_{gq}^{\mu\nu}(z, \hat{k}_t;\epsilon)   &=  P_{gq}(z;\epsilon)(-g^{\mu\nu}) - 2G_{gq}(z) \biggl( -g^{\mu\nu} - 2(1-\epsilon)\hat{k}_t^{\mu}\hat{k}_t^{\nu}   \biggr) 
  \end{aligned}
\end{equation}
where the spin-averaged splitting kernels are
\begin{equation}
  \begin{aligned}
  	P_{qq}(z;\epsilon) &= C_F\left( \frac{z^2+1}{(1-z)_+} -\epsilon(1-z) \right)+\frac{3}{2}C_F\delta(1-z)  \\
  	P_{qg}(z;\epsilon) &= T_R\left( 1-\frac{2z(1-z)}{1-\epsilon} \right)  \\
  	P_{gg}(z;\epsilon) &= 2C_A\left( \frac{z}{(1-z)_+} + \frac{1-z}{z} + z(1-z)  \right) +2\beta_0\delta(1-z)  \\
  	P_{gq}(z;\epsilon) &= C_F\left( \frac{(1-z)^2+1}{z} -\epsilon z \right) \, ,
  \end{aligned}
\end{equation}
with
\begin{equation}
  \beta_0=\frac{\left(11 C_A-4 n_f T_R\right)}{12}\, .
\end{equation}
The Altarelli-Parisi splitting kernels are obtained from the above by setting $\epsilon=0$. One can obtain the unregularized tree-level splitting kernels $\hat{P}$ by replacing $(1-z)_+\to (1-z)$ and dropping terms proportional to $\delta(1-z)$.
The functions $G_{ab}(z)$ are~\cite{Catani:2010pd}
\begin{equation}
  \begin{aligned}
  	G_{gg}(z) = C_A \frac{1-z}{z}  &&
  	G_{gq}(z) =  C_F \frac{1-z}{z}  \, .
  \end{aligned}
\end{equation}
In the final-state collinear limit, we use the unregularized and spin-polarized splitting kernels
\begin{equation}
  \begin{aligned}
  	\hat{P}_{q \to qg}^{ss'}(z, \hat{k}_\bot;\epsilon) &= \hat{P}_{q\to qg}(z;\epsilon)\, \delta^{ss'} = \hat{P}_{q\to gq}^{ss'}(1-z, k_\bot;\epsilon)  \\
  	\hat{P}_{g \to gg}^{\mu\nu}(z, \hat{k}_{\bot};\epsilon)  &= \hat{P}_{g \to gg}(z;\epsilon)(-g^{\mu\nu}) - 2G_{g \to gg}(z, \epsilon) \biggl( -g^{\mu\nu} - 2(1-\epsilon)\hat{k}_\bot^{\mu}\hat{k}_\bot^{\nu}  \biggr)  \\
  	\hat{P}_{g \to q\bar{q}}^{\mu\nu'}(z, \hat{k}_{\bot};\epsilon)   &=  \hat{P}_{g \to q\bar{q}}(z;\epsilon)(-g^{\mu\nu}) - 2G_{g \to q\bar{q}}(z, \epsilon) \biggl( -g^{\mu\nu} - 2(1-\epsilon)\hat{k}_\bot^{\mu}\hat{k}_\bot^{\nu}  \biggr) 
  \end{aligned}
\end{equation}
where the spin-averaged splitting kernels and the functions $G_{g\to gg}(z, \epsilon)$ and $G_{g\to q{\bar q}}(z, \epsilon)$ read
\begin{equation}
  \begin{aligned}
  	\hat{P}_{q \to qg}(z;\epsilon) &= C_F\left( \frac{z^2+1}{1-z} -\epsilon(1-z) \right) = \hat{P}_{q \to gq}(1-z;\epsilon)  \\
  	\hat{P}_{g \to gg}(z;\epsilon) &= 2C_A\left( \frac{z}{1-z} + \frac{1-z}{z} + z(1-z)  \right) \\
  	\hat{P}_{g \to q\bar{q}}(z;\epsilon) &= T_R\left( 1-\frac{2z(1-z)}{1-\epsilon}\right) \\
  	G_{g \to gg}(z, \epsilon) &= C_Az(1-z) \\
  	G_{g \to q\bar{q}}(z, \epsilon) &=  -\frac{T_R}{1-\epsilon} z(1-z)  \, .
  \end{aligned}
\end{equation}

We also list the triple-collinear splitting kernels required for quark jet functions:
\begin{equation}
  \left\langle\hat{P}_{q\to \bar{q}_1^{\prime} q_2^{\prime} q_3}\right\rangle=\frac{1}{2} C_F T_R \frac{s_{123}}{s_{12}}\left[-\frac{t_{12, 3}^2}{s_{12} s_{123}}+\frac{4 z_3+\left(z_1-z_2\right)^2}{z_1+z_2}+(1-2 \epsilon)\left(z_1+z_2-\frac{s_{12}}{s_{123}}\right)\right]
\end{equation}
\begin{equation}
  \begin{aligned}
  \left\langle\hat{P}_{q\to \bar{q}_1 q_2 q_3}^{(\mathrm{id})}\right\rangle & =C_F\left(C_F-\frac{1}{2} C_A\right)\left\{(1-\epsilon)\left(\frac{2 s_{23}}{s_{12}}-\epsilon\right)\right. \\
  & +\frac{s_{123}}{s_{12}}\left[\frac{1+z_1^2}{1-z_2}-\frac{2 z_2}{1-z_3}-\epsilon\left(\frac{\left(1-z_3\right)^2}{1-z_2}+1+z_1-\frac{2 z_2}{1-z_3}\right)-\epsilon^2\left(1-z_3\right)\right] \\
  & \left.-\frac{s_{123}^2}{s_{12} s_{13}} \frac{z_1}{2}\left[\frac{1+z_1^2}{\left(1-z_2\right)\left(1-z_3\right)}-\epsilon\left(1+2 \frac{1-z_2}{1-z_3}\right)-\epsilon^2\right]\right\}+(2 \leftrightarrow 3) 
  \end{aligned}
\end{equation}
\begin{equation}
  \left\langle\hat{P}_{q\to g_1 g_2 q_3}\right\rangle=C_F^2\left\langle\hat{P}_{q\to g_1 g_2 q_3}^{(\mathrm{ab})}\right\rangle+C_F C_A\left\langle\hat{P}_{q\to g_1 g_2 q_3}^{(\mathrm{nab})}\right\rangle
\end{equation}
\begin{equation}
  C_F^2\left\langle\hat{P}_{q\to g_1 g_2 q_3}^{(\mathrm{ab})}\right\rangle = \Biggl(\frac{s_{123}\hat{P}_{q\to gq}(z_1)\hat{P}_{q\to gq}\left(\frac{z_2}{z_2+z_3}\right)}{s_{23}}+(1\leftrightarrow 2)\Biggr)+C_F^2\left\langle\hat{R}_{q\to g_1 g_2 q_3}^{(\mathrm{ab})}\right\rangle
\end{equation}
\begin{equation}
  \label{eq_Rab}
  \begin{aligned}
    \left\langle\hat{R}_{q\to g_1 g_2 q_3}^{(\mathrm{ab})}\right\rangle &=\frac{s_{123} \left(z_1 z_2 \epsilon ^2+\left(z_1^2+z_2 z_1+z_2^2\right) \epsilon -z_3^2-1\right) (z_1 s_{13}+z_1 s_{23}-z_3 s_{12})}{z_1 z_2 s_{13} (s_{13}+s_{23})}\\
    &+\frac{(\epsilon -1)^2 (z_3 (s_{12}+s_{13})-z_1 s_{23})}{(z_1+z_3) s_{13}}-\epsilon  (\epsilon -1) +(1\leftrightarrow 2)
  \end{aligned}
\end{equation}
\begin{equation}
\begin{aligned}
\left\langle\hat{P}_{q\to g_1 g_2 q_3}^{(\mathrm{nab})}\right\rangle & =\left\{(1-\epsilon)\left(\frac{t_{12, 3}^2}{4 s_{12}^2}+\frac{1}{4}-\frac{\epsilon}{2}\right)+\frac{s_{123}^2}{2 s_{12} s_{13}}\left[\frac{\left(1-z_3\right)^2(1-\epsilon)+2 z_3}{z_2}\right.\right. \\
& \left.+\frac{z_2^2(1-\epsilon)+2\left(1-z_2\right)}{1-z_3}\right]-\frac{s_{123}^2}{4 s_{13} s_{23}} z_3\left[\frac{\left(1-z_3\right)^2(1-\epsilon)+2 z_3}{z_1 z_2}+\epsilon(1-\epsilon)\right] \\
& +\frac{s_{123}}{2 s_{12}}\left[(1-\epsilon) \frac{z_1\left(2-2 z_1+z_1^2\right)-z_2\left(6-6 z_2+z_2^2\right)}{z_2\left(1-z_3\right)}+2 \epsilon \frac{z_3\left(z_1-2 z_2\right)-z_2}{z_2\left(1-z_3\right)}\right] \\
& +\frac{s_{123}}{2 s_{13}}\left[(1-\epsilon) \frac{\left(1-z_2\right)^3+z_3^2-z_2}{z_2\left(1-z_3\right)}-\epsilon\left(\frac{2\left(1-z_2\right)\left(z_2-z_3\right)}{z_2\left(1-z_3\right)}-z_1+z_2\right)\right. \\
& \left.\left.-\frac{z_3\left(1-z_1\right)+\left(1-z_2\right)^3}{z_1 z_2}+\epsilon\left(1-z_2\right)\left(\frac{z_1^2+z_2^2}{z_1 z_2}-\epsilon\right)\right]\right\}+(1 \leftrightarrow 2)\, ,
\end{aligned}
\end{equation}
where we used the kinematic quantity 
\begin{equation}
t_{i j, k} \equiv 2 \frac{z_i s_{j k}-z_j s_{i k}}{z_i+z_j}+\frac{z_i-z_j}{z_i+z_j} s_{i j}\, .
\end{equation}

The one-loop $1\to 2$ spin-averaged splitting kernels have the structure
    \begin{equation}
      \hat{P}^{(1)}_{a\to a_1a_2}(z)= \cos(\pi \epsilon) c_P \left(  \hat{P}^{(1)}_{a\to a_1a_2, H}(z)+ \hat{P}^{(1)}_{a\to a_1a_2, C}(z) \right)\, ,
    \end{equation}
    where 
    \begin{equation}
      \hat{P}^{(1)}_{q\to gq, C}(z)=\frac{1}{2}\hat{P}^{(0)}_{q\to gq}(z)\Biggl\{-\frac{C_A}{\epsilon^2}-\frac{(2 C_F-C_A) f(\epsilon ;1-z)+C_A f(\epsilon
      ;z)}{\epsilon }\Biggr\}
    \end{equation}

    \begin{equation}
      \hat{P}^{(1)}_{g\to gg, C}(z)=\frac{1}{2}\hat{P}^{(0)}_{g\to gg}(z)\Biggl\{-\frac{C_A}{\epsilon^2}-\frac{C_A f(\epsilon , 1-z)+C_A f(\epsilon , z)}{\epsilon
      }\Biggr\}
    \end{equation}
   
    \begin{equation}
      \hat{P}^{(1)}_{g\to q\bar{q}, C}(z)=\frac{1}{2}\hat{P}^{(0)}_{g\to q\bar{q}}(z)\Biggl\{\frac{C_A-2 C_F}{\epsilon ^2}+\frac{4 \beta_0-3 C_F}{\epsilon }-\frac{C_A f(\epsilon , 1-z)+C_A f(\epsilon , z)}{\epsilon }\Biggr\}\, ,
    \end{equation}
    which can be summarized in the compact formula
    \begin{equation}
     \begin{aligned}
       &\hat{P}^{(1)}_{a\to a_1a_2, C}(z_1, z_2)=\frac{1}{2}\hat{P}^{(0)}_{a\to a_1a_2}(z_1, z_2)\Biggl\{\frac{1}{\epsilon^2}\left(C_{12}-C_1-C_2\right)+\frac{2}{\epsilon}\left(\gamma_{12}-\gamma_1-\gamma_2+\beta_0\right)\\
       &-\frac{1}{\epsilon}\left[\left(C_{12}+C_1-C_2\right) f\left(\epsilon ; z_1\right)+\left(C_{12}+C_2-C_1\right) f\left(\epsilon ; z_2\right)\right]\Biggr\}\, ,
     \end{aligned}
    \end{equation} 
    where 
\begin{equation}
  \gamma_q=\gamma_{\bar{q}}=\frac{3C_F}{4} \quad \gamma_g=\beta_0\, .
\end{equation}
    We have defined 
    \begin{equation}
     c_{P}=\frac{e^{\epsilon \gamma_E } \Gamma^2(1-\epsilon ) \Gamma (1+\epsilon)}{\Gamma (1-2 \epsilon )}=1-\frac{\pi ^2 \epsilon ^2}{12}-\frac{7 \zeta_3 \epsilon ^3}{3}-\frac{47 \pi ^4 \epsilon ^4}{1440}+O\left(\epsilon ^5\right)
  \end{equation}
  and the function
  \begin{equation}
    \begin{aligned}
      f(\epsilon ; 1 / x) &\equiv \frac{1}{\epsilon}\left[{ }_2 F_1(1, -\epsilon ; 1-\epsilon ; 1-x)-1\right] \mytext {. }\\
      &=\log x-\epsilon\left[\mathrm{Li}_2(1-x)+\sum_{k=1}^{\infty} \epsilon^k \mathrm{Li}_{k+2}(1-x)\right]
    \end{aligned}
  \end{equation}
  The $\hat{P}^{(1)}_{a\to a_1a_2, H}$ splitting kernel relevant for the two-loop quark jet function reads 
\begin{equation}
      \begin{aligned}
        \hat{P}^{(1)}_{q\to qg, H}(z)&=\frac{1}{4}\frac{N_c^2+1}{N_c}\frac{1-\epsilon \delta_{R, I}}{(1-2\epsilon)(1-\epsilon)}C_F\big(1-\delta_{R, E}\epsilon(1-z)\big)\\
          &=\frac{1}{2}C_F(C_A-C_F)\frac{1-(1-z)\epsilon}{1-2\epsilon}\, ,
      \end{aligned}
\end{equation}
    where $\delta_{R, I}=1$ if internal gluons have $2-2\epsilon$ dimensions and $0$ if they have $2$. The same goes for $\delta_{R, E}$ for external gluons. For CDR, we have $\delta_{R, I}=\delta_{R, E}=1$, which leads to the second line.

\section{Infrared Singularities in Massless QCD Amplitudes} \label{sec_IRStructureLoops}
The IR singularities of massless QCD amplitudes can be subtracted multiplicatively as \cite{Catani:1998bh, Sterman:2002qn, Becher:2009qa}
\begin{equation}
  \label{eq_ProductFactorizationOfIRPoles}
 \ket{\amp_n(\{p_i\}; \mu)}=\lim _{\epsilon \rightarrow 0} \boldsymbol{Z}^{-1}(\epsilon, \{p_i\}, \mu)\ket{\amp_n(\{p_i\}, \epsilon)}\, .
\end{equation}

The operator $\boldsymbol{Z}$  acts in color space and can be expressed through the hard anomalous dimension $\boldsymbol{\Gamma}$ as 
\begin{equation}
  \label{eq_ZFactorFromAnomalousDimension}
  \begin{aligned}
      \boldsymbol{Z}(\epsilon, \{p_i\}, \mu)&=\mathbf{P} \exp \left[\int_\mu^{\infty} \frac{\rd \mu^{\prime}}{\mu^{\prime}} \boldsymbol{\Gamma}\left(\{p_i\}, \mu^{\prime}\right)\right]\\
      \Rightarrow\log \boldsymbol{Z}(\epsilon, \{p_i\}, \mu)&=\int_0^{\alpha_s(\mu^2)} \frac{\rd \alpha}{2\alpha} \frac{1}{\epsilon-\beta(\alpha) / \alpha}\left[\boldsymbol{\Gamma}(\{p_i\}, \mu, \alpha)+\int_0^\alpha \frac{\rd \alpha^{\prime}}{2\alpha^{\prime}} \frac{\Gamma^{\prime}\left(\alpha^{\prime}\right)}{ \epsilon-\beta\left(\alpha^{\prime}\right) / \alpha^{\prime}}\right]
  \end{aligned}
\end{equation}
where the second line holds to NNLO, where the anomalous dimension satisfies the \emph{dipole formula} \cite{Gardi:2009qi, Gardi:2009zv}
\begin{equation}
  \label{eq_HardAnomalousDimensionMSbar}
  \boldsymbol{\Gamma}(\{p_i\}, \mu)=\sum_{\alpha\neq \beta} \frac{\boldsymbol{T}_\alpha \cdot \boldsymbol{T}_\beta}{2} \gamma_{\mytext {cusp}}\left(\alpha_s\right) \log \frac{\mu^2}{-s_{\alpha \beta}}+\sum_\alpha \gamma^\alpha\left(\alpha_s\right)\, ,
 \end{equation} in which case the path ordering symbol can be dropped\footnote{Note that \eqref{eq_ZFactorFromAnomalousDimension} does not exactly look the same as (2.9) in~\cite{Becher:2009qa} because our conventions for the $\beta$-funtion differ by a factor of 2. }. Explicit formulas for the $\boldsymbol{Z}$-factor in terms of $\gamma_{\mytext{cusp}}$ and $\gamma^\alpha$ are listed in~\cite{Becher:2009qa}.

 \section{Mass Factorization Counter Terms}
\label{sec_MassFactorizationCounterTerms}
The bare parton distribution functions can be expressed in terms of the renormalized ones as 
\begin{equation}
  \label{eq_RenormalizedPDFs}
  f_{b/h}(x, \epsilon)=\left[ (\Gamma^{-1})_{ba}(\mu_F^2)\otimes f_{a/h}(\mu_F^2) \right](x)\, ,
\end{equation}
where the operator $\Gamma^{-1}_{ba}$ can be expressed through the DGLAP splitting kernels~\cite{Altarelli:1977zs, Gribov:1972ri, Dokshitzer:1977sg} as 
\begin{equation}
  \label{eq_GammaOperator}
\begin{aligned}
    &\Gamma^{-1}_{ba}(x, \mu^2)= \exp\left( \int_{\mu^2}^\infty \frac{\rd \tilde{\mu}^2}{\tilde{\mu}^2}P(\as(\tilde{\mu}^2)) \right)_{ba}=\exp\left( \int_{0}^{\as(\mu^2)} \frac{\rd \tilde{\alpha}_S}{ \tilde{\alpha}_S\left( \epsilon-\frac{\beta(\tilde{\alpha}_S)}{\tilde{\alpha}_S} \right)}P(\as(\tilde{\mu}^2)) \right)_{ba}\\
    &= 1+ \frac{\as(\mu^2)}{\pi}\frac{P_{ba}^{(0)}}{2\epsilon}+\left(\!\frac{\as(\mu^2)}{\pi} \right)^{\!\!2}\!\!\left[ \frac{1}{8\epsilon^2}\left(   P_{bc}^{(0)}\otimes P_{ca}^{(0)} -2\beta_0P_{ba}^{(0)}\right)+\frac{P_{ba}^{(1)}}{8\epsilon}\right]\! + \mathcal{O}(\as^3)\, ,
\end{aligned}
\end{equation}
where the splitting kernel in the exponential has to be understood as a matrix in flavor space and an operator acting as a convolution kernel on functions $f:[0, 1]\to \mathbb{R}$. The Mellin convolution of two functions $f_1$ and $f_2$ is defined as 
\begin{equation}
  \left[ f_1\otimes f_2 \right](x)\equiv \int_0^1 \rd z_1 \int_0^1 \rd z_2 \delta(x-z_1z_2)f_1(z_1)f_2(z_2)\, .
\end{equation}

\section{Limits of Kinematic Functions}
\label{sec_LimitsOfKinematicFunctions}
The beam, jet, and soft subtracted functions defined in this paper depend on limits $\qft$ of the resolution variable $\qf$ in certain regions. The soft subtracted functions also depend on approximations of the resolution variable, where one takes multiple limits, one after the other. In this appendix, we explain how these limits are performed, and we explain some notation in more detail.

For a kinematic function $\qf$ of the massless final-state colored parton momenta $\{k_i\}$ and a color-singlet momentum $p_F$, we write $\qf(k_1, k_2, \ldots, k_n,p_F)=\qf\left( \{k_i\} ,p_F\right)$. Whenever we make an approximation we will use the symbol $\qft$ instead. In general, we want to approximate in a region $\pf_n$, where the final-state momenta are partitioned into sectors according to their scaling,
\begin{equation}
\begin{aligned}
    I_1&=\{k_{1_1},k_{1_2},\dots\},\quad I_2=\{k_{2_1},k_{2_2},\dots\},\quad  F_3=\{k_{3_1},k_{3_2},\dots\},\dots,\\
    F_{n_J+2}&=\{k_{(n_J+2)_1},k_{(n_J+2)_2},\dots\},\quad S=\{k_{S_1},k_{S_2},\dots\}\, .
\end{aligned}
\end{equation} 
We write the approximation of $\qf$ in this region as $\qft_{\pf_n}(\{k\},p_F)$. To obtain the approximation, we perform the following steps:
\begin{enumerate}
  \item Write the observable in terms of scalar products of the momenta $k_i$ and $p_F$.
  \item Express all collinear momenta through the hard momenta $\{p_i\}=\{p_1,p_2,\dots,p_{n_J+2}\}$, $p_F$ and homogeneously scaling variables as 
  \begin{equation}
    \label{Momentum_With_Homogeneous_ScalingAppendix}
    \begin{aligned}
      k_{1_j}^\mu&=\frac{\greenmath{z_{1_j}}}{z_1}p_1^\mu+\greenmath{k_{t,1_j}^\mu}+\frac{z_1\lvert \greenmath{k_{t,1_j}}\rvert^2}{\greenmath{z_{1_j}}Q^2}p_2^\mu\\
      k_{2_j}^\mu&=\frac{\greenmath{z_{2_j}}}{z_2}p_2^\mu+\greenmath{k_{t,2_j}^\mu}+\frac{z_2\lvert \greenmath{k_{t,2_j}}\rvert^2}{\greenmath{z_{2_j}}Q^2}p_1^\mu\\
      k_{i_j}^\mu&=\greenmath{z_{i_j}} p_i^\mu+k_{\perp,i_j}^\mu+\frac{Q^2\lvert k_{\perp,i_j}\rvert^2}{4\greenmath{z_{i_j}}\left( p_i\cdot q \right)^2}\bar{p}_i^\mu, \quad \text{for }i>2\\
      k_{\perp,i_j}&=\greenmath{\tilde{k}_{\perp,i_j}}-\greenmath{z_{i_j}} \frac{p_i\cdot q}{Q^2}k_\mathrm{rec,\perp}^\mu =\greenmath{\tilde{k}_{\perp,i_j}}-\greenmath{z_{i_j}} \frac{p_i\cdot q}{Q^2}\left( k_\mathrm{rec}^\mu-\frac{\bar{p}_i\cdot k_\mathrm{rec}}{p_i\cdot \bar{p}_i}p_i-\frac{p_i\cdot k_\mathrm{rec}}{p_i\cdot \bar{p}_i}\bar{p}_i \right)\\
      k_\mathrm{rec}&=\greenmath{k_{t,I_1}}+\greenmath{k_{t,I_1}}+\greenmath{k_S}=\sum_{j}\greenmath{k_{t,1_j}}+\sum_{j}\greenmath{k_{t,2_j}}+\sum_{j}\greenmath{k_{S_j}}\\
      \,
    \end{aligned}
  \end{equation}
  where we defined $\bar{p}_i=E_i \bar{n}_i=2\frac{p_i\cdot q}{Q}\frac{q}{Q}-p_i$, and we highlighted the homogeneous scaling variables that we want to use as integration variables in the radiative functions in {\color{ReadableGreen}{green}}.\footnote{Typically, we can rewrite $p_F$ as $q-\sum_i k_i$, which allows us to eliminate the scalar products of QCD momenta with color singlet momenta. In some complicated cases, if the resolution variable depends non-trivially on multiple color-singlet momenta, one might need to express each $p_F$ in terms of the corresponding Born momentum $p_F^\prime$ using the equations in \eqref{Gen_eq_LP-relations-for-ktilde-LC}, or more generally, using \eqref{Gen_eq_BoostedMomentum}. }
  \item Rescale the momenta with their respective scaling dimensions,
\begin{equation}
 \begin{aligned}
   k_{t,i_j}
    \to \lambda^a k_{t,i_j}\\
    k_{\perp,i_j}
    \to \lambda^b k_{t,i_j}\\
    k_{S_j}\to \lambda^c k_{S_j}\, ,\\
 \end{aligned}
\end{equation}
expand $\qf$ in $\lambda\to 0$, and keep only the order $\lambda$ term.
\end{enumerate}

We now know how to construct an approximation $\qft_{\pf_n}$, but to construct the soft subtracted function, we need to take a secondary limit in a different region $\pf_n^\prime$ where collinear momenta in $\pf_n$ are now soft instead. The approximation in this region is denoted as $\qft_{\pf_n,\pf_n^\prime}$. While $\qft_{\pf_n}$ is a function of the $z_{i_j},k_{t,i_j},\tilde{k}_{\perp,i_j},k_{S_j}$ as explained in \eqref{Momentum_With_Homogeneous_ScalingAppendix}, the function $\qft_{\pf_n,\pf_n^\prime}$ is written in terms of the new homogeneous collinear scaling variables $\tilde{z}_{i_j}^\prime,k_{t,i_j}^\prime,\tilde{k}_{\perp,i_j}^\prime$ and the new soft momenta $S^\prime$, where the $j$-indices now have different ranges. In particular, if the partition $\pf_n$ yields the sectors
\begin{equation}
  \begin{aligned}
    I_i&=\{k_{i_1},k_{i_2},\dots,k_{i_{n_i^\prime}},k_{i_{n_i^\prime+1}},\dots,k_{i_{n_i}}\}\\
    F_i&=\{k_{i_1},k_{i_2},\dots,k_{i_{n_i^\prime}},k_{i_{n_i^\prime+1}},\dots,k_{i_{n_i}}\}\\
    S&=\{k_{S_1},k_{S_2},\dots,k_{S_{n_S}}\}\, ,
  \end{aligned}
\end{equation}
and the partition $\pf_n^\prime$ yields
\begin{equation}
  \begin{aligned}
    I_i^\prime&=\{k_{i_1},k_{i_2},\dots,k_{i_{n_i^\prime}}\}\\
    F_i^\prime&=\{k_{i_1},k_{i_2},\dots,k_{i_{n_i^\prime}}\}\\
    S^\prime&=\{k_{S_1},k_{S_2},\dots,k_{S_{n_S}},\}\cup\bigcup_{i=1}^{n_J+2} \{k_{i_{n_i^\prime+1}},\dots,k_{i_{n_i}}\}\, ,
  \end{aligned}
\end{equation}
we define the change of variables
\begin{equation}
  \label{eq_zerobin_change_of_variables}
  \begin{aligned}
    \greenmath{z_{i_j}}&=z_i\frac{\bluemath{k_{i_j}}\cdot \bar{p}_i}{p_i\cdot\bar{p}_i}, \quad \text{for } j=n_i^\prime+1,\dots,n_i  \land i\leq 2\\
    \greenmath{z_{i_j}}&=\frac{\bluemath{k_{i_j}}\cdot \bar{p}_i}{p_i\cdot\bar{p}_i}, \quad \text{for } j=n_i^\prime+1,\dots,n_i  \land i>2\\
    \greenmath{z_{i_j}}&=\left( 1-z_i \right)^{-1}\bluemath{z_{i_j}^\prime} \left( 1-z_i-z_i\sum_{k=n_i^\prime+1}^{n_i}\frac{\bluemath{k_{i_k}}\cdot \bar{p}_i}{p_i\cdot\bar{p}_i} \right)\sim \bluemath{z_{i_j}^\prime }, \quad \text{for } j=1,\dots,n_i^\prime\land i\leq2\\
   \greenmath{z_{i_j}}&=\bluemath{z_{i_j}^\prime} \left( 1-\sum_{k=n_i^\prime+1}^{n_i}\frac{\bluemath{k_{i_k}}\cdot \bar{p}_i}{p_i\cdot\bar{p}_i} \right)\sim \bluemath{z_{i_j}^\prime} , \quad \text{for } j=1,\dots,n_i^\prime\land i>2\\
    \greenmath{k_{t,i_j}}&=\bluemath{k_{t,i_j}^\prime}, \quad \text{for } j=1,\dots,n_i^\prime\\
    \greenmath{k_{t,i_j}}&=\bluemath{k_{t,i_j}}, \quad \text{for }  j=n_i^\prime+1,\dots,n_i\\
    \greenmath{\tilde{k}_{\perp,i_j}}&=\bluemath{k_{\perp,i_j}}, \quad \text{for }  j=n_i^\prime+1,\dots,n_i\\
    \greenmath{\tilde{k}_{\perp,i_j}}&=\bluemath{\tilde{k}_{\perp,i_j}^\prime}- \bluemath{z_{i_j}^\prime}\sum_{k=n_i^\prime+1}^{n_i}  \bluemath{ k_{\perp,i_k}}, \quad \text{for } j=1,\dots,n_i^\prime\, ,
  \end{aligned}
\end{equation}
where now the momenta $k_{i_j},k_{t,i_j},k_{\perp,i_j}$ with $j=n_i^\prime+1,\dots,n_i$ have soft scaling, and, it is sometimes useful to express the transverse components in terms of the four vectors $k_{i_j}$ as
\begin{equation}
  \begin{aligned}
    k_{t,1_j}&=k_{1_j}-\frac{k_{1_j}\cdot p_2}{p_1\cdot p_2}p_1-\frac{ k_{1_j}\cdot p_1}{p_1\cdot p_2}p_2\\
    k_{t,2_j}&=k_{2_j}-\frac{k_{2_j}\cdot p_1}{p_1\cdot p_2}p_2-\frac{ k_{2_j}\cdot p_2}{p_1\cdot p_2}p_1\\
    k_{\perp,i_j}&=k_{i_j}-\frac{k_{i_j}\cdot p_i}{p_i\cdot \bar{p}_i}\bar{p}_i-\frac{ k_{i_j}\cdot \bar{p}_i}{p_i\cdot \bar{p}_i}p_i\, .
  \end{aligned}
\end{equation}
Note in particular that now
\begin{equation}
  \sum_{j=1}^{n_i^\prime} \bluemath{\tilde{k}_{\perp,i_j}^\prime}=0\, ,
\end{equation}
i.e., the $\bluemath{\tilde{k}_{\perp,i_j}^\prime}$ are the boost invariant transverse momenta relevant for the collinear particles in sector $F_i^\prime$. In \eqref{eq_zerobin_change_of_variables} we highlighted the homogeneous scaling variables entering the radiative functions for the region $\pf_n$ in green and the homogeneous scaling variables entering the radiative functions for the zero bin region $\pf_n^\prime$ in blue. 
Making these replacements, we have expressed $\qf_{\pf_n}$ in terms of the homogeneous scaling variables of the region $\pf_n^\prime$, and we can again expand to leading non-vanishing order in $\lambda$ to obtain the approximation $\qft_{\pf_n,\pf_n^\prime}$. One proceeds similarly to get a zero-zero bin in a region $\pf_n^{\prime\prime}$ and so on.

\section{Shorthand Notation for Regions at NLO and NNLO}
\label{sec_ShorthandNotationRegions}
At NLO and NNLO it would be too cumbersome to specify every sector $I_1, I_2, F_3,\dots, S$ to define a region $\pf_n$, because most of them are empty or contain only one collinear momentum that can be expressed though a hard momentum and the recoil from the soft and initial-state collinear sectors. Instead, we do not assign variable names for the momenta in the sectors $F_i$ that contain only one particle. In the following, we list how the regions defined at NLO and NNLO distribute the non-trivial momenta into the sectors $I_1, I_2, F_3,\dots, S$. The regions in the one-emission phase space, $\mathfrak{P}_{n_J+1}$, are
\begin{equation}
  \begin{aligned}
    C_1&: \{k_1\}\to I_1=\{k_1\}, \quad I_2=\emptyset,\quad \tilde{F}_i=\{p_i\}, \quad S=\emptyset\\
    C_i&: \{k_1,k_2\}\to I_1=\emptyset, \quad I_2=\emptyset,\quad F_i=\{k_1,k_2\},\quad \tilde{F}_j=\{p_j\} \text{ for }i\neq j, \quad S=\emptyset \\
    S&:\{k_1\}\to I_1=\emptyset, \quad I_2=\emptyset,\quad \tilde{F}_i=\{p_i\},\quad S=\{k_1\}\, .
  \end{aligned}
\end{equation}
The regions in the two-emission phase space, $\mathfrak{P}_{n_J+2}$, are
\begin{equation}
  \begin{aligned}
    C_1&: \{k_1,k_2\}\to I_1=\{k_1,k_2\}, \quad I_2=\emptyset,\quad \tilde{F}_i=\{p_i\}, \quad S=\emptyset\\
    C_1C_2&: \{k_1,k_2\}\to I_1=\{k_1\}, \quad I_2=\{k_2\},\quad \tilde{F}_i=\{p_i\}, \quad S=\emptyset\\
    C_i &: \{k_1,k_2,k_3\}\to I_1=\emptyset, \quad I_2=\emptyset,\quad F_i=\{k_1,k_2,k_3\},\quad \tilde{F}_j=\{p_j\} \text{ for }i\neq j, \quad S=\emptyset \\
    C_1C_i &: \{k_1,k_2,k_3\}\to I_1=\{k_1\}, \quad I_2=\emptyset,\quad F_i=\{k_2,k_3\},\quad \tilde{F}_j=\{p_j\} \text{ for }i\neq j, \quad S=\emptyset \\
    C_iC_j &: \{k_1,k_2,k_3,k_4\}\to I_1=\emptyset, \quad I_2=\emptyset,\quad F_i=\{k_1,k_2\},\quad F_j=\{k_3,k_4\},\\
    &\hspace{2.43mm}\tilde{F}_k=\{p_k\} \text{ for }k\notin \{i,j\}, \quad S=\emptyset \\
    S&:\{k_1,k_2\}\to I_1=\emptyset, \quad I_2=\emptyset,\quad \tilde{F}_i=\{p_i\},\quad S=\{k_1,k_2\}\\
    C_1S&: \{k_1,k_2\}\to I_1=\{k_1\}, \quad I_2=\emptyset,\quad \tilde{F}_i=\{p_i\},\quad S=\{k_2\}\\
    SC_1&: \{k_1,k_2\}\to I_1=\{k_2\}, \quad I_2=\emptyset,\quad \tilde{F}_i=\{p_i\},\quad S=\{k_1\}\\
    C_iS &: \{k_1,k_2,k_3\}\to I_1=\emptyset, \quad I_2=\emptyset,\quad F_i=\{k_1,k_2\},\quad \tilde{F}_j=\{p_j\} \text{ for }i\neq j, \quad S=\{k_3\}  \\
    S C_i &: \{k_1,k_2,k_3\}\to I_1=\emptyset, \quad I_2=\emptyset,\quad F_i=\{k_2,k_3\},\quad \tilde{F}_j=\{p_j\} \text{ for }i\neq j, \quad S=\{k_1\}\, ,
  \end{aligned}
\end{equation}
for $\mathfrak{P}_{n_J+2}$, where $i,j>2$, and we left out regions obtained by the replacement $1\to 2$. Above, we replaced the trivial final state collinear sets $F_i$ with only one particle with the sets $\tilde{F}_i$ that contain the corresponding boosted and rescaled momentum, which agrees with the born jet momentum $p_i$. When specifying a limit of a resolution variable, we only put the non-trivial momenta into the arguments of the function, we write the functions in terms of the boosted and rescaled momenta $\tilde{k}_i$ for the final state collinear sectors, and we group particles in the same sector together into lists to improve readability if there are multiple sectors. For instance, we would write $\qft_{C_1}(\tilde{k}_1)$ or $\qft_{C_1C_i}(\{k_1\},\{\tilde{k}_2,\tilde{k}_3\})$, where the number of specified momenta also defines whether we are considering a region in $\mathfrak{P}_{n_J+1}$ or $\mathfrak{P}_{n_J+2}$. 

When taking zero bins of regions, we specify the corresponding approximation as
\begin{equation}
  \qft_{\pf_n,\pf_n^\prime,\dots}
\end{equation}
where $\qf$ is first approximated in the region $\pf_n$, then in the region $\pf_n^\prime$, and so on. We only specify the momenta that appear non-trivially in the last region as arguments of the function $\qft_{\pf_n,\pf_n^\prime,\dots}$.

For instance, the approximation $\qft_{C_iC_j, C_iS, S}(k_1, k_2)$, for $i,j>2$, could be obtained by starting from the region $C_iC_j$ with momenta $\{k_3,k_1,k_4,k_2\}$ with the approximation $\qft_{C_iC_j}(\{\tilde{k}_3,\tilde{k}_1\},\{\tilde{k}_4,\tilde{k}_2\})$ going over to the region $C_iS$ with non-trivial momenta $\{k_3,k_1,k_2\}$ and approximation $\qft_{C_iC_j,C_iS}(\{\tilde{k}_3,\tilde{k}_1\},\{k_2\})$ and finally going over to the region $S$ with non-trivial momenta $\{k_1,k_2\}$.

\section{\texorpdfstring{\boldmath Cumulant NLO Jet Functions in the $z_N$-Prescription\unboldmath}{Cumulant NLO Jet Functions with and without the zN-Prescription}}
\label{App_NLOJetFunctions}
In this section, we will calculate the NLO jet functions $\hat{\J}_{N, p}(\qfcut)$ and $\hat{\J}_{p}(\qfcut)$ for a quite generic definition of the resolution variable with and without the $z_N$-prescription respectively. We assume that the resolution variable in the collinear limit, where two momenta $k_1$ and $k_2$ are collinear to the hard jet $p$, approximates as 
\begin{equation}
  \label{NLOJet:qft}
  \qft_C(k_1, k_2) = Q \left(\frac{\lvert k_\perp \rvert}{Q}\right)^{1/b}z^{1-\frac{1}{b}}  \bar{z}^{1-\frac{1}{b}} g_1 g(\lvert \cos\phi \rvert)\, ,
\end{equation}
where $z$ and $\bar{z}$ are the longitudinal momentum fractions of $k_1$ and $k_2$ respectively, and $k_\perp$ is the transverse momentum of $k_1$ (and $-k_2$). $g(\lvert \cos\phi \rvert)$ encodes a possible azimuthal dependence of the resolution variable---we assume that the variable only depends on the azimuthal angle between $k_1$ and some direction $v_\perp$ that could, for instance, be specified by another jet or the beam direction. The variable only depends on $\lvert \cos\phi \rvert$ because of the symmetry between $k_1$ and $k_2$. The positive number $g_1$ is chosen such that 
\begin{equation}
  \int_0^\pi \frac{\rd \phi}{\pi} \log\left( g(\lvert \cos\phi \rvert) \right) = 0 \, .
\end{equation}  The approximation \eqref{NLOJet:qft} is consistent with a resolution variable with the scaling exponents $c=1$\footnote{One can always go from a variable $\qf$ with generic $c$ to a new variable $\qf^\prime =Q \left( \frac{\qf}{Q} \right)^c$ with $c=1$, where $Q$ is any IR-safe function of the momenta that reduces to the center of mass of the hard system for the leading order kinematics. The cumulant jet function $\hat{\J}_{N, p}$ for $\qf$ is then obtained from the cumulant jet function $\hat{\J}^\prime_{N, p}$ of $\qf^\prime$ as
\begin{equation}
  \hat{\J}_{N, p}(\qfcut)=\hat{\J}^\prime_{N, p}\left(Q\left( \frac{\qfcut}{Q} \right)^c\right)
\end{equation}}. For our calculation, we need the integrals
\begin{equation}
  \begin{aligned}
   I_{N, -1}&\equiv \int_0^1 \rd z \int_0^\infty \rd \lvert k_\perp\rvert  \lvert k_\perp\rvert^{-1-2\epsilon}\frac{1}{z_N}\theta\!\left(\qfcut- Q \left(\frac{\lvert k_\perp \rvert}{Q}\right)^{1/b}z^{1-\frac{1}{b}}  \bar{z}^{1-\frac{1}{b}} g_1 g(\lvert \cos\phi \rvert)  \right)\\
   I_n&\equiv\int_0^1 \rd z \int_0^\infty \rd \lvert k_\perp\rvert  \lvert k_\perp\rvert^{-1-2\epsilon}z^n\theta\!\left(\qfcut- Q \left(\frac{\lvert k_\perp \rvert}{Q}\right)^{1/b}z^{1-\frac{1}{b}}  \bar{z}^{1-\frac{1}{b}} g_1 g(\lvert \cos\phi \rvert)  \right)\, .
  \end{aligned}
\end{equation}
 The integral $I_n$ is trivial, and we find
\begin{equation}
  \begin{aligned}
    I_n &=-\left( Q \left( \frac{\qfcut}{Q g_1 g\left( \lvert \cos\phi \rvert \right)} \right)^b \right)^{-2 \epsilon } \frac{\Gamma (2 (b-1) \epsilon +1) \Gamma (n+2 (b-1) \epsilon +1)}{2 \epsilon  \Gamma (n+4 (b-1) \epsilon +2)}\\
    &=\left( Q \left( \frac{\qfcut}{Q g_1 g\left( \lvert \cos\phi \rvert \right)} \right)^b \right)^{-2 \epsilon }\begin{cases}
      \frac{1}{4 (1-b) \epsilon ^2}-\frac{1}{6} \pi ^2 (1-b)+\mathcal{O}(\epsilon) &\mytext{for }n=-1\\
      -\frac{1}{2 \epsilon }-2 (1-b) +\mathcal{O}(\epsilon) &\mytext{for }n=\phantom{-}0\\
      -\frac{1}{4 \epsilon }-(1-b)+\mathcal{O}(\epsilon) &\mytext{for }n=\phantom{-}1\\
      -\frac{1}{6 \epsilon }-\frac{13}{18}  (1-b)+\mathcal{O}(\epsilon) &\mytext{for }n=\phantom{-}2\, .
    \end{cases}
  \end{aligned}
\end{equation}
Note that $I_{-1}$ is not well-defined for $b=1$, where rapidity divergences need to be regularized. For the calculation of $I_{N, -1}$ we remember that 
\begin{equation}
  z_N=z+\frac{N^2 k_\perp^2}{4(p\cdot N)^2}\frac{1}{z}\, .
\end{equation}
We then find
\begin{equation}
\begin{aligned}
    I_{N, -1}&=\left(Q \left(\frac{\qfcut}{g_1 g(\lvert \cos\phi \rvert)Q}\right)^b\right)^{-2 \epsilon }\Bigg(\frac{1}{4 \epsilon ^2}+\frac{L_N}{4 \epsilon }\\
    &\quad-\frac{\pi^2 (5-4 (1-b)) (1-b)}{24 b}-\frac{(1-b) L_N^2}{8 b}+\mathcal{O}(\epsilon)\Bigg)\, ,
\end{aligned}\end{equation}
where we defined
\begin{equation}
  L_N\equiv \log \left(\frac{Q^2 N^2 \left(\frac{\qfcut}{g_1 Q g(\lvert \cos\phi \rvert )}\right)^{2 b}}{4 (p\cdot N)^2}\right)\, .
\end{equation}
Using these ingredients, we find that the quark jet function without the $z_N$-prescription becomes
\begin{equation}
  \begin{aligned}
    \hat{\J}_q^{(1)}&= \mu^{2\epsilon} \frac{ e^{\epsilon \gamma_E}}{\Gamma(1-\epsilon)}\int_0^1 \rd z \frac{\rd^{d-2}k_\perp}{\Omega_{d-2}}\frac{\hat{P}_{q\to gq}}{\lvert k_{\perp} \rvert^2}\theta\!\left( \qfcut- \qft_C(k_1, k_2)\right)\\
    &= \left(\frac{\mu}{Q} \left(\frac{g_1 Q}{\qfcut}\right)^b\right)^{2 \epsilon } \frac{C_Fe^{\epsilon \gamma_E}}{\Gamma(1-\epsilon)}\braket{g(\lvert \cos\phi \rvert)^{2\epsilon b}}\\
    &\quad\times\bigg(\frac{1}{2 (1-b) \epsilon ^2}+\frac{3}{4 \epsilon }- \left(\frac{\pi ^2}{3}-3\right) (1-b)+\frac{1}{4} +\mathcal{O}(\epsilon)\bigg)\, ,
  \end{aligned}
\end{equation}
where we defined 
\begin{equation}
 \begin{aligned}
   \braket{g(\lvert \cos\phi \rvert)^{2\epsilon b}}&\equiv \int \frac{\rd \Omega}{\Omega_{2-2\epsilon}}g(\lvert \cos\phi \rvert)^{2\epsilon b}\\
   &= \frac{\Gamma(1-\epsilon)^2}{\Gamma(1-2\epsilon)}\int_0^{\pi}\frac{\rd \phi}{\pi}\left( 2\sin\phi  \rvert \right)^{-2\epsilon}g(\lvert \cos\phi \rvert)^{2\epsilon b}\\
   &\equiv 1+2\epsilon^2 b^2\tilde{g}_2+\mathcal{O}(\epsilon^3)\, ,
 \end{aligned}
\end{equation}
where 
\begin{equation}
  \tilde{g}_2=\int_0^\pi\frac{\rd \phi}{\pi}\log\left( g(\lvert\cos\phi\rvert) \right)\log\left( \frac{g(\lvert\cos\phi\rvert)}{(4\sin^2\phi)^{\frac{1}{b}}} \right)\, .
\end{equation}
The quark jet function with the $z_N$-prescription becomes
\begin{equation}
  \begin{aligned}
    \hat{\J}_{N, q}^{(1)}&= \mu^{2\epsilon} \frac{e^{\epsilon \gamma_E}}{\Gamma(1-\epsilon)}\int_0^1 \rd z \frac{\rd^{d-2}k_\perp}{\Omega_{d-2}}\frac{\hat{P}_{N, q\to gq}}{\lvert k_{\perp} \rvert^2}\theta\!\left( \qfcut- \qft_C(k_1, k_2)\right)\\
    &= \left(\frac{\mu}{Q} \left(\frac{g_1 Q}{\qfcut}\right)^b\right)^{2 \epsilon } \frac{C_F e^{\epsilon \gamma_E}}{\Gamma(1-\epsilon)}\Bigg(\frac{1}{2 \epsilon ^2}+\frac{\frac{1}{2} L_N+\frac{3}{4}}{\epsilon } \\
    &\quad-\frac{(1-b) L_N^2}{4 b}  -b (g_2+3)+\frac{1}{12} \pi ^2 \left(4 b-\frac{1}{b}-3\right)+\frac{13}{4} + \mathcal{O}(\epsilon)\Bigg)\, ,
  \end{aligned}
\end{equation}
where we defined
\begin{equation}
  g_2=\int_0^\pi\frac{\rd \phi}{\pi}\log^2\left( g(\lvert\cos\phi\rvert) \right)\, .
\end{equation}
Before giving the analogous results for the gluon jet functions, we remind the reader that the gluon jet functions are operators in the spin space of the parent gluon. Instead of writing the jet function as an operator in the $d-2$-dimensional space of physical polarizations, it is more convenient to write it as an operator in Lorentz space. We write 
\begin{equation}
  \hat{\J}_g^{\mu\nu}=\braket{\mu|  \hat{\J}_g|\nu}\, ,
\end{equation}
where $\mu$ and $\nu$ are Lorentz indices of the parent gluon in the hard (conjugate) amplitudes $\bra{M}$ and $\ket{M}$, respectively.

The gluon jet function without the $z_N$-prescription then becomes
\begin{equation}
  \label{App_NLOJetFunctions:Jg}
  \begin{aligned}
    \hat{\J}_g^{\mu\nu, (1)}&= \mu^{2\epsilon} \frac{e^{\epsilon \gamma_E}}{\Gamma(1-\epsilon)}\int_0^1 \rd z \frac{\rd^{d-2}k_\perp}{\Omega_{d-2}}\frac{\left( \frac{1}{2}\tilde{P}^{\mu\nu}_{g\to gg} +n_f\tilde{P}^{\mu\nu}_{g\to q\bar{q}} \right)}{\lvert k_{\perp} \rvert^2}\theta\!\left( \qfcut-\qft_C(k_1, k_2)\right)\\
    &= \left(\frac{\mu}{Q} \left(\frac{g_1 Q}{\qfcut}\right)^b\right)^{2 \epsilon }\frac{e^{\epsilon \gamma_E}}{\Gamma(1-\epsilon)}\Biggl[ d^{\mu\nu}(p, \bar{n})\Biggl(\frac{C_A}{2 (1-b) \epsilon ^2} +\frac{\beta_0}{ \epsilon }+\frac{n_f T_R}{6} +\frac{b^2 C_A \tilde{g}_2}{1-b}\\
    &\quad -\frac{1}{18} (1-b) \left[\left(6 \pi ^2-67\right) C_A+26 n_f T_R\right] \Biggr)-\Amunu\frac{1}{3}b g_c \left( C_A-2n_f T_R \right)+\mathcal{O}(\epsilon)\Biggr]\, ,
  \end{aligned}
\end{equation}
where 
\begin{equation}
  g_c=\int_0^\pi\frac{\rd \phi}{\pi}\log\left( g(\lvert\cos\phi\rvert) \right)\cos^2\phi
\end{equation}
and
\begin{equation}
  \Amunu=2 v_\perp^\mu v_\perp^\nu -d^{\mu\nu}(p, \bar{n})\, .
\end{equation}
For the $z_N$-prescription jet function, we find
\begin{equation}
  \label{App_NLOJetFunctions:JgzN}
  \begin{aligned}
    \hat{\J}_{N, g}^{\mu\nu, (1)}&= \mu^{2\epsilon} \frac{e^{\epsilon \gamma_E}}{\Gamma(1-\epsilon)}\int_0^1 \rd z \frac{\rd^{d-2}k_\perp}{\Omega_{d-2}}\frac{\left( \frac{1}{2}\tilde{P}^{\mu\nu}_{N, g\to gg} +n_f\tilde{P}^{\mu\nu}_{N, g\to q\bar{q}} \right)}{\lvert k_{\perp} \rvert^2}\theta\!\left( \qfcut-\qft_C(k_1, k_2)\right)\\
    &=  \left(\frac{\mu}{Q} \left(\frac{g_1 Q}{\qfcut}\right)^b\right)^{2 \epsilon }\frac{e^{\epsilon \gamma_E}}{\Gamma(1-\epsilon)}\Biggl[ d^{\mu\nu}(p, \bar{n})\Biggl( \frac{C_A}{2 \epsilon ^2} + \frac{ C_A L_N}{2\epsilon }+ \frac{ \beta_0}{\epsilon }\\
    &+\frac{b-1}{4 b} C_A L_N^2 +C_A \left(-b g_2+\frac{1}{12} \pi ^2 \left(4 b-\frac{1}{b}-3\right)+\frac{67 (1-b)}{18}\right)\\
    &+\left(\frac{1}{6}-\frac{13 (1-b)}{9}\right) n_fT_R \Biggr)-\Amunu\frac{1}{3}b g_c \left( C_A-2n_f T_R \right)\Biggr]\, .
  \end{aligned}
\end{equation}

Following the example for the definition of the QCD beam functions in~\cite{Catani:2022sgr},  we used the notation
\begin{equation}
  \tilde{P}_{(N), g\to *}^{\mu\nu}\equiv d^\mu{}_{\mu^\prime}(p, \bar{n}) \hat{P}_{(N), g\to *}^{{\mu^\prime}{\nu^\prime}}d_{\nu^\prime}{}^\nu (p, \bar{n})
\end{equation}
in \eqref{App_NLOJetFunctions:Jg} and \eqref{App_NLOJetFunctions:JgzN}, 
where 
\begin{equation}
  d^{\mu\nu}(p, \bar{n})=-g^{\mu \nu}+\frac{p^\mu \bar{n}^\nu+\bar{n}^\mu p^\nu}{\bar{n}\cdot p}=-g^{\mu \nu}+\frac{p^\mu n^\nu+N^\mu p^\nu}{N \cdot p}-\frac{N^2 p^\mu p^\nu}{(N\cdot p)^2}\, .
\end{equation}
The reason why we use $ \tilde{P}_{(N), g\to *}^{\mu\nu}$ instead of $ \hat{P}_{(N), g\to *}^{\mu\nu}$ in the definition of the jet function is to get rid of unphysical longitudinal polarizations in the jet function. This should not matter if the hard function $\Hb$ is properly defined such that 
\begin{equation}
  \Hb^{\mu\nu}=\braket{\nu|\Hb|\mu}\, 
\end{equation}
satisfies
\begin{equation}
  p^\mu \Hb_{\mu\nu}=p^\nu \Hb_{\mu\nu}=0\, , 
\end{equation}
but can help to avoid mistakes, in particular in cases where the hard function contains multiple gluons that split collinearly in different collinear sectors. 

Finally, we can also summarize that the difference between the jet functions with and without the $z_N$-prescription is  
\begin{equation}
\begin{aligned}
    \hat{\J}_{N, a}^{ss', (1)}-\hat{\J}_{a}^{ss', (1)}&=\delta^{ss'}\left( \frac{\mu g_0}{\qfcut} \right)^{2\epsilon}C_a \frac{e^{\epsilon \gamma_E}}{\Gamma(1-\epsilon)}\left(\frac{(2 p\cdot N)^2}{Q^2 N^2}\right)^{-\frac{(1-b) \epsilon }{b}}\\
    &\quad\quad\times\Bigg[-\frac{b}{1-b}\frac{1}{2\epsilon^2} -\frac{1-b}{b}\frac{\pi^2}{12}
   -b\left( g_2+\frac{b}{1-b}\tilde{g}_2 \right)+\mathcal{O}(\epsilon)\Bigg]\, .
\end{aligned}\end{equation}

To conclude this section, we point out that although the parametrization \eqref{NLOJet:qft} is not completely general, one can easily also obtain more general jet functions by taking the difference between the jet function one wants to calculate and a jet function in the class \eqref{NLOJet:qft} with the same $b$. For example, for $\ktness$ definitions that contain a minimum in $M_{ij}$ in \eqref{eq_ktn_distances} the collinear approximation of $\ktness$ reads 
\begin{equation}
  \qft_C(k_1, k_2) = \frac{\lvert k_\perp \rvert }{\max (z, 1-z)}
\end{equation}
while the approximation for $\ktness$ with any other $M_{ij}$ choice in \eqref{eq_ktn_distances} is 
\begin{equation}
  \qft_C(k_1, k_2) = \lvert k_\perp \rvert \, .
\end{equation} 
The difference this induces on our ingredients $I_n$ is 
\begin{align}
    \Delta I_n &\equiv\int_0^1 \rd z \int_0^\infty \rd \lvert k_\perp\rvert  \lvert k_\perp\rvert^{-1-2\epsilon}z^n\left[ \theta\!\left(\qfcut-\frac{\lvert k_\perp \rvert }{\max (z, 1-z)}\right) - \theta\!\left(\qfcut-\lvert k_\perp \rvert \right)\right]\notag\\
    &=\qfcut^{-2\epsilon}\int_0^1 \rd z z^n\log\left( \max\left( {z, 1-z} \right) \right)+\mathcal{O}(\epsilon)\notag\\
    &=\qfcut^{-2\epsilon}\begin{cases}
      -\frac{\pi ^2}{12} +\mathcal{O}(\epsilon) &\mytext{for }n=-1\\
      \log (2)-1 +\mathcal{O}(\epsilon)&\mytext{for }n=0\\
      \frac{1}{2} (\log (2)-1) +\mathcal{O}(\epsilon)&\mytext{for }n=1\\
      \frac{1}{72} (24 \log (2)-23)+\mathcal{O}(\epsilon) &\mytext{for }n=2\, .
    \end{cases}
\end{align}
Note also that for $n=-1$, the difference is the same, whether one uses the $z_N$-prescription or not, because the two approximations for the different $\ktness$ definitions agree in the collinear-soft limit.

\section{\texorpdfstring{\boldmath Approximations of the $y_{23}$ Resolution Variable \unboldmath}{Approximations of the y23 Resolution Variable}}
\label{App_Approximationsy23}
In this appendix, we give the explicit expressions for the approximations of $y_{23}$ in dijet production at lepton colliders. In this paper, we define $y_{23}$ as $2-\ktness$ with E-scheme and distances
\begin{equation}
  M_{ij}=\frac{E_i E_j}{E_i+E_j}, \quad R_{ij}^2=2(1-\cos\theta_{ij})\, .
\end{equation}
The relevant approximations for the resolution variable at NLO are
  \begin{equation}
    \begin{aligned}
      \qft_{C_i}(\tilde{k}_1, \tilde{k}_2)&=d_{c, 12}=\lvert \tilde{k}_{1, \perp}\rvert=\lvert \tilde{k}_{2, \perp}\rvert\\
      \qft_{C_i, S}(k)&=\lvert k_{\perp}\rvert\\
      \qft_{S}(k)&=E_k\sqrt{2\left( 1-\lvert \cos\theta \rvert \right)}\, ,
    \end{aligned}
  \end{equation}
   where $\theta$ is the angle between the momentum of the soft gluon and the jet with momentum $p_3$. We point out that $d_{c, 12}=\lvert \tilde{k}_{1, \perp}\rvert$ only holds if $k_1$ and $k_2$ are the only collinear momenta in the sector, more generally, \eqref{Fac:eq_dijc} needs to be used.  At NNLO, we also need the approximations
  \begin{equation}
    \begin{aligned}
     \qft_{C_iC_j}(\{\tilde{k}_1, \tilde{k}_2\}, \{\tilde{k}_3, \tilde{k}_4\})&=\max\left( \qft_{C_i}(\tilde{k}_1, \tilde{k}_2), \qft_{C_j}(\tilde{k}_3, \tilde{k}_4) \right)\\
     \qft_{C_i}(\tilde{k}_1, \tilde{k}_2, \tilde{k}_3)&=\theta\!\left( \min\left( d_{c, 13}, d_{c, 23} \right)-d_{c, 12} \right)d_{c, (1+2)3}+(\mathrm{perm.})\\
     \qft_{C_iS}(\{\tilde{k}_1, \tilde{k}_2\}, \tilde{k}_3)&= \max\left(\qft_{C_i}(\tilde{k}_1, \tilde{k}_2), \qft_S(k_3)\right)\\
     &\hspace{-3.5cm} +\theta\!\left(\pm\cos\theta-\max\left( 1-\frac{\lvert \tilde{k}_{1, \perp}\rvert^2}{2E_3^2}, 0 \right) \right)\left[ \theta\!\left( \tilde{d}_{13}-\tilde{d}_{23} \right)\left( d_{c, 1(2+3)}-\lvert\tilde{k}_{1, \perp}\rvert \right)+ \left( 1\leftrightarrow 2 \right)\right]
    \end{aligned}\end{equation}
     \begin{align}
     \qft_{C_iC_j, C_iS}(\{\tilde{k}_1, \tilde{k}_2\}, k_3)&=\max\left( \qft_{C_i}(\tilde{k}_1, \tilde{k}_2), \lvert k_{\perp 3}\rvert \right)\notag\\
     \qft_{C_i, C_iS}(\{\tilde{k}_1, \tilde{k}_2\}, k_3)&=\max \left(\qft_{C_i}(\tilde{k}_1, \tilde{k}_2), \lvert k_{3, \perp}\rvert  \right)\notag\\
    &+\theta\!\left( \lvert \tilde{k}_{1, \perp}\rvert -\lvert k_{3, \perp}\rvert\right)\left[ \theta(\tilde{d}_{c, 13}-\tilde{d}_{c, 23} )\left( d_{c, (2+3)1} -\lvert \tilde{k}_{1, \perp}\rvert \right)+\left( 1\leftrightarrow 2 \right)\right]\notag\\
    \qft_S(k_1, k_2)&=\max(\qft_{S}(k_1), \qft_{S}(k_2))\notag\\
    &\hspace{-1.5cm} + \theta\!\left( \min\left(  \qft_S(k_1), \qft_S(k_2)\right)-d_{12}\right)\left[\qft_S(k_{12})-\max\left( \qft_S(k_1), \qft_S(k_2) \right) \right]\notag\\
    \qft_{C_i, S}(k_2, k_3)&=\theta(\min(\lvert k_{2, \perp}\rvert, \lvert k_{3\perp}\rvert)-d_{c, 23})\lvert k_{\perp, 2} +k_{\perp, 3s}  \rvert\notag\\
        &+\theta(d_{c, 23}-\min(\lvert k_{2, \perp}\rvert, \lvert k_{3\perp}\rvert)) \max(\lvert k_{2, \perp}\rvert, \lvert k_{3\perp}\rvert)\notag\\
    \qft_{C_iS, S}(k_2, k_3)&=\max\left(k_{\perp, 2}, \qft_S(k_3)\right)\notag\\
    &\hspace{-1.5cm}+\theta\!\left(\pm\cos\theta-\max\left( 1-\frac{\lvert k_{2, \perp}\rvert^2}{2E_3^2}, 0 \right) \right) \theta\!\left(\tilde{ \tilde{d}}_{13}-\tilde{\tilde{d}}_{23} \right)\left( \lvert k_{2, \perp}+k_{3, \perp}\rvert -\lvert k_{2, \perp}\rvert \right)\notag\\
    \qft_{C_iC_j, C_iS, S}(k_2, k_3)&=\qft_{C_iC_j, SC_j, S}(k_2, k_3)=\max\left( \lvert k_{\perp 2}\rvert, \lvert k_{\perp 3}\rvert \right)\notag\\
    \qft_{C_i, C_iS, S}(k_2, k_3)&=\qft_{C_i, SC_i, S}(k_3, k_2)=\max\left( \lvert k_{\perp 2}\rvert, \lvert k_{\perp 3}\rvert \right)+\notag\\
    &+\theta\!\left( \lvert k_{2, \perp}\rvert -\lvert k_{3, \perp}\rvert\right)\theta(\tilde{\tilde{d}}_{c, 13}-\tilde{\tilde{d}}_{c, 23} )\left( \lvert k_{2, \perp}+k_{3, \perp}\rvert -\lvert k_{2, \perp}\rvert \right)\, ,
  \end{align}
  where the `$\pm$ 'in the fourth line is a `$+$' for $C_3$ and a `$-$' for $C_4$.  Furthermore, we have
  \begin{equation}
   \begin{aligned}
    \theta\!\left( \tilde{d}_{13}-\tilde{d}_{23} \right)&= \theta\!\left( \left(z_1-z_2\right) \left(k_{3, -}^2-k_{3, \perp}\cdot k_{3, \perp}\right)+k_{3, \perp}\cdot \tilde{k}_{1, \perp} \right)\\
    &=\theta\!\left( \left(e^{-2 y_3}+1\right) \left(z_1-z_2\right)+\cos\phi\frac{ \lvert \tilde{k}_{\perp 1}\rvert}{\lvert \tilde{k}_{\perp 3}\rvert} \right)\\
      \theta(\tilde{d}_{c, 13}-\tilde{d}_{c, 23} )&=\theta\!\left( \tilde{k}_{\perp 1}\cdot k_{\perp3}+\left( z_1-z_2 \right)\lvert k_{\perp3} \rvert^2\right)=\theta\!\left(\left( z_1-z_2 \right)+\cos\phi\frac{\lvert \tilde{k}_{\perp1} \rvert}{\lvert k_{\perp3} \rvert}\right)\\
      \theta(\tilde{\tilde{d}}_{c, 13}-\tilde{\tilde{d}}_{c, 23} )&=\theta\!\left(1+\cos\phi\frac{ \lvert k_{\perp 2}\rvert}{\lvert k_{\perp 3}\rvert}\right)\\
      \theta\!\left(\tilde{ \tilde{d}}_{13}-\tilde{\tilde{d}}_{23} \right)&=\theta\!\left(1+e^{-2 y_3}+\cos\phi\frac{ \lvert k_{\perp 2}\rvert}{\lvert k_{\perp 3}\rvert}\right)\\
     d_{c, 1(2+3)}&=\sqrt{\lvert \tilde{k}_{\perp 1}\rvert^2+2 \cos\phi \lvert \tilde{k}_{\perp 1}\rvert \lvert k_{\perp 3}\rvert z_1+\lvert k_{\perp 3}\rvert^2 z_1^2}\\
     d_{c, 2(1+3)}&=\sqrt{\lvert \tilde{k}_{\perp 1}\rvert^2-2 \cos\phi \lvert \tilde{k}_{\perp 1}\rvert \lvert k_{\perp 3}\rvert z_2+\lvert k_{\perp 3}\rvert^2 z_2^2}\, ,
   \end{aligned}
  \end{equation}
  where $\phi$ is the azimuthal angle between $ \tilde{k}_{\perp2}$ and $k_{\perp3}$ ($\pi-\phi$ is the azimuthal angle between $ \tilde{k}_{\perp1}$ and $k_{\perp3}$) on the soft-collinear phase space and the angle between $ k_{\perp2}$ and $k_{\perp3}$ on the double-soft phase space. The definition of $d_{c,ij}$ was provided in \eqref{Fac:eq_dijc}.
\end{document}